\pdfoutput=1
\documentclass[12pt,preprint]{aastex}
\usepackage{graphicx}        

\shorttitle{Abrupt changes in photospheric magnetic and Lorentz-force vectors during neutral-line flares}
\shortauthors{Petrie}

\begin{document}


\title{The abrupt changes in the photospheric magnetic and Lorentz-force vectors during six major neutral-line flares}


\author{G.J.D.~Petrie}
\affil{National Solar Observatory, 950 N. Cherry Avenue, Tucson, AZ 85719}



\begin{abstract}
We analyze the spatial and temporal variations of the abrupt photospheric magnetic changes associated with six major flares using 12-minute, $0.\!\!^{\prime\prime}5$~pixel$^{-1}$ vector magnetograms from NASA's Helioseismic and Magnetic Imager instrument (HMI) on the Solar Dynamics Observatory satellite. The six major flares occurred near the main magnetic neutral lines of four active regions, NOAA 11158, 11166, 11283 and 11429. During all six flares the neutral line field vectors became stronger and more horizontal, in each case almost entirely due to strengthening of the horizontal field components parallel to the neutral line. In all six cases the neutral line pre-flare fields were more vertical than the reference potential fields, and collapsed abruptly and permanently closer to potential field tilt angles during every flare, implying that the relaxation of magnetic stress associated with non-potential tilt angles plays a major role during major flares. The shear angle with respect to the reference potential field did not show such a pattern, demonstrating that flare processes do not generally relieve magnetic stresses associated with photospheric magnetic shear. The horizontal fields became significantly and permanently more aligned with the neutral line during the four largest flares, suggesting that the collapsing field is on average more aligned with the neutral line than the pre-flare neutral line field. The vertical Lorentz force had a large, abrupt, permanent downward change during each of the flares, consistent with loop collapse. The horizontal Lorentz force changes acted mostly parallel to the neutral line in opposite directions on each side, a signature of the fields contracting during the flare, pulling the two sides of the neutral line towards each other. The greater effect of the flares on field tilt than on shear may be explained by photospheric line-tying.
\end{abstract}



\section{Introduction}
\label{s:introduction} 

Solar Cycle 24 has produced several major flares since NASA's Helioseismic and Magnetic Imager (HMI) instrument (Schou et al.~2011) on NASA's Solar Dynamics Observatory (SDO) satellite (Pesnell et al.~2012) began observing continuously in March 2010. HMI vector magnetogram series covering some of these major flares have been released to the community since late 2011\footnote{http://jsoc.stanford.edu/jsocwiki/ReleaseNotes2}. SDO/HMI produces full-disk vector magnetograms with $0.\!\!^{\prime\prime}5$ pixels every 12 minutes. From filtergrams in six polarization states at six wavelengths on the Fe~{\sc I} 617.3~nm spectral line, images for the Stokes parameters, $I$, $Q$, $U$ and $V$ are derived, which are inverted for the magnetic vector components by the Very Fast Inversion of the Stokes Algorithm (VFISV) code (Borrero et al.~2010). The $180^{\circ}$ azimuthal field ambiguity is resolved using the ``minimum energy'' method (Metcalf~1994, Leka et al.~2009). In this paper we analyze six 12-hour series of vector magnetograms covering six major flares occurring near the main magnetic neutral lines of four active regions, NOAA 11158, 11166, 11283 and 11429. We describe the abrupt and permanent field changes that occurred during each flare and characterize the associated Lorentz force vector changes near the main neutral line of the region and within the neighboring sunspots. Here a change is deemed ``permanent'' if its effects last until at least several hours after the flare.

Abrupt photospheric field changes have been observationally linked to flares in the past two decades; see the discussions in Sudol and Harvey~(2005) and Wang~(2006).  Wang~(2006) found an unshearing movement parallel to the neutral lines in flare-related longitudinal magnetic field changes in all five $\delta$-spot flares that he studied, implying an overall release of shear, but that the two polarities converged towards the neutral line during some events and diverged during others. Wang and Liu~(2010) studied 11 X-class flares for which vector magnetograms were available, and found in each case an increase of transverse field at the polarity inversion line. Wang et al.~(2012), Sun et al.~(2012) and Petrie~(2012) analyzed the HMI vector data for the 2011 February 15 X2.2 flare, and found similar behavior, as did Liu et al.~(2012) for the 2011 February 13 M6.6 flare. The HMI vector data for these two major flares from AR~11158 have already been studied in several papers using a variety of methods (Wang et al.~2012, Gosain~2012, Sun et al.~2012, Liu et al.~2012, Petrie~2012, Jing et al.~2012). Sun et al.~(2012) calculated nonlinear force-free field models for the coronal field from the HMI vector measurements and argued that the increase in magnetic shear observed at the photosphere is localized at low heights and the shear decreases above a certain height in the corona (see also Jing et al.~2008). Petrie~(2012) found an increase in strength of the field vector at the neutral line at the time of the flare, particularly its horizontal component parallel to the neutral line, accompanied by a large, abrupt, downward vertical Lorentz force change and a horizontal Lorentz force change acting in opposite directions on each side of neutral line, with the two sunspots at each end subject to abrupt torsional un-twisting forces. The downward and un-shearing forces were consistent with a collapse and contraction of fields near the neutral line. These observations support the coronal implosion interpretation (Hudson 2000, Hudson, Fisher and Welsch 2008, Fisher et al.~2012) where, after a coronal magnetic eruption, the remaining coronal field contracts downward resulting in the field become more horizontal at the photospheric level. Petrie and Sudol~(2010) analyzed one-minute GONG longitudinal magnetograms covering 77 flares of GOES class at least M5 and, exploring the relationship between increasing/decreasing longitudinal fields and azimuth and tilt angles at various positions on the disk, found that the overall distributions of longitudinal increases and decreases at different parts of the disk was found to be consistent with Hudson, Fisher and Welsch's~(2008) loop-collapse scenario. Fletcher and Hudson's~(2008) physical description of flaring field changes remains the only detailed explanation of how a coronal event could cause permanent change in the photospheric field.

The goal of this paper is to use the high-cadence HMI vector data covering six major flares to extend and clarify the above results. For example, if the transverse field component generally increases near neutral lines during flares, do the magnetic shear changes also follow a general pattern, and, if not, why not? We expect a flare to involve the relaxation of magnetic stresses built up during the preceding hours and days. Can the photospheric field measurements shed light on how magnetic stresses are relieved during flares?

The paper is organized as follows. In Section~\ref{s:magch} we will present the vector fields observed by HMI before and after the main flare-related field changes took place, discussing the differences between these vector fields in each spatial dimension and plotting the vector field evolution in time. In Section~\ref{s:potential} we will discuss the magnetic changes with reference to potential fields. Section~\ref{s:current} will describe the associated electric current changes that occurred during the flares. We will derive the accompanying Lorentz force changes in Section~\ref{s:lorentzfch}. We will conclude in Section~\ref{s:conclusion}.

\section{The magnetic field vector changes}
\label{s:magch}

\begin{figure} 
\begin{center}
\resizebox{0.49\textwidth}{!}{\includegraphics*{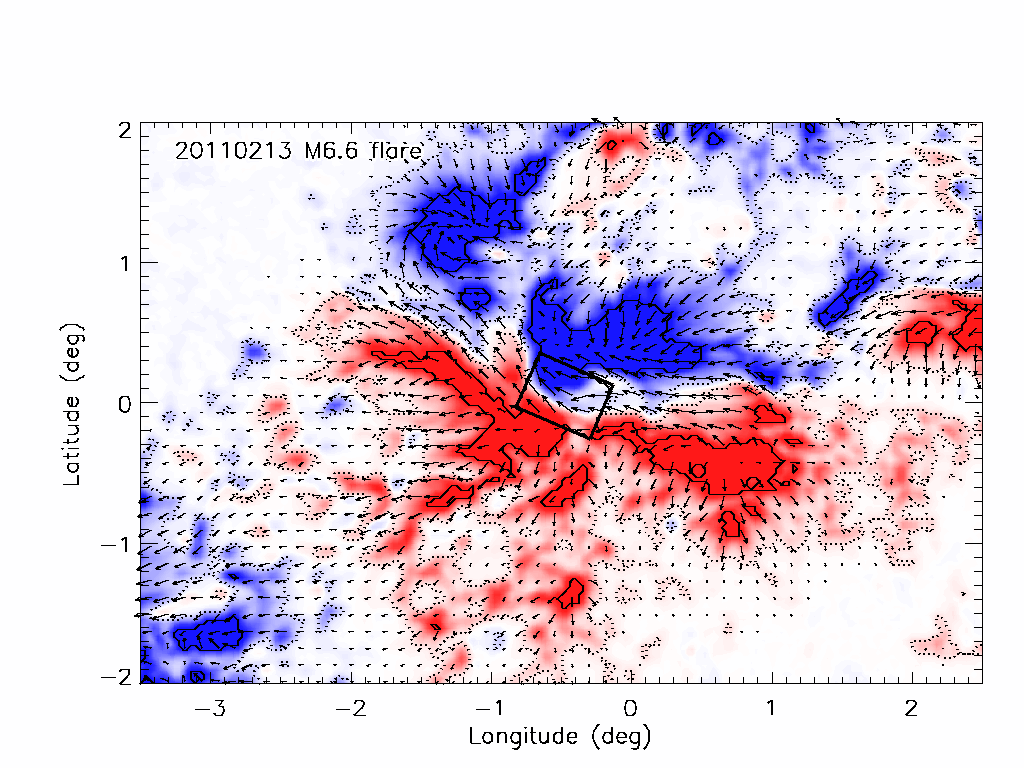}}
\resizebox{0.49\textwidth}{!}{\includegraphics*{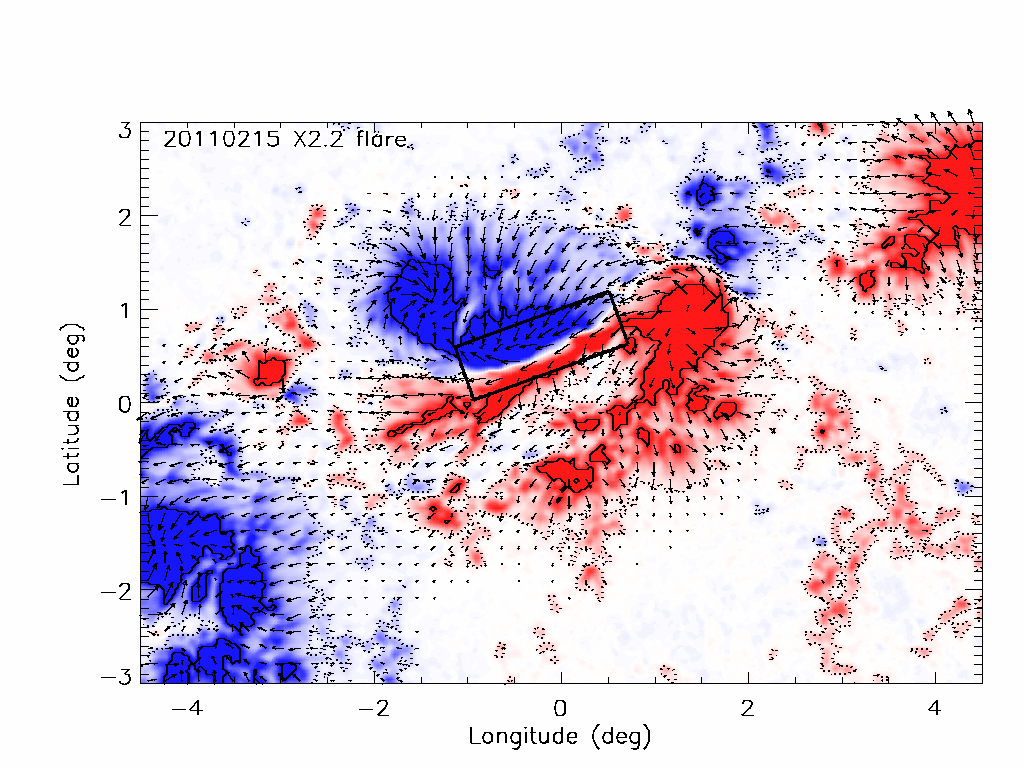}}
\resizebox{0.49\textwidth}{!}{\includegraphics*{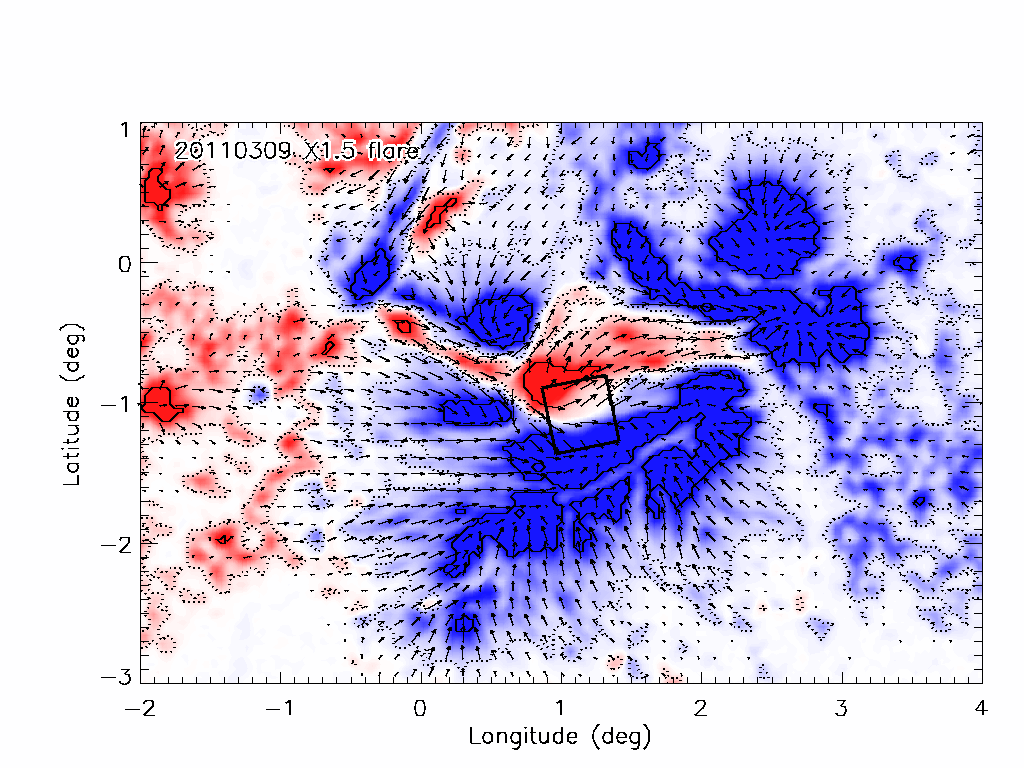}}
\resizebox{0.49\textwidth}{!}{\includegraphics*{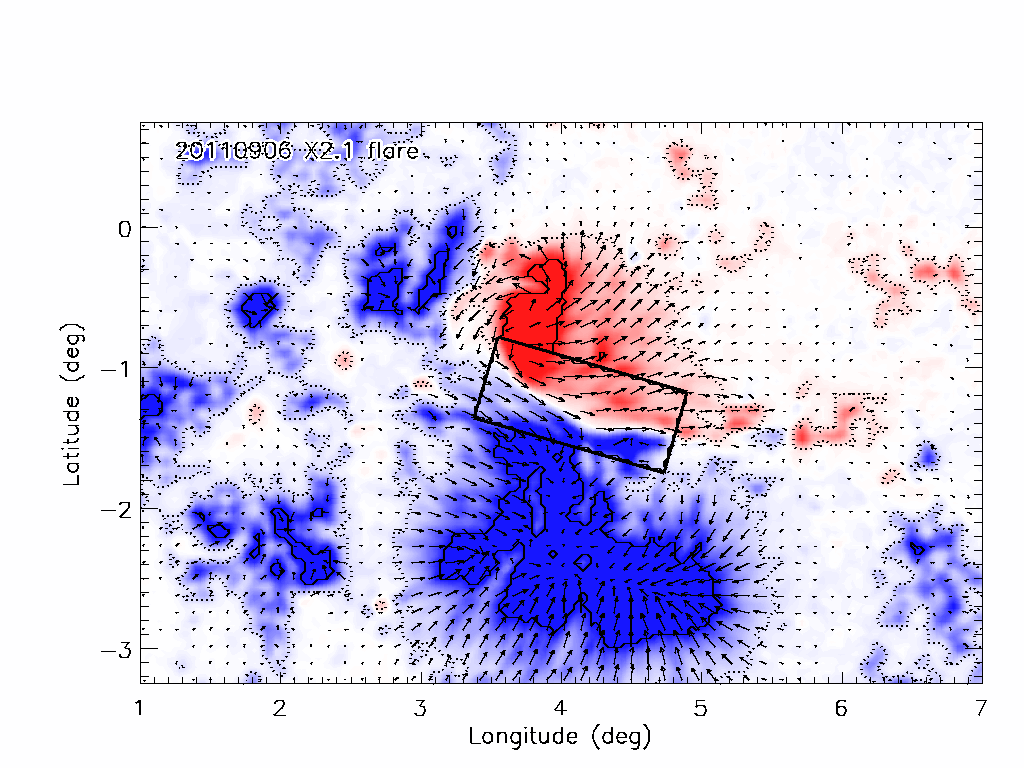}}
\resizebox{0.49\textwidth}{!}{\includegraphics*{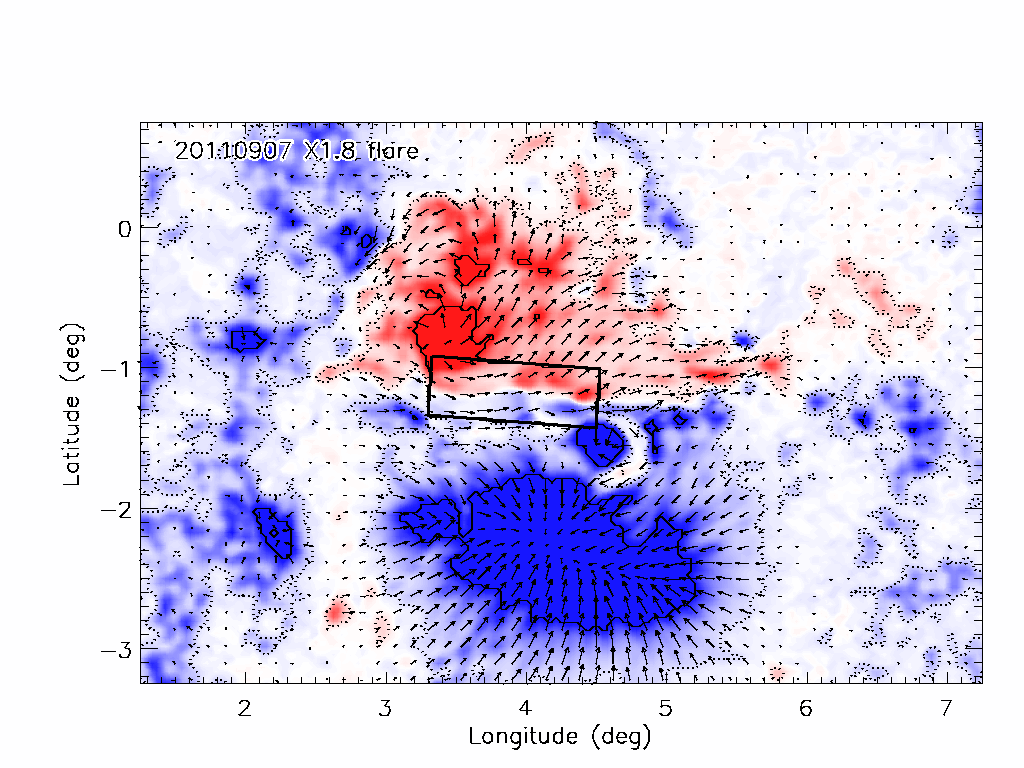}}
\resizebox{0.49\textwidth}{!}{\includegraphics*{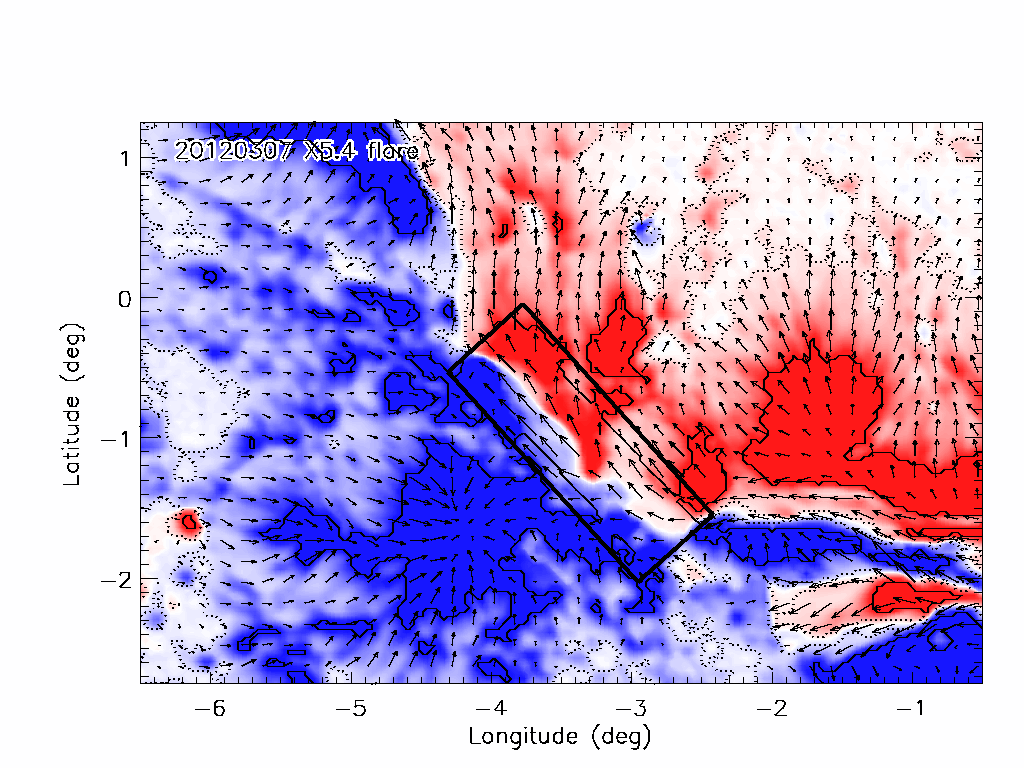}}
\end{center}
\caption{The vector magnetic field before each of the six flares. The vertical field component, $B_r$, is indicated by the color scale and the horizontal component by the arrows, with saturation values $\pm 1000$~G. Red/blue coloring represents positive/negative vertical field. The black rectangles mark the regions of major field change near the neutral lines that are used in subsequent analysis. The solid and dotted contours indicate strong ($\mathrm{|}B_r\mathrm{|}>1000$~G) and quite strong  ($\mathrm{|}B_r\mathrm{|}>100$~G) fields, respectively. }
\label{fig:br}
\end{figure}

\begin{figure} 
\begin{center}
\resizebox{0.49\textwidth}{!}{\includegraphics*{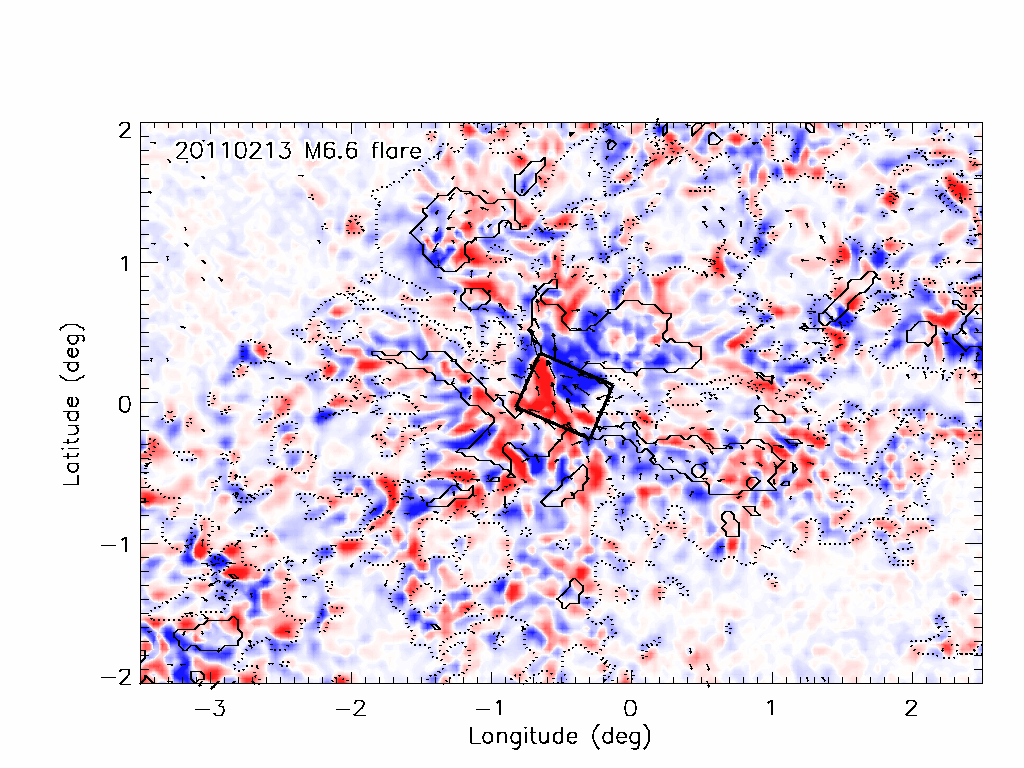}}
\resizebox{0.49\textwidth}{!}{\includegraphics*{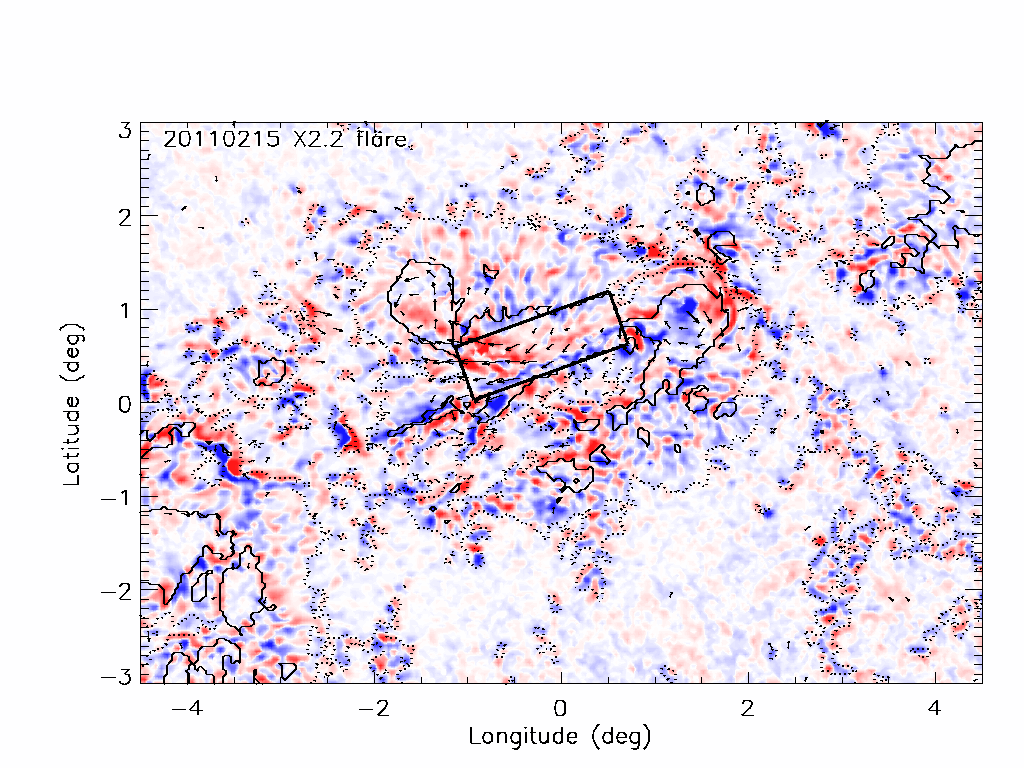}}
\resizebox{0.49\textwidth}{!}{\includegraphics*{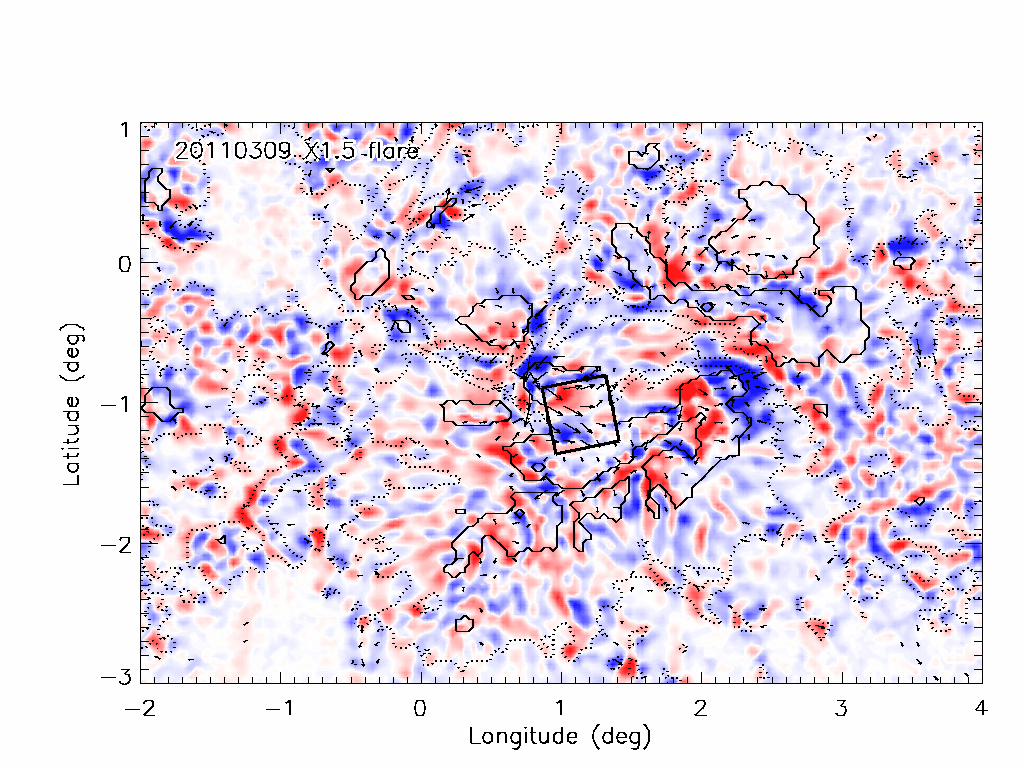}}
\resizebox{0.49\textwidth}{!}{\includegraphics*{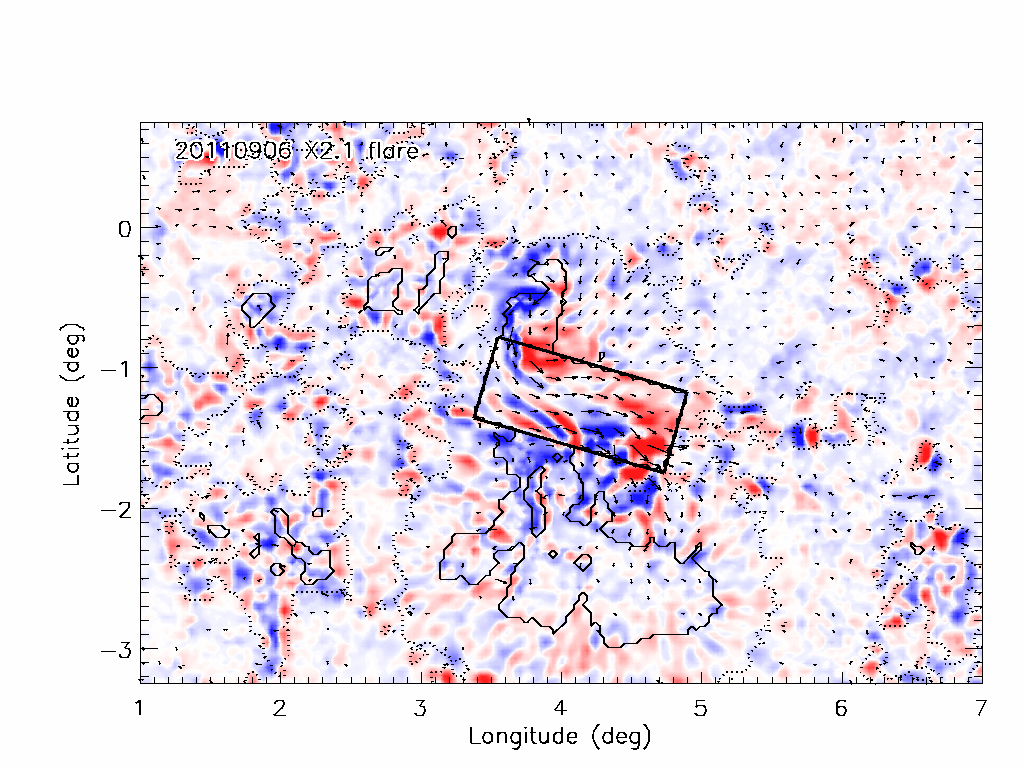}}
\resizebox{0.49\textwidth}{!}{\includegraphics*{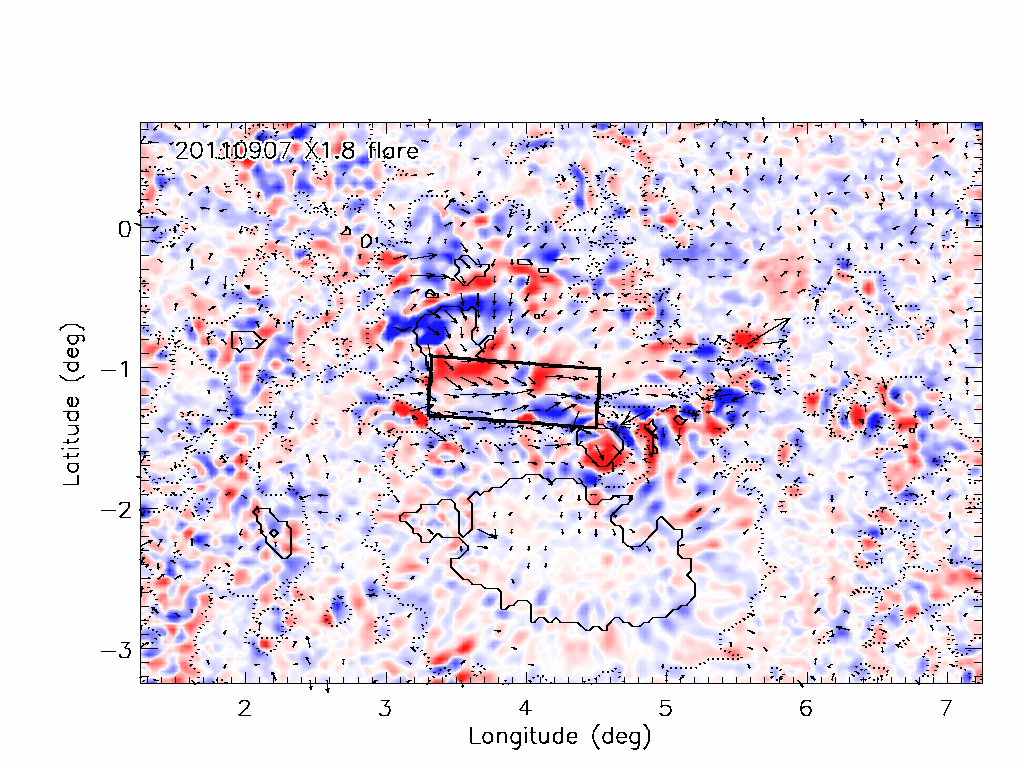}}
\resizebox{0.49\textwidth}{!}{\includegraphics*{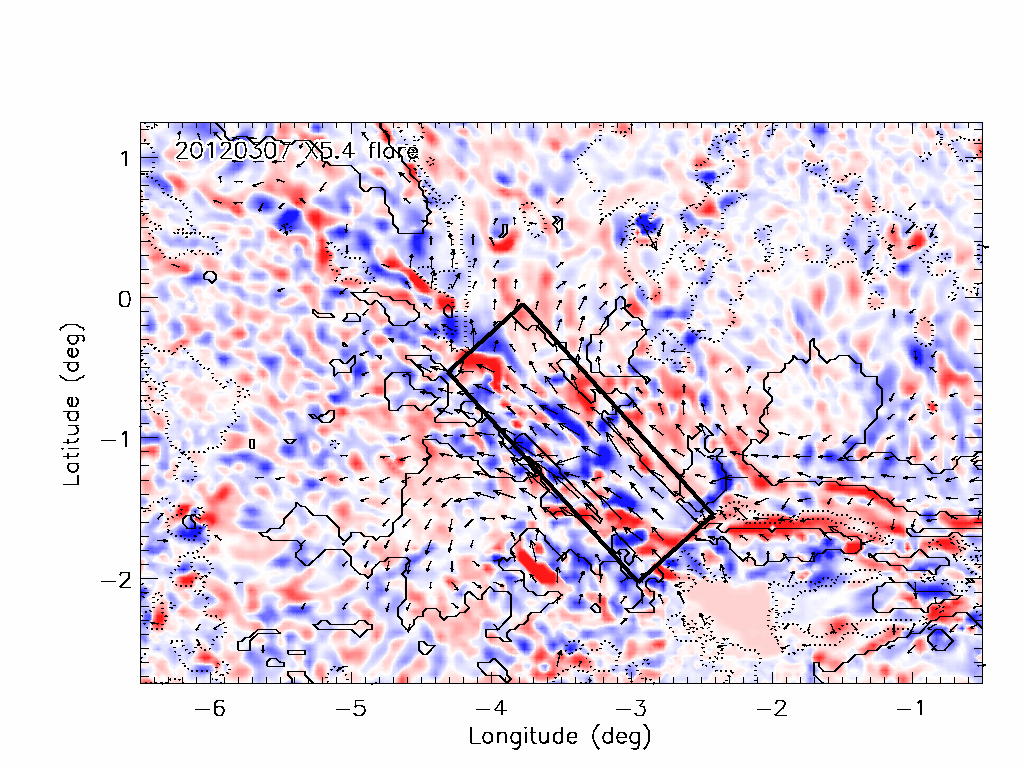}}
\end{center}
\caption{The vector magnetic field changes during each of the six flares. The vertical field change component, $\delta B_r$, is indicated by the color scale and the horizontal component by the arrows, with saturation values $\pm 300$~G. Red/blue coloring represents positive/negative vertical field changes. The black rectangles mark the regions of major field change near the neutral lines that are used in subsequent analysis. The solid and dotted contours indicate strong ($\mathrm{|}B_r\mathrm{|}>1000$~G) and quite strong  ($\mathrm{|}B_r\mathrm{|}>100$~G) fields, respectively. }
\label{fig:dbr}
\end{figure}

\begin{table}
\scriptsize
\caption{Flares studied in this paper.}\
\label{flaretable}
\\
\begin{tabular}{lcccccc}
\hline\hline
 & {\it GOES} Start & {\it GOES} Peak & {\it GOES} End & {\it GOES} & NOAA & Location \\
Date (UT) & Time (UT) & Time (UT) & Time (UT) & Class & Number & on Disk \\
\hline
2011 Feb 13 & 1728 & 1738 & 1747 & M6.6 & 11158 & S20E05 \\
2011 Feb 15 & 0144 & 0156 & 0206 & X2.2 & 11158 & S20W10 \\
2011 Mar 9 & 2313 & 2323 & 2329 & X1.5 & 11166 & N08W11 \\
2011 Sep 6 & 2212 & 2220 & 2224 & X2.1 & 11283 & N13W18 \\
2011 Sep 7 & 2232 & 2238 & 2244 & X1.8 & 11283 & N14W31 \\
2012 Mar 7 & 0002 & 0024 & 0040 & X5.4 & 11429 & N18E31 \\
\hline
\end{tabular}
\end{table}

\begin{table}
\scriptsize
\caption{Summary of changes in magnetic tilt and shear angles. The tilt angle is calculated with respect to the local vertical direction and the azimuthal angular displacement with respect to the direction normal to the neutral line. The shear and dip angles are calculated with respect to the reference potential field.}\
\label{fieldtable}
\\
\begin{tabular}{lccccc}
\hline\hline
 & Tilt Angle & Azim. Angle & Shear Angle & Horiz. Shear & Dip Angle \\
Date (UT) & Inc/Dec & Inc/Dec & Inc/Dec & Inc/Dec & Inc/Dec \\
\hline
2011 Feb 13 & Inc & $-$ & $-$ & $-$ & Inc \\
2011 Feb 15 & Inc & Inc & Inc & Inc & Inc \\
2011 Mar 9 & Inc & Inc & $-$ & $-$ & Inc \\
2011 Sep 6 &  Inc & Inc & $-$ & $-$ & Inc \\
2011 Sep 7 & $-$ & Inc & $-$ & $-$ & Inc \\
2012 Mar 7 & Inc & Inc & Inc & Inc & Inc \\
\hline
\end{tabular}
\end{table}

The six flares and their published GOES times and classes and NOAA active region numbers are given in Table~\ref{flaretable}. The vector field measurements covering these flares were released by the HMI team in the form of 12-minute vector magnetogram images $(B_r, B_{\theta} ,B_{\phi})$ in heliographic coordinates $(r, \theta ,\phi )$ on grids with pixel size $0.03^{\circ}$. Figure~\ref{fig:br} shows spatial maps of the vertical magnetic field component, $B_r$, before the flares with the corresponding horizontal field, ${\bf B}_\mathrm{h} = (B_{\phi}, -B_{\theta})$, indicated by arrows. In all cases the active region field structure is complex but includes a prominent, highly sheared magnetic neutral line, indicated in each plot by a black rectangle. These rectangles were chosen balancing the following priorities: to align the each rectangle satisfactorily with an approximately straight section of each neutral line shown in Figure~\ref{fig:br}, and to capture and characterize the significant horizontal and vertical field changes near the neutral lines shown in Figure~\ref{fig:dbr}. Different rectangles were tried and the results were stable, although some of the patterns in the results became stronger or weaker for different choices of rectangle. Regular rectangular domains were chosen because they define directions parallel and perpendicular to the neutral lines and because the estimates of Lorentz force change in Section~\ref{s:lorentzfch} are expected only to apply to well-resolved field changes in major magnetic structures (Petrie~2012).

It is in this central portion of the active regions near the main magnetic neutral lines that most of the organized magnetic changes occurred during the flares, as the difference maps in Figure~\ref{fig:dbr} show. If we have observations of the photospheric vector field at two times, $t=0$ before the field changes begin, and $t=\delta t$ after the main field changes have occurred, the magnetic vector changes due to the flare can be represented by the difference $\delta{\bf B} = {\bf B}(\delta t) - {\bf B}(0)$. Figure~\ref{fig:dbr} shows spatial maps of the vertical magnetic field change, $\delta B_r$, with the horizontal field changes $\delta {\bf B}_\mathrm{h} = (\delta B_{\phi}, -\delta B_{\theta})$ indicated by arrows. Each difference map was constructed by subtracting the last pre-flare 12-minute image from the image with time stamp 24 minutes later. This is analogous to the differencing of 10-minute averages of MDI one-minute data by Sudol and Harvey~(2005), which these authors used to validate their GONG field change results. These authors showed that most field changes occur during the first 10 minutes or so of major flares, thus most of the field changes are accounted for in the difference calculation. The photospheric plasma acts continuously on the field, while the coronal field is believed to evolve through series of nearly force-free equilibria, punctuated by brief dynamical episodes such as the major flares studied here. The major permanent photospheric field changes that occur during flares correspond to the abrupt and permanent restructuring of the coronal field due to the flare.

The difference maps in Figure~\ref{fig:dbr} for the 2011 February 13 M6.6, 2011 September 6 X2.1 and 2011 September 7 X1.8 flares show that the vertical changes were mostly positive/negative on the positive/negative side of the neutral line, weakening the vertical field on both sides of the neutral line. The map for the 2011 February 15 X2.2 flare shows the opposite pattern and the map for the 2011 March 9 and 2012 March 7 X5.4 flares show a less organized mixtures of positive and negative changes in the vertical field. The vertical field changes therefore do not show a pattern that generalizes across the six-flare data set.  In contrast, the horizontal changes point in approximately the same direction as the field itself near the neutral lines of all of the six flaring region, strengthening the horizontal field there. Spatial maps of the changes in the field tilt angle $\gamma =\tan^{-1} ([B_{\theta}^2+B_{\phi}^2]^{1/2}/|B_r|)$ and the total field strength $B=(B_r^2+B_{\theta}^2+B_{\phi}^2)^{1/2}$ during the flare (not shown) indicate clear increase in the tilt angle and strength of the vector field near the neutral lines during the flares. We will discuss these increases in field strength in more detail below.


\begin{figure} 
\begin{center}
\resizebox{0.49\textwidth}{!}{\includegraphics*{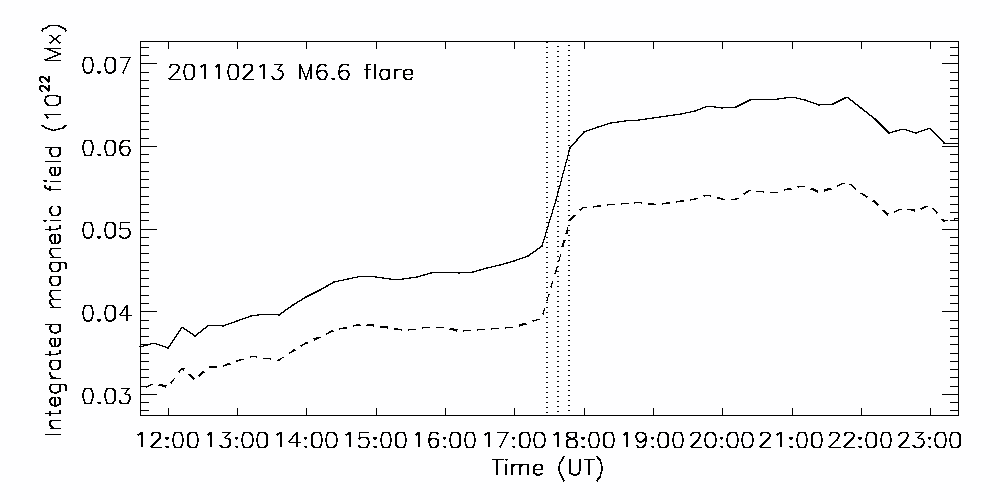}}
\resizebox{0.49\textwidth}{!}{\includegraphics*{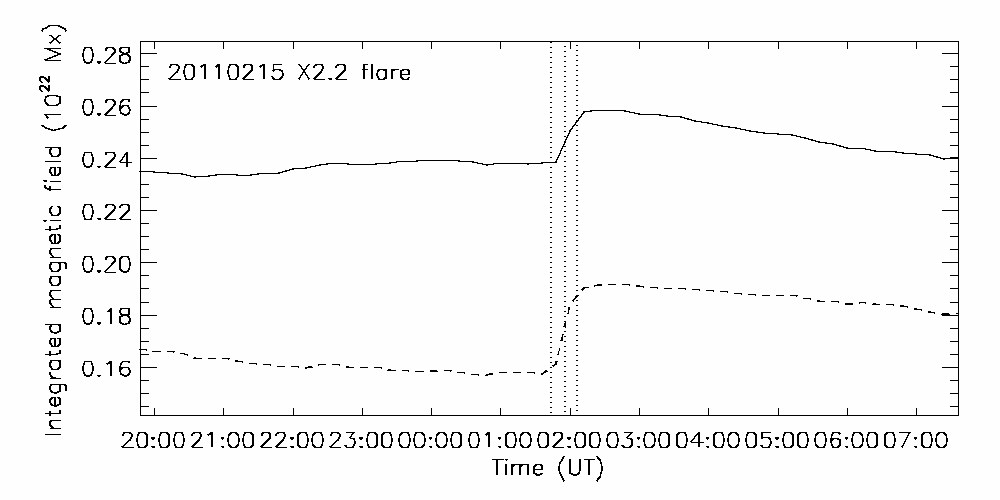}}
\resizebox{0.49\textwidth}{!}{\includegraphics*{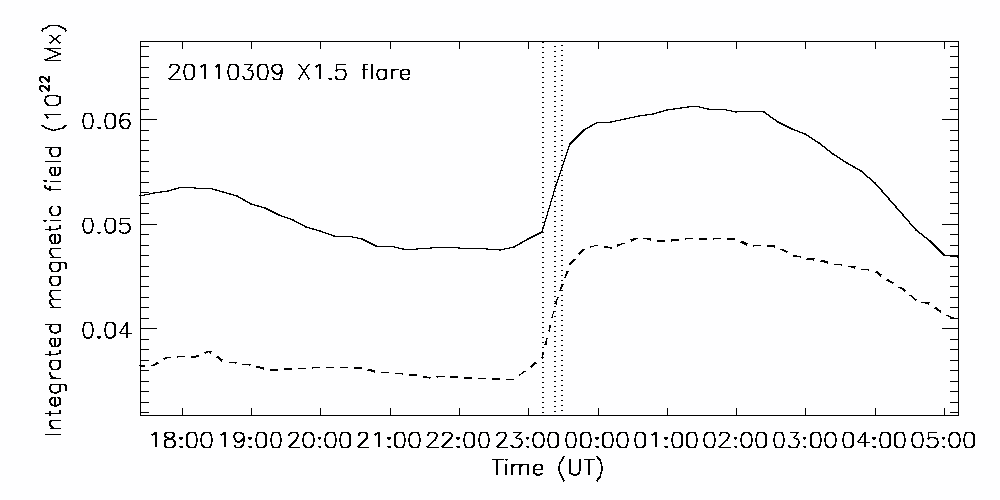}}
\resizebox{0.49\textwidth}{!}{\includegraphics*{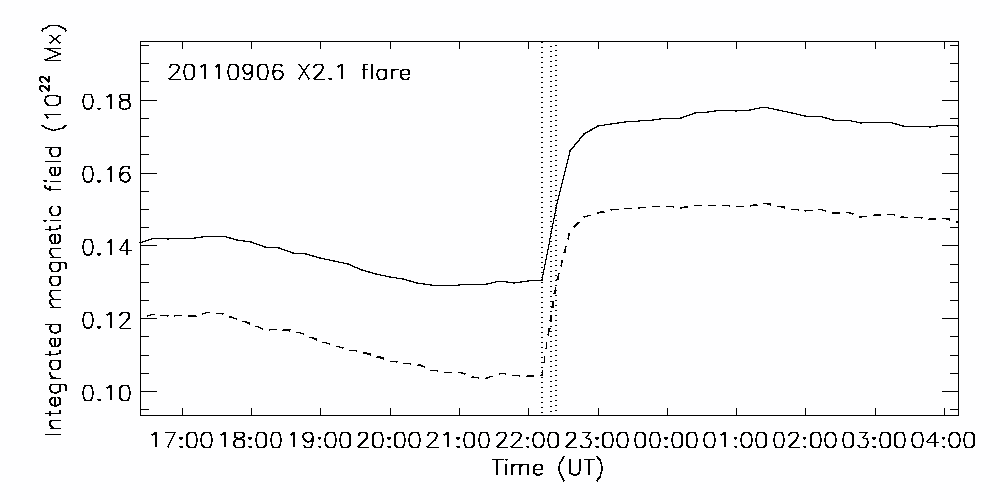}}
\resizebox{0.49\textwidth}{!}{\includegraphics*{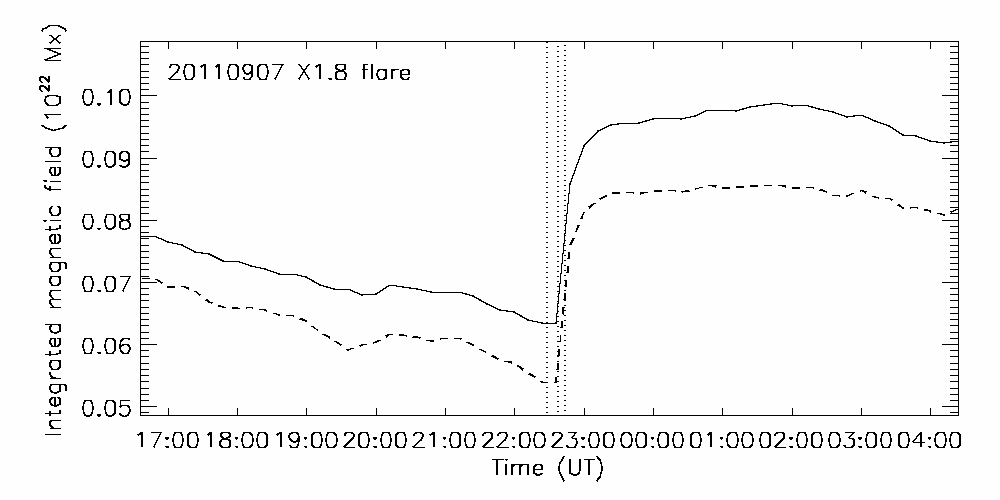}}
\resizebox{0.49\textwidth}{!}{\includegraphics*{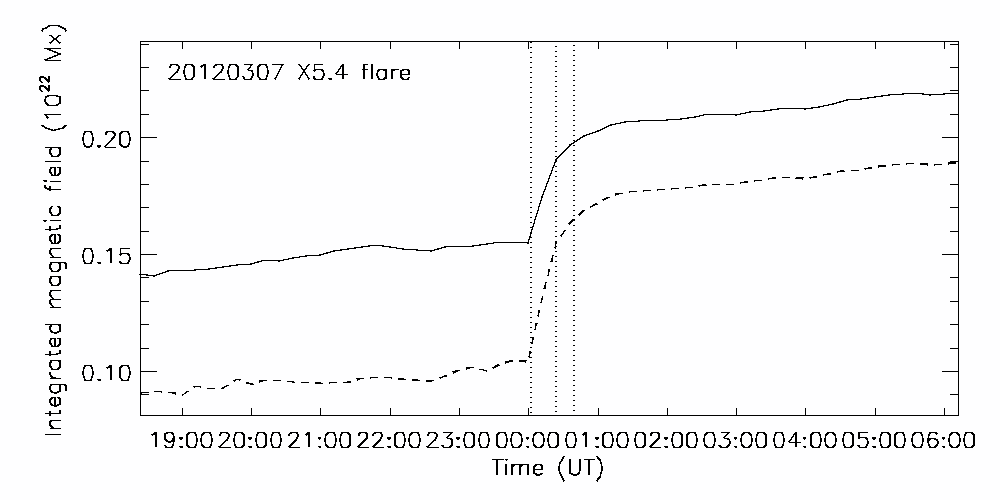}}
\end{center}
\caption{Shown here are the integrated magnetic field strength $B^\mathrm{NL}$ (solid lines) and $B_\mathrm{h}^\mathrm{NL}$ (dashed lines) near each neutral line plotted against time. The areas of integration are indicated by the black rectangles in Figure~\ref{fig:br}. The vertical lines represent the GOES flare start, peak and end times. }
\label{fig:fmodnlt}
\end{figure}

\begin{figure} 
\begin{center}
\resizebox{0.49\textwidth}{!}{\includegraphics*{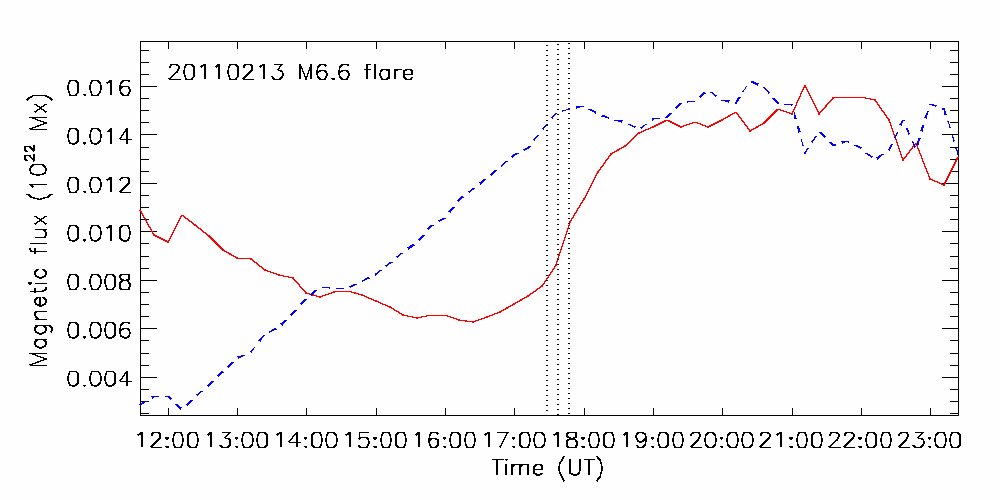}}
\resizebox{0.49\textwidth}{!}{\includegraphics*{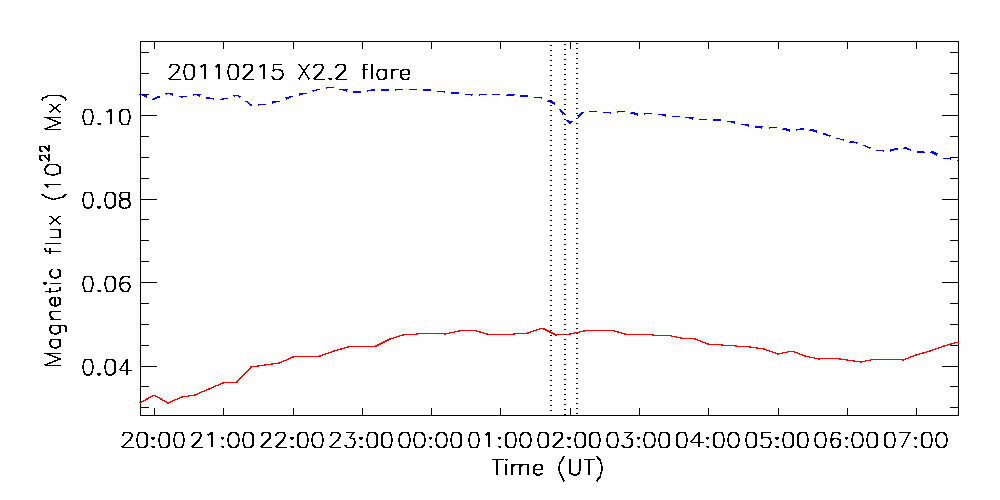}}
\resizebox{0.49\textwidth}{!}{\includegraphics*{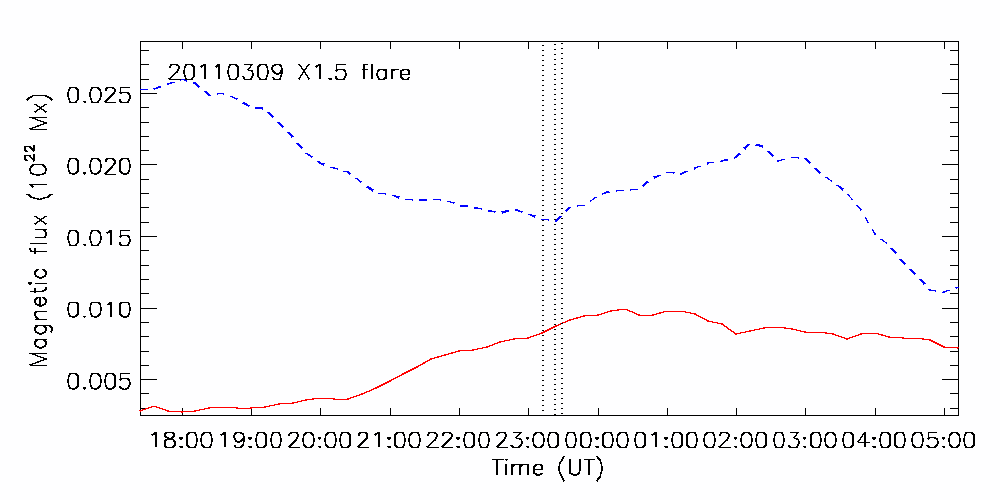}}
\resizebox{0.49\textwidth}{!}{\includegraphics*{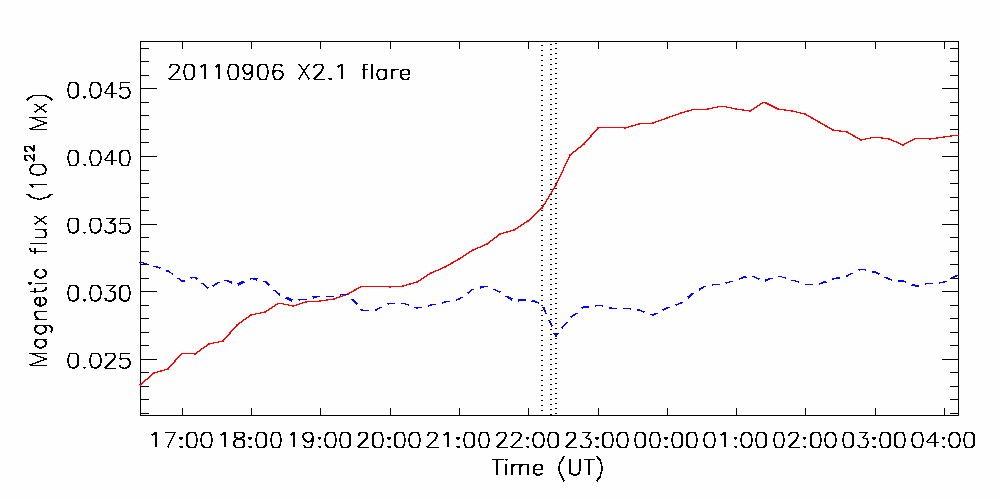}}
\resizebox{0.49\textwidth}{!}{\includegraphics*{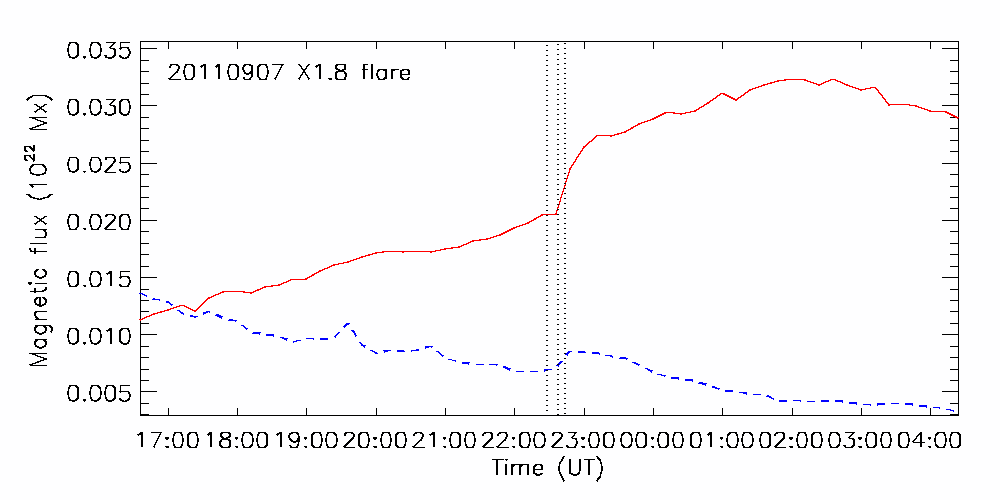}}
\resizebox{0.49\textwidth}{!}{\includegraphics*{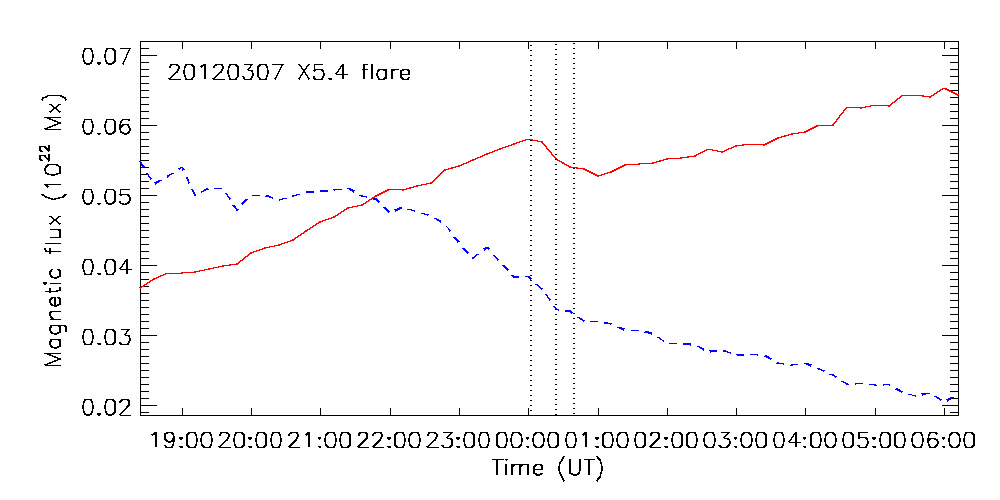}}
\end{center}
\caption{Shown here are the total vertical magnetic flux $B_r^\mathrm{NL}$ near each neutral line plotted against time. The red/blue solid/dashed lines represent positive/negative flux. The areas of integration are indicated by the black rectangles in Figure~\ref{fig:br}. The vertical lines represent the GOES flare start, peak and end times. }
\label{fig:frnlt}
\end{figure}

\begin{figure} 
\begin{center}
\resizebox{0.49\textwidth}{!}{\includegraphics*{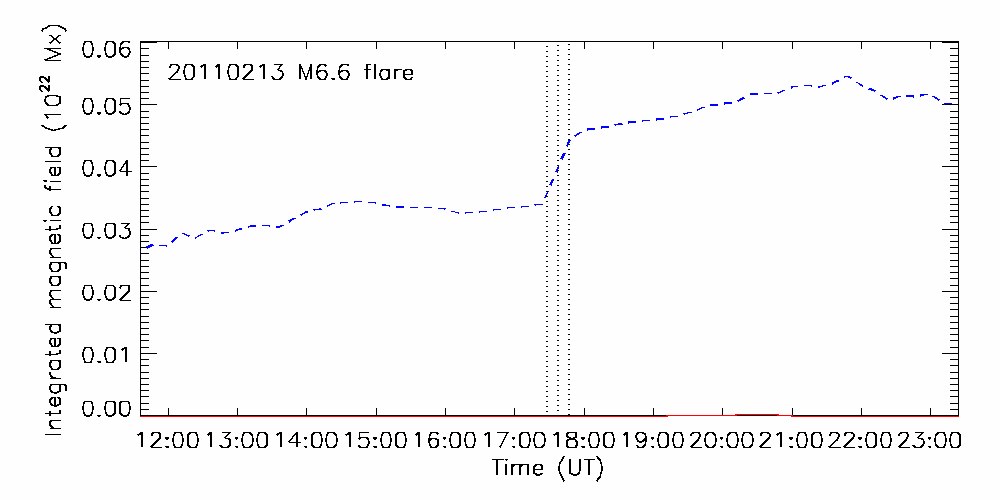}}
\resizebox{0.49\textwidth}{!}{\includegraphics*{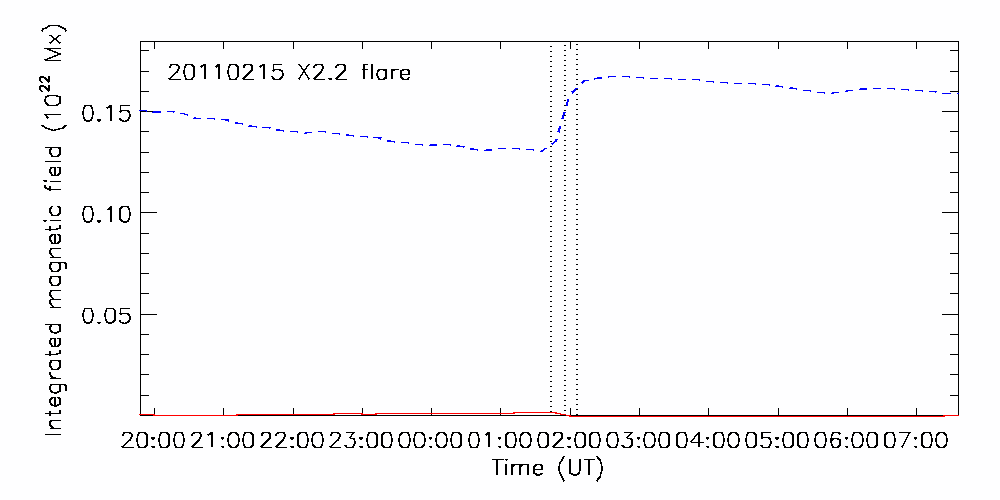}}
\resizebox{0.49\textwidth}{!}{\includegraphics*{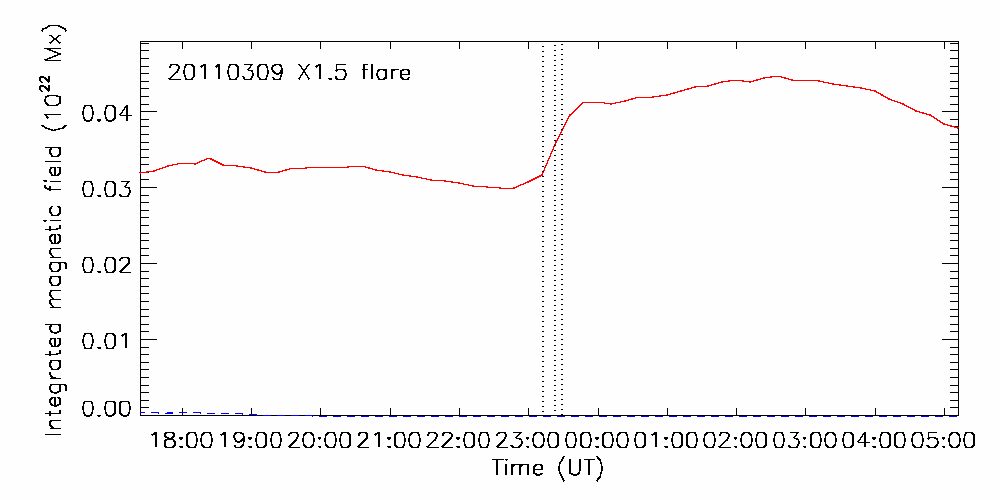}}
\resizebox{0.49\textwidth}{!}{\includegraphics*{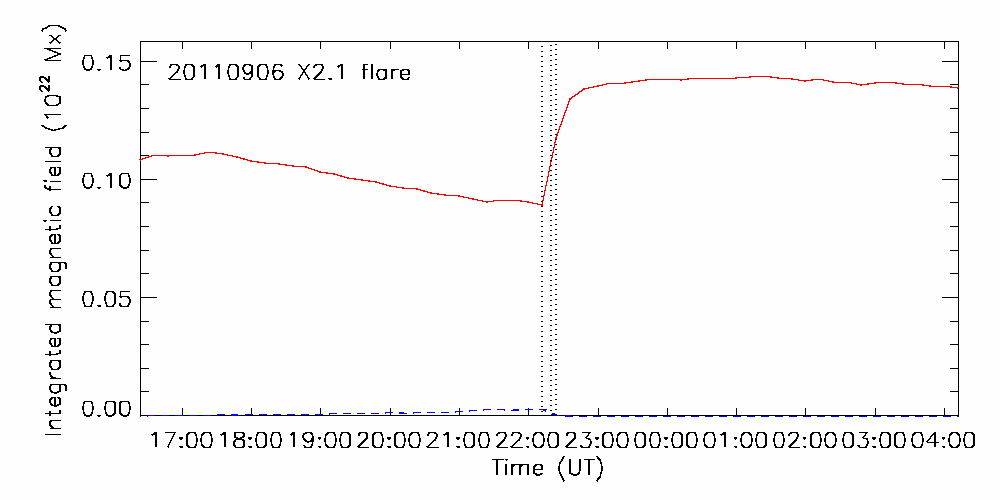}}
\resizebox{0.49\textwidth}{!}{\includegraphics*{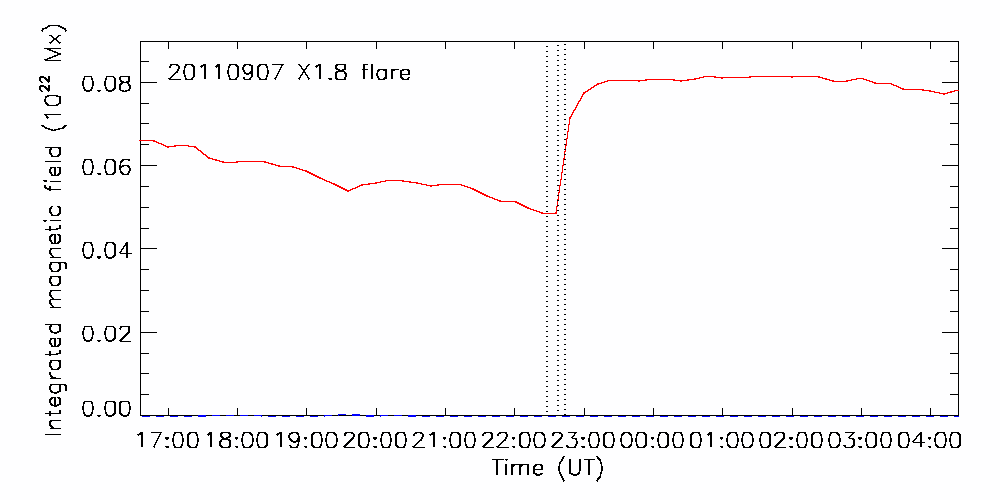}}
\resizebox{0.49\textwidth}{!}{\includegraphics*{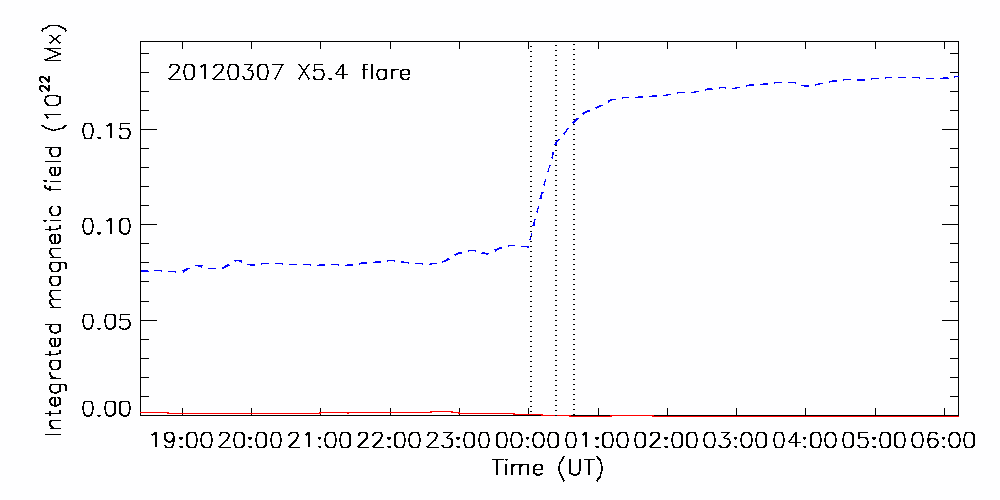}}
\end{center}
\caption{Shown here are the integrated horizontal field parallel to each neutral line, $B_\parallel^\mathrm{NL}$, near each neutral line plotted against time. The red/blue solid/dashed lines represent positive/negative field, i.e., approximately westward/eastward field. The areas of integration are indicated by the black rectangles in Figure~\ref{fig:br}. The vertical lines represent the GOES flare start, peak and end times. }
\label{fig:fparanlt}
\end{figure}

\begin{figure} 
\begin{center}
\resizebox{0.49\textwidth}{!}{\includegraphics*{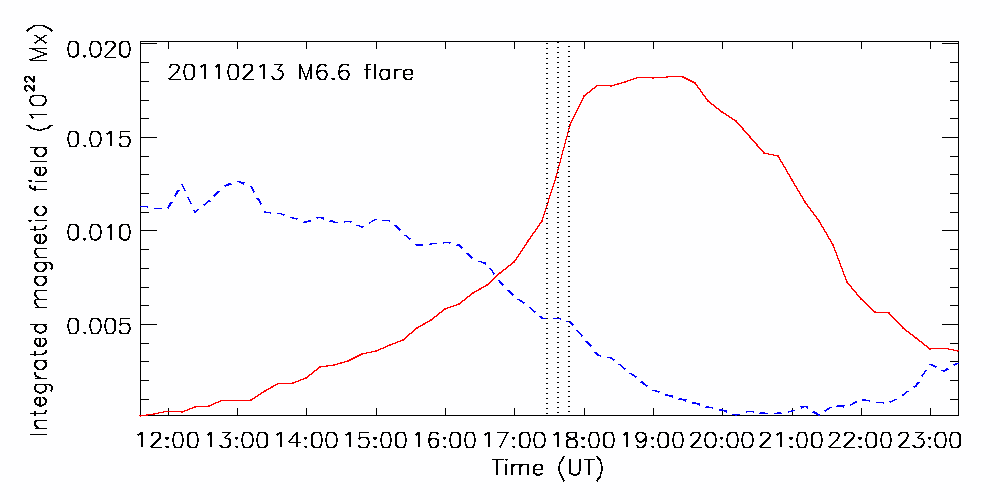}}
\resizebox{0.49\textwidth}{!}{\includegraphics*{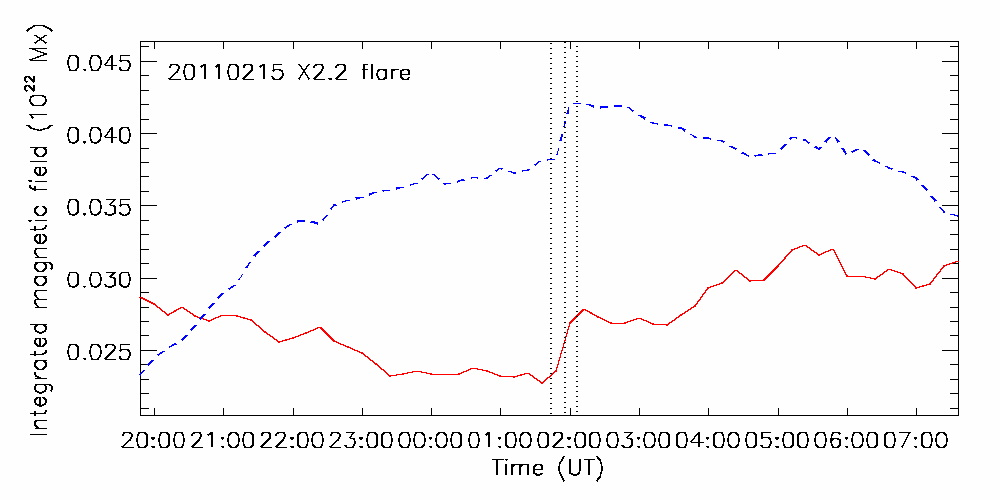}}
\resizebox{0.49\textwidth}{!}{\includegraphics*{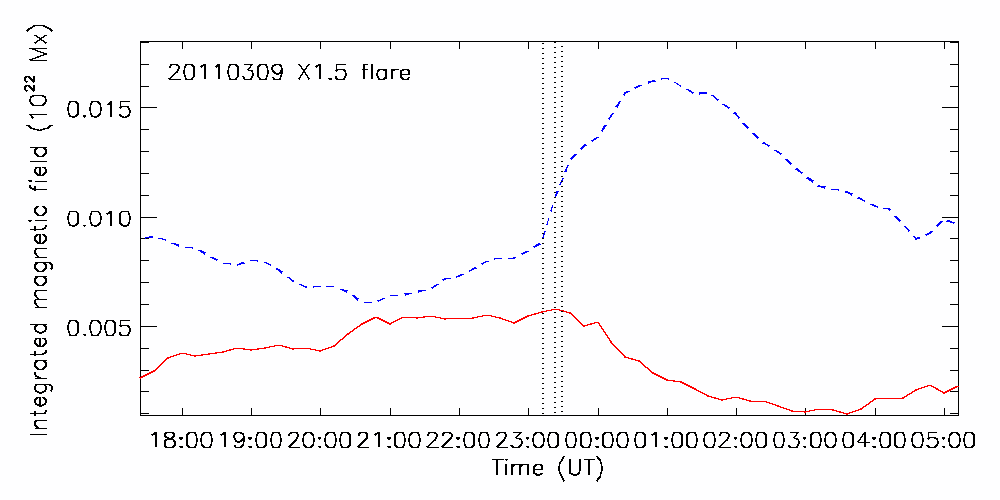}}
\resizebox{0.49\textwidth}{!}{\includegraphics*{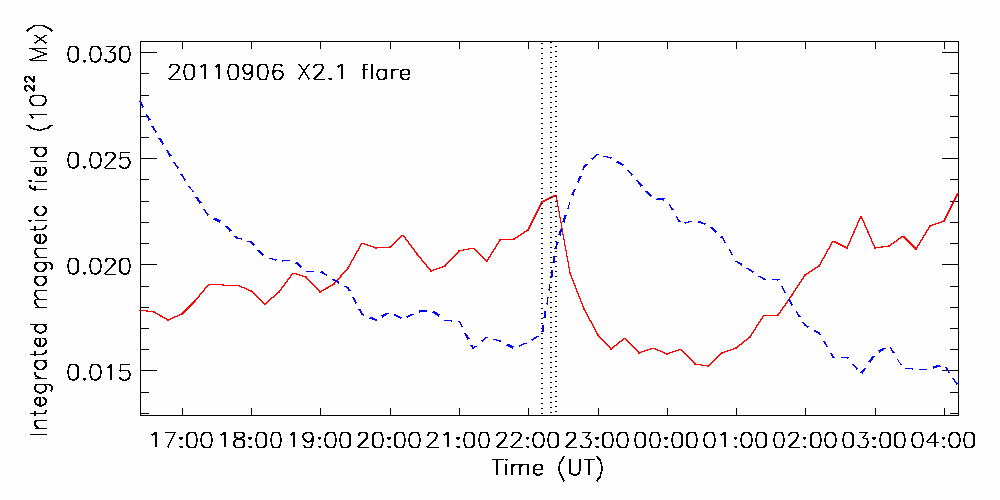}}
\resizebox{0.49\textwidth}{!}{\includegraphics*{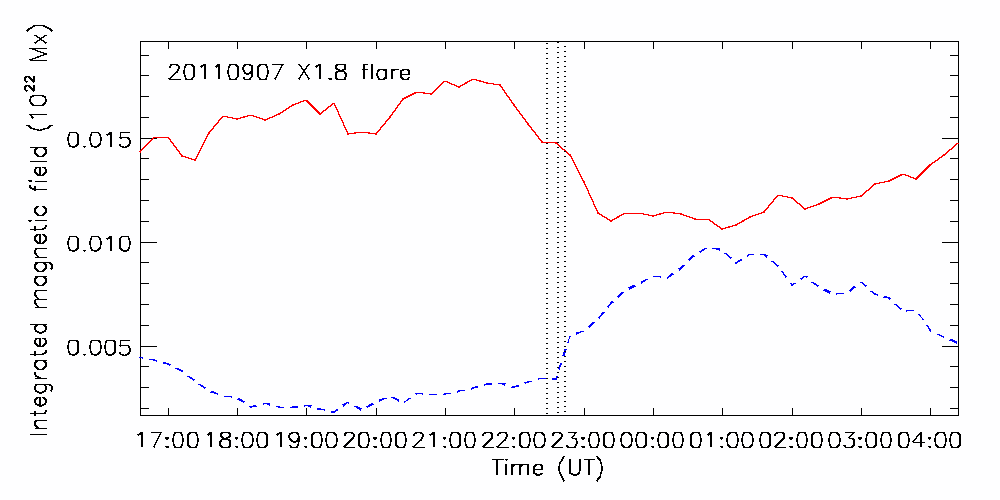}}
\resizebox{0.49\textwidth}{!}{\includegraphics*{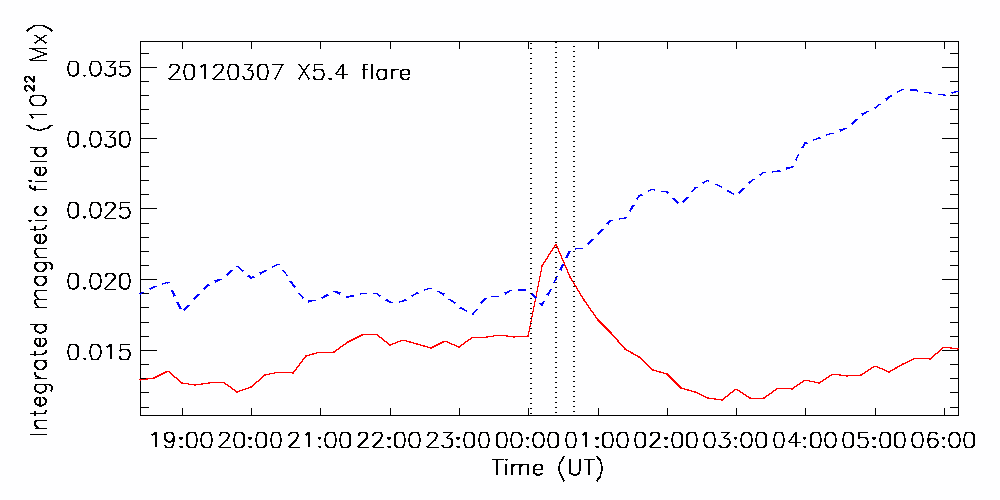}}
\end{center}
\caption{Shown here are the integrated horizontal field perpendicular to each neutral line, $B_\perp^\mathrm{NL}$, near each neutral line plotted against time. The red/blue solid/dashed lines represent positive/negative field, i.e., approximately northward/southward field. The areas of integration are indicated by the black rectangles in Figure~\ref{fig:br}. The vertical lines represent the GOES flare start, peak and end times. }
\label{fig:fperpnlt}
\end{figure}

As the difference maps in Figure~\ref{fig:dbr} show, solar flares are extremely complex physical events and can involve diverse magnetic changes occurring throughout the flaring region, most notably near major neutral lines and in sunspots. Here our goal is to shed light on the changes occurring near neutral lines. We now discuss the temporal profiles of the neutral line magnetic changes, shown in Figures \ref{fig:fmodnlt}-\ref{fig:fperpnlt}. These and subsequent plots of temporal changes were derived by calculating area integrals of the field components over the chosen photospheric areas in the 60-image 12-hour series, represented by the black rectangles in Figures~\ref{fig:br} and \ref{fig:dbr}. Figure~\ref{fig:fmodnlt} shows the evolution of the integrated magnetic field strength

\begin{equation}
B^\mathrm{NL}=\int_{A_\mathrm{NL}} B\ \mathrm{d}A,
\label{eq:BNLint}
\end{equation}

\noindent near the neutral line over each of the 12-hour time intervals. The area $A_\mathrm{NL}$ corresponds to the rectangular region near the neutral line of each flaring region in Figures~\ref{fig:br} and \ref{fig:dbr}. Figure~\ref{fig:fmodnlt} also shows the equivalent integral $B_\mathrm{h}^\mathrm{NL}$ of $B_\mathrm{h}$, where $B_\mathrm{h}=(B_{\theta}^2+B_{\phi}^2)^{1/2}$. Near each neutral line, according to Figure~\ref{fig:fmodnlt}, the average field strength $B^\mathrm{NL}$ increased abruptly and permanently during each flare because of an increase in the horizontal field component $B_\mathrm{h}^\mathrm{NL}$ there, also shown in Figure~\ref{fig:fmodnlt}, where the two quantities are seen to track each other closely in every case.

The behavior of the integrated vertical flux $B_r^\mathrm{NL}$ was more diverse in character. According to Figure~\ref{fig:frnlt}, around the times of the 2011 September 6 and 7 X-class flares the vertical flux increased somewhat abruptly near the neutral line of AR~11283. In all other cases the vertical field changes were either small or gradual. In all cases the vertical changes were less significant than the horizontal changes. From the temporal plots it is therefore clear that the changes in the horizontal field dominated the behavior of the total field strength near the neutral line around the time of each flare, strengthening it.  However, the vertical field changes did have a bearing on the physics of the 2011 September 7 X1.8 flare, as we will see in Section~\ref{s:potential}.

Figures~\ref{fig:fparanlt} and \ref{fig:fperpnlt} show the evolution of the integrated horizontal magnetic field components parallel and perpendicular to the neutral line, $B_{\parallel}^\mathrm{NL}$ and $B_{\perp}^\mathrm{NL}$. These directions are defined by the black rectangles in Figure~\ref{fig:br}. The parallel direction is the direction of the long edges of the rectangles, pointing approximately west. The perpendicular direction is the direction of the short edges of the rectangle, pointing approximately north. The fields pointing in the positive (north or west) and negative (south or east) directions are plotted separately. The parallel fields had a clearly dominant direction in each case and this field increased during all six flares. The perpendicular field changes were generally not as significant as the parallel changes: contrast Figures~\ref{fig:fparanlt} and \ref{fig:fperpnlt}. The perpendicular field strength increased abruptly and permanently during the 2011 February 15 X2.2 flare but even in this case the perpendicular increase was dwarfed by the parallel field increase. The pre- and post-flare evolution of the horizontal field was more steady in the parallel than in the perpendicular component and the flare-related changes much more significant.

Flare-induced line profile changes can produce signatures that do not indicate real changes in the magnetic field as discussed by Sudol \& Harvey~(2005) - see the bottom part of their Figure~1. Working with GONG 1-minute longitudinal field images, Sudol \& Harvey~(2005) and Petrie \& Sudol~(2010) identified artifacts by fitting a $\tan^{-1}$ step-like function to the time profile of each pixel, applying selection criteria based on the quality of the function fits, and inspecting the results for representative pixels by eye. For the HMI vector data this approach is not as useful because the HMI vector field inversions are not as sensitive as the GONG data. 

Because the difference images in Figure~\ref{fig:dbr} are derived from measurements taken before and after the main field changes have taken place, thereby excluding the times when most of the flare artifacts are expected to have occurred, these difference images are not expected to be significantly compromised by flare emission artifacts. The time series plots in Figures~\ref{fig:fmodnlt}-\ref{fig:fperpnlt} do not avoid these times when most of the flare artifacts are expected to have occurred. Comparing Figures~\ref{fig:fmodnlt} and \ref{fig:fparanlt} with the bottom part of Sudol \& Harvey's~(2005) Figure~1, the profiles of $B^\mathrm{NL}$, $B_\mathrm{h}^\mathrm{NL}$ and $B_{\parallel}^\mathrm{NL}$ in Figures~\ref{fig:fmodnlt} and \ref{fig:fparanlt} have clear, stepwise changes with no sign of an emission artifact. This serves as evidence that, while emission transients may have affected some of the pixels, the calculations of the integrated quantities plotted in Figures~\ref{fig:fmodnlt}-\ref{fig:fperpnlt} were not significantly compromised by artifacts.

\begin{figure} 
\begin{center}
\resizebox{0.49\textwidth}{!}{\includegraphics*{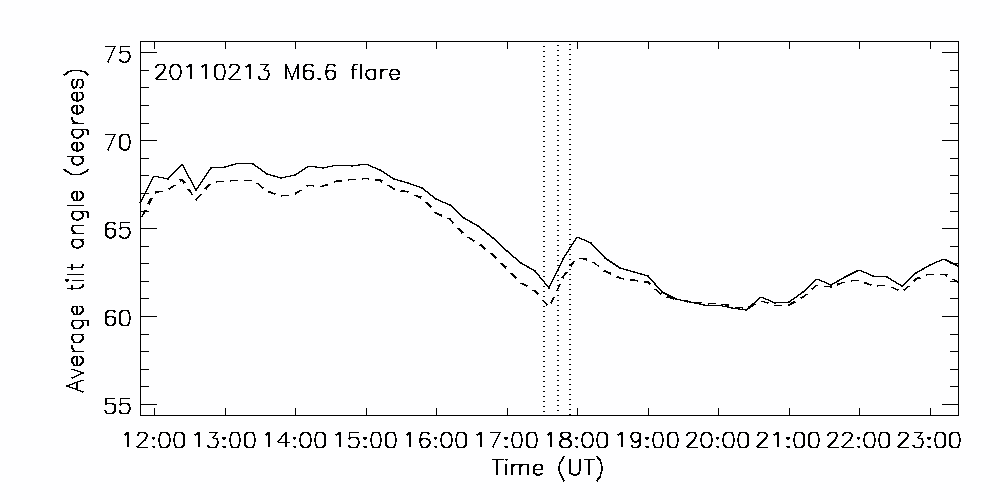}}
\resizebox{0.49\textwidth}{!}{\includegraphics*{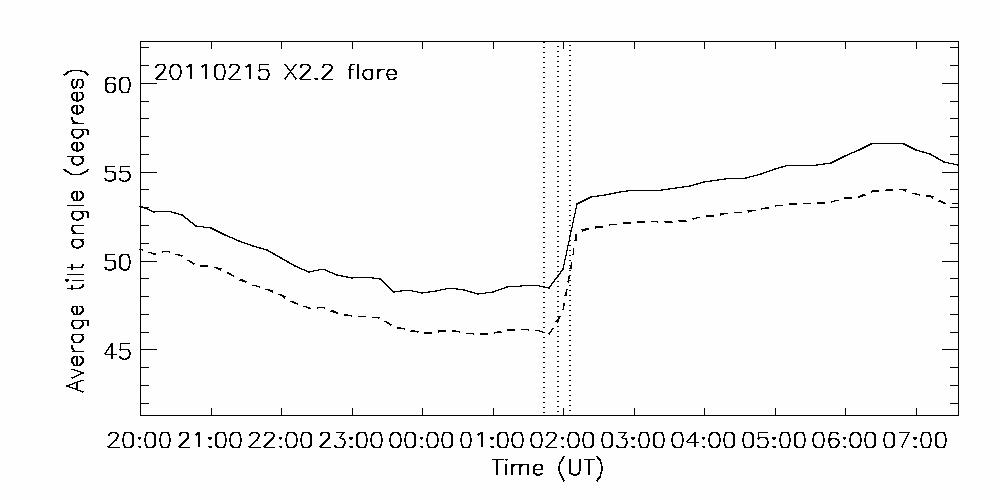}}
\resizebox{0.49\textwidth}{!}{\includegraphics*{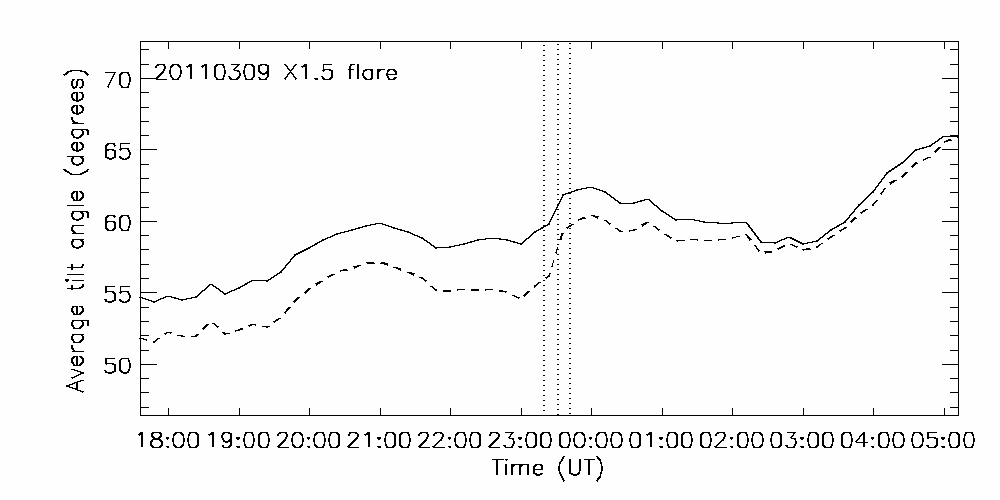}}
\resizebox{0.49\textwidth}{!}{\includegraphics*{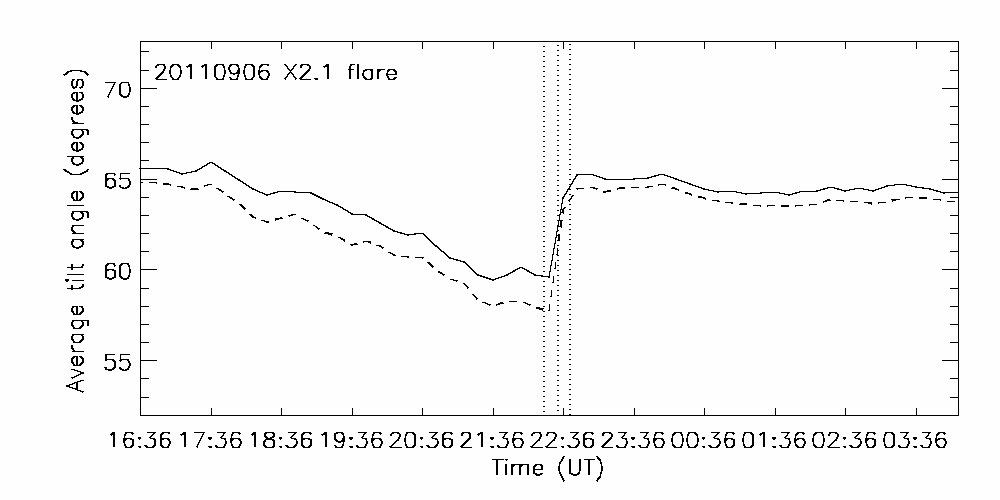}}
\resizebox{0.49\textwidth}{!}{\includegraphics*{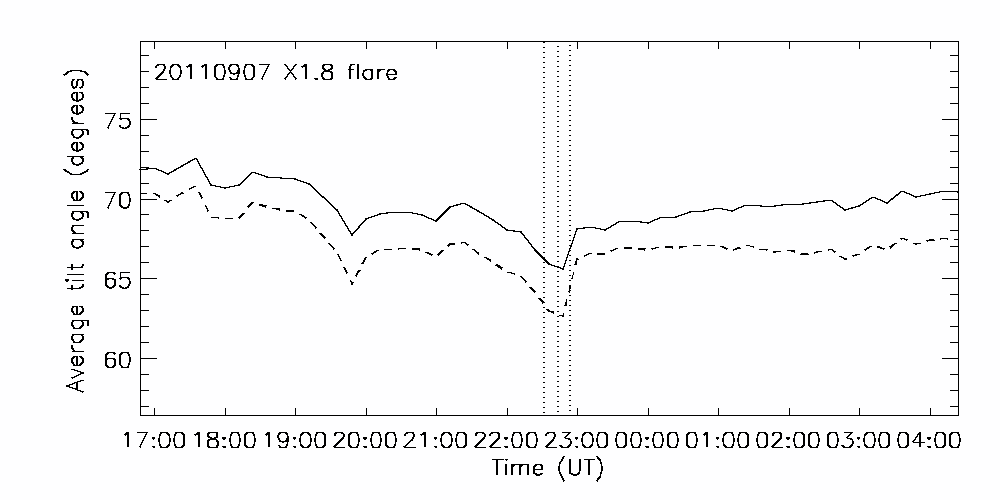}}
\resizebox{0.49\textwidth}{!}{\includegraphics*{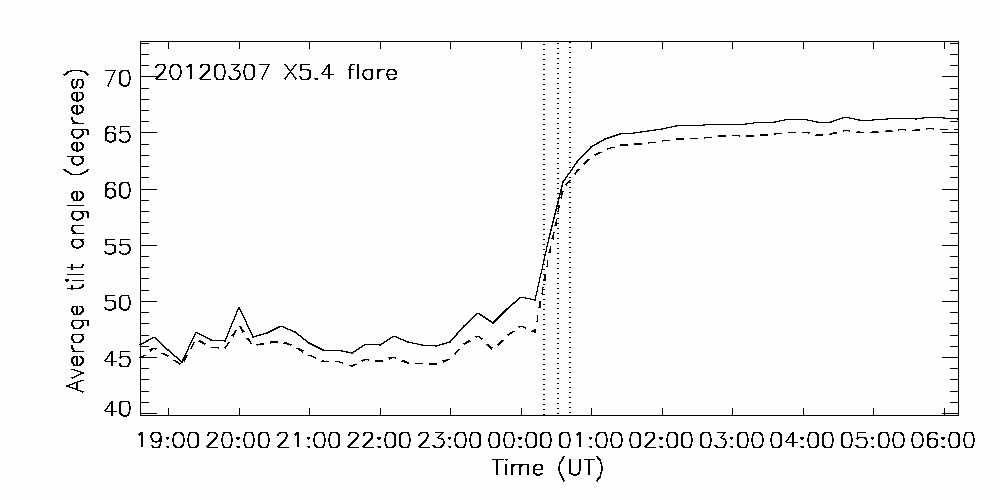}}
\end{center}
\caption{Shown here are the average tilt angles (solid lines) and the field-weighted average tilt angles (dashed lines) near each neutral line plotted against time. These angles are averaged over the areas of integration indicated by the black rectangles in Figure~\ref{fig:br}. The vertical lines represent the GOES flare start, peak and end times. }
\label{fig:tiltnlt}
\end{figure}

\begin{figure} 
\begin{center}
\resizebox{0.49\textwidth}{!}{\includegraphics*{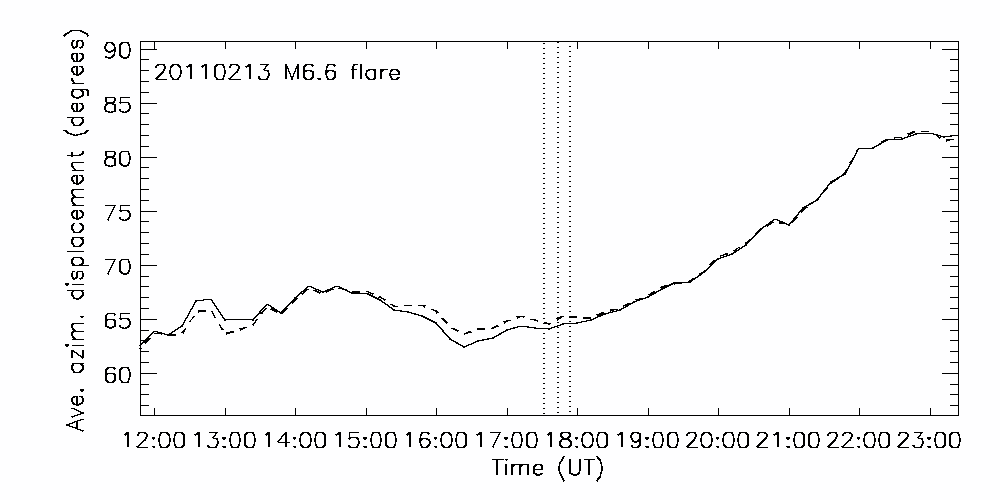}}
\resizebox{0.49\textwidth}{!}{\includegraphics*{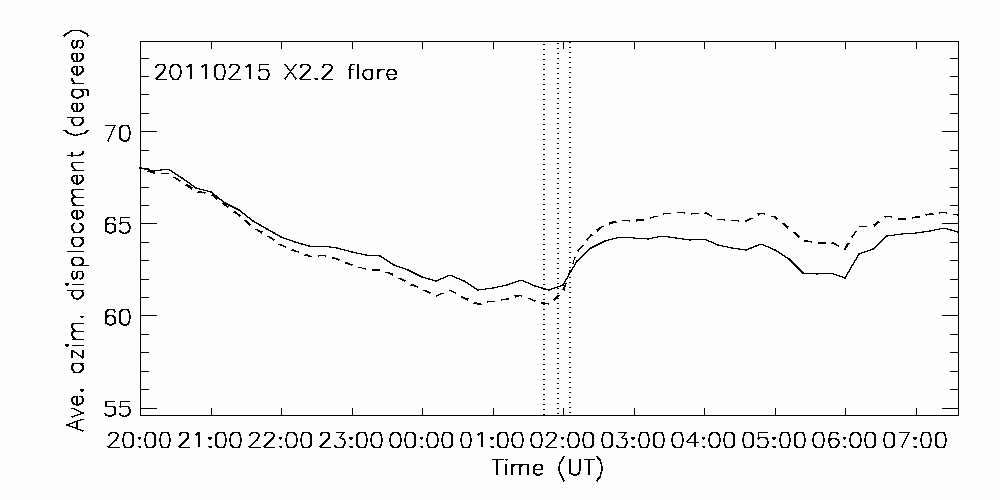}}
\resizebox{0.49\textwidth}{!}{\includegraphics*{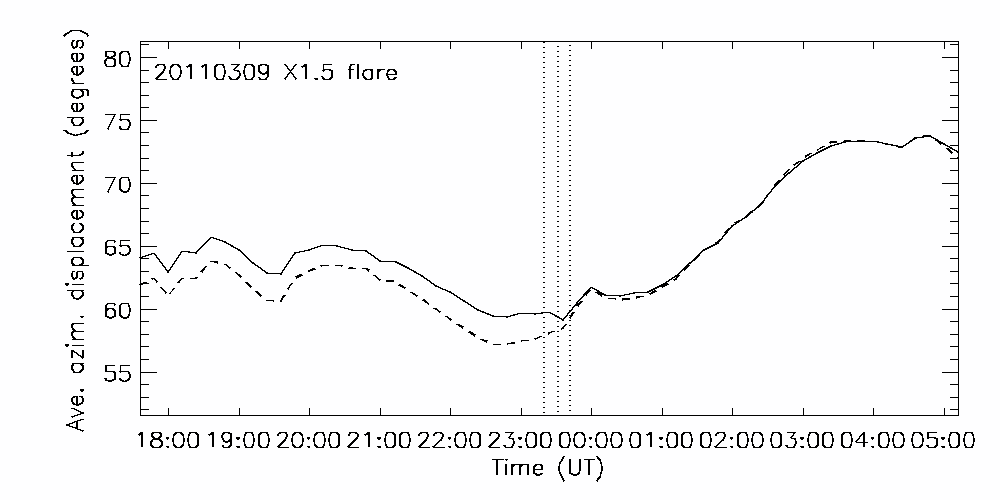}}
\resizebox{0.49\textwidth}{!}{\includegraphics*{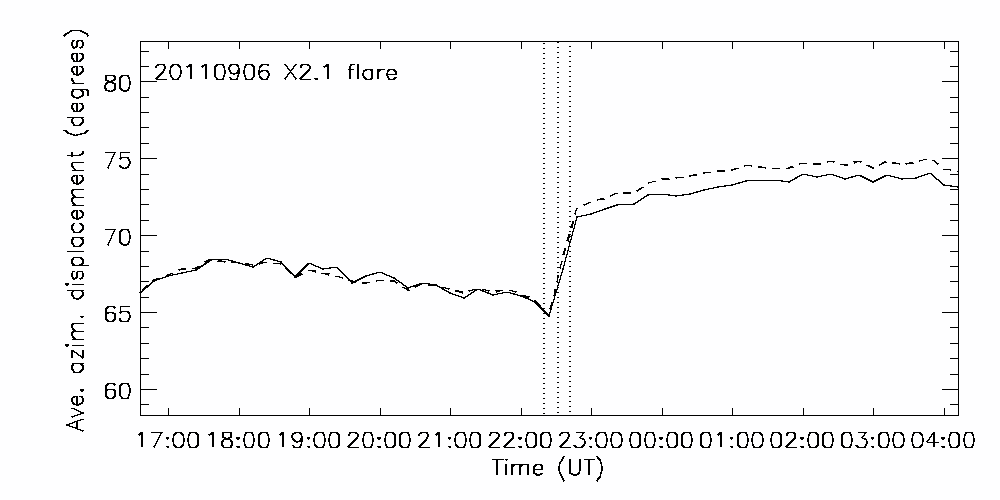}}
\resizebox{0.49\textwidth}{!}{\includegraphics*{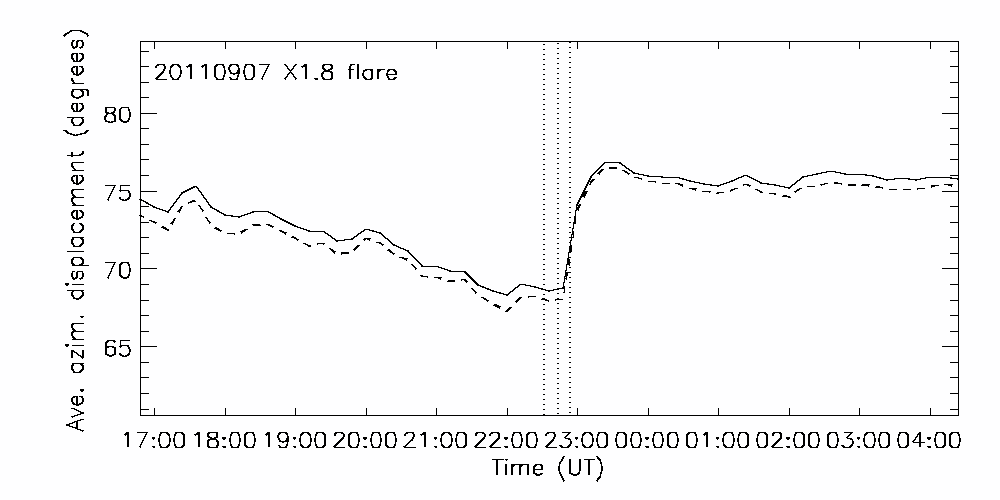}}
\resizebox{0.49\textwidth}{!}{\includegraphics*{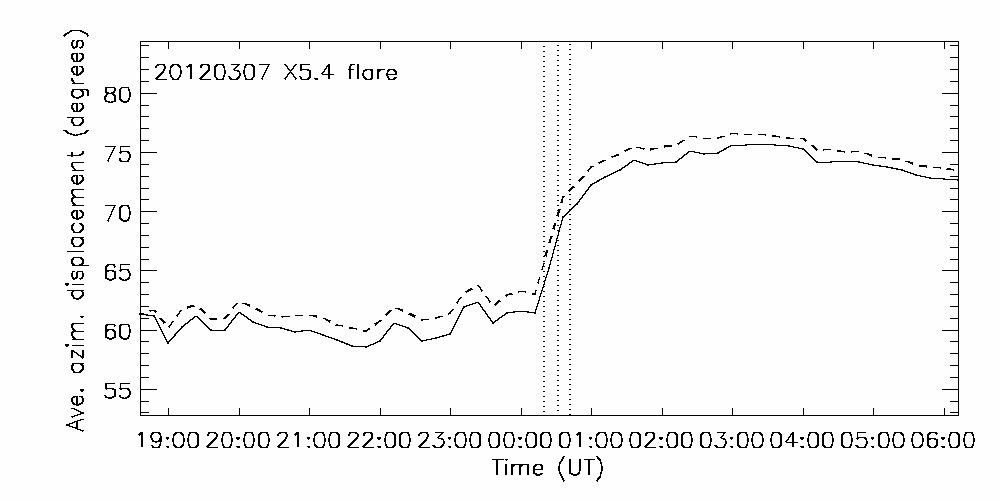}}
\end{center}
\caption{Shown here are the average azimuthal angular displacements from the normal to the neutral lines (solid lines) and the field-weighted average azimuthal angular displacements (dotted lines), at each neutral line plotted against time. These quantities are averaged over the areas of integration indicated by the black rectangles in Figure~\ref{fig:br}. The vertical lines represent the GOES flare start, peak and end times. }
\label{fig:azimdiffnlt}
\end{figure}

Figure~\ref{fig:tiltnlt} shows the temporal behavior of the average tilt angle for each flare, the integral of $\gamma =\tan^{-1} ([B_{\theta}^2+B_{\phi}^2]^{1/2}/|B_r|)$ over the boxes around the neutral lines shown in Figure~\ref{fig:br} divided by the area of the box in each case. Also shown is the field-strength-weighted average $\tilde{\gamma}$ of $\gamma$, of the form

\begin{equation}
\tilde{\gamma} = \int_{A_\mathrm{NL}} B \gamma \ \mathrm{d}A \left/ \int_{A_\mathrm{NL}} B \ \mathrm{d}A \right. .
\label{eq:tiltwt}
\end{equation}

\noindent The changes in the tilt angle for the six flares, and other magnetic changes discussed below, are summarized in Table~\ref{fieldtable}. In all cases the tilt angle increased on average (the field became more horizontal), except, arguably, for the 2011 September 7 X1.8 flare. Figures~\ref{fig:fmodnlt} and \ref{fig:frnlt} show that $B^\mathrm{NL}_\mathrm{h}$ and $B^\mathrm{NL}_r$ both increased during the 2011 September 7 X1.8 flare, making the increase in the tilt profile less strong for this flare than for the other flares, whose vertical field changes are not so significant. The increasing pre-flare trend of $B^\mathrm{NL}_r$ and the decreasing pre-flare trend of $B^\mathrm{NL}_\mathrm{h}$ give the tilt profile a steep decline before the flare, that shows the lack of real significance of the tilt increase during this flare.

The temporal behavior of the average angle $\alpha = \tan^{-1} (|B_{\parallel}|/|B_{\perp}|)$ between the horizontal field vector and the direction normal to the neutral line (approximated by the orientations of the short edges of the boxes around the neutral lines in Figure~\ref{fig:br}) are shown in Figure~\ref{fig:azimdiffnlt}. Also shown is the field-strength-weighted angular displacement $\tilde{\alpha}$, calculated in a manner analogous to Equation~\ref{eq:tiltwt}. In the plots in Figure~\ref{fig:azimdiffnlt}, zero angular displacement would be consistent with an unsheared arcade whereas $90^{\circ}$ would represent field aligned with the neutral line. The azimuthal angular displacement has been identified by some authors (e.g., Aulanier et al.~2012) with magnetic shear. This is because an unstressed magnetic field is expected to cross the main neutral line at an angle approximately normal to the neutral line. During the four largest flares the horizontal neutral-line fields abruptly became less aligned with the normal direction, i.e., more aligned with the neutral line. The angular changes range from a few to nearly $10^{\circ}$.  The 2011 March 9 X1.5 flare fields show some evidence of the same pattern but there is no such signature above the background evolution for the 2011 February 13 M6.6 flare.

Based on the evidence of Figures~\ref{fig:tiltnlt} and \ref{fig:azimdiffnlt}, the fields show a strong tendency to become more tilted and more aligned with the neutral line as a result of flares. This is because the main effect of the flare on the photospheric neutral-line field is to strengthen the horizontal component parallel to the neutral line.

\section{Comparison with potential fields}
\label{s:potential}

\begin{figure} 
\begin{center}
\resizebox{0.49\textwidth}{!}{\includegraphics*{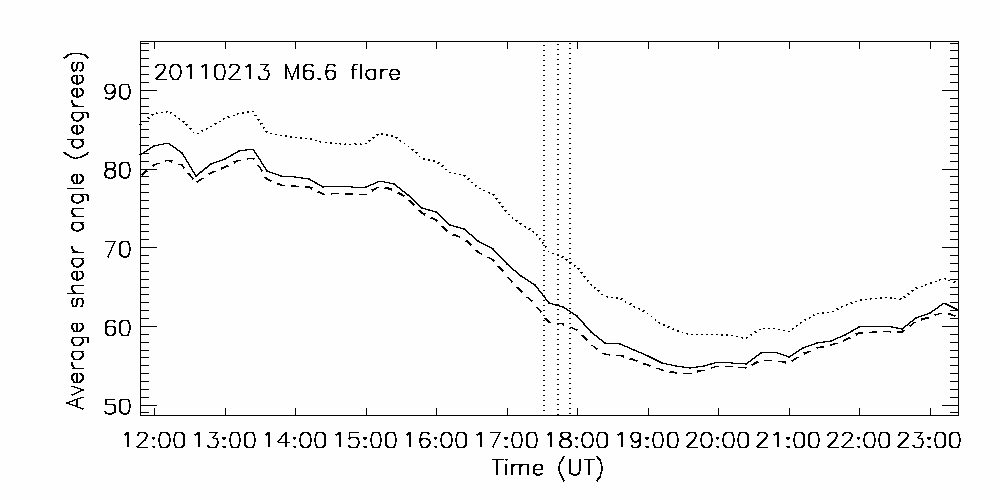}}
\resizebox{0.49\textwidth}{!}{\includegraphics*{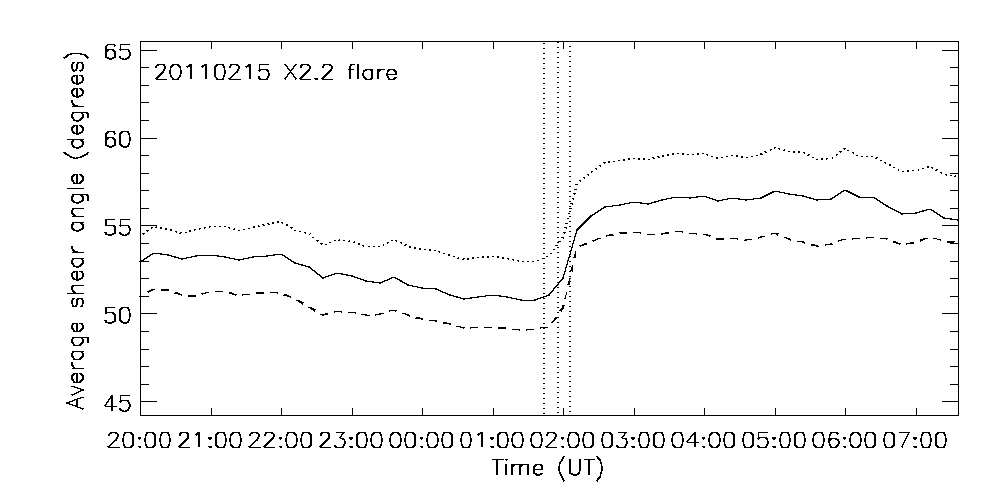}}
\resizebox{0.49\textwidth}{!}{\includegraphics*{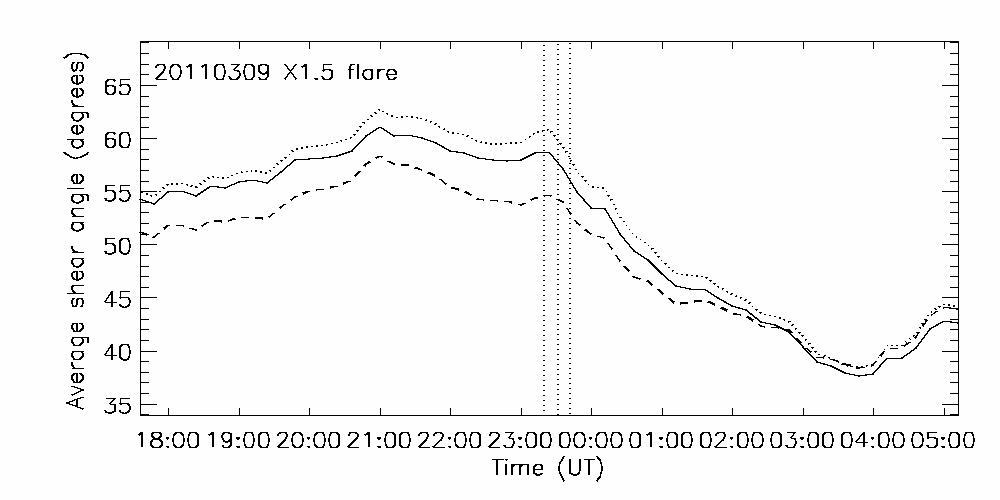}}
\resizebox{0.49\textwidth}{!}{\includegraphics*{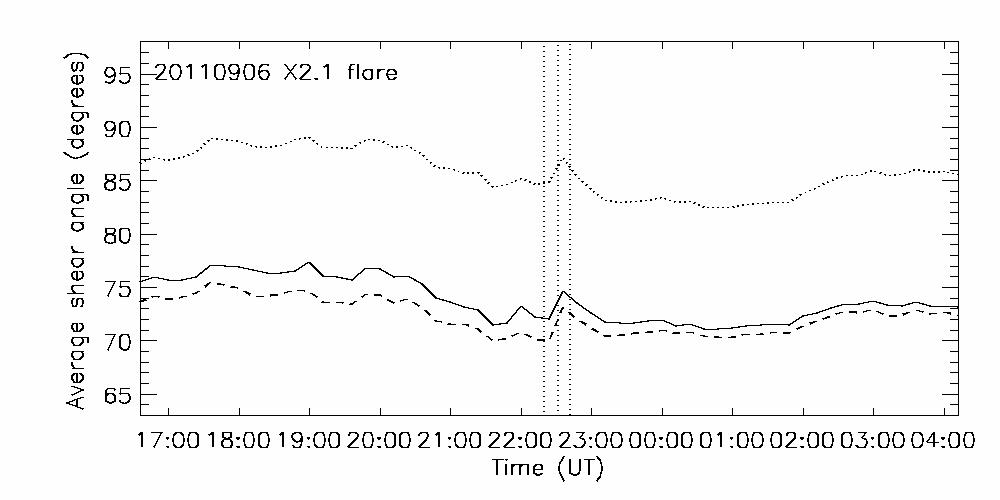}}
\resizebox{0.49\textwidth}{!}{\includegraphics*{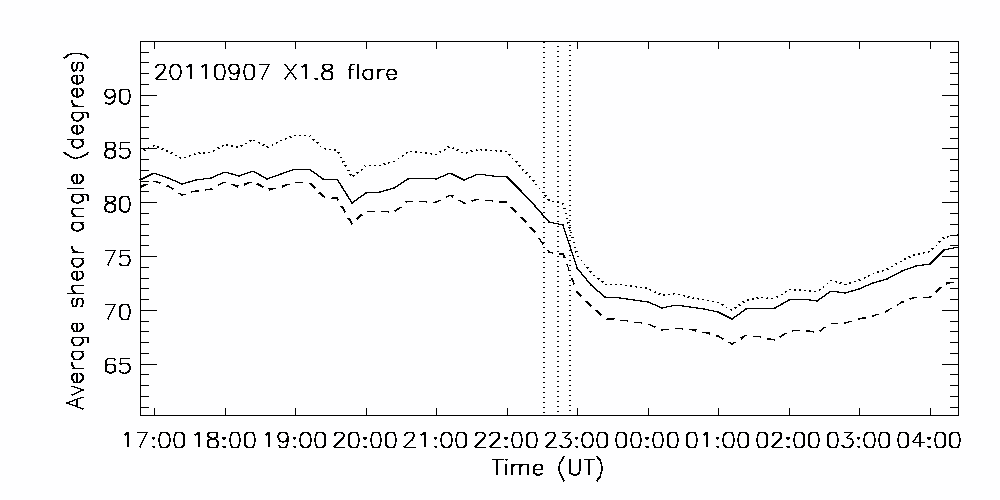}}
\resizebox{0.49\textwidth}{!}{\includegraphics*{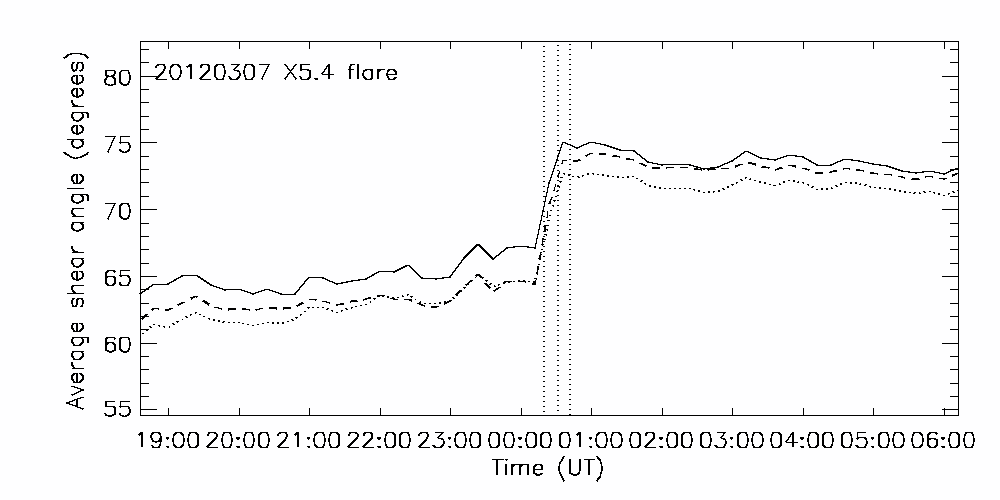}}
\end{center}
\caption{Shown here are the average shear angles (solid lines) and the field-weighted average shear angles (dashed lines) near each neutral line plotted against time. The average of the magnetic shear is also plotted (dotted lines) in units of G$^{\circ}$, divided by $1.5\times 10^3$ for the 2011 February 15 and March 9 flares and $10^3$ for the other flares. These quantities are averaged over the areas of integration indicated by the black rectangles in Figure~\ref{fig:br}. The vertical lines represent the GOES flare start, peak and end times. }
\label{fig:shear}
\end{figure}

\begin{figure} 
\begin{center}
\resizebox{0.49\textwidth}{!}{\includegraphics*{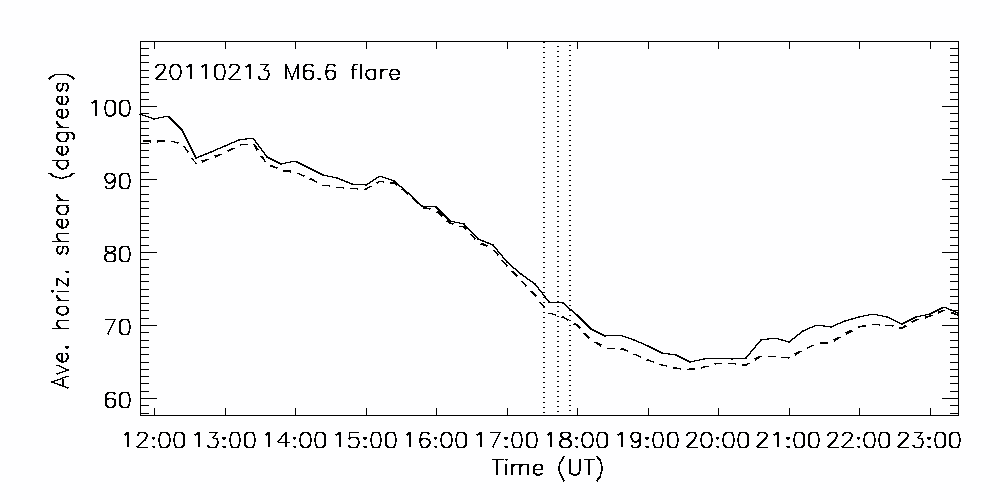}}
\resizebox{0.49\textwidth}{!}{\includegraphics*{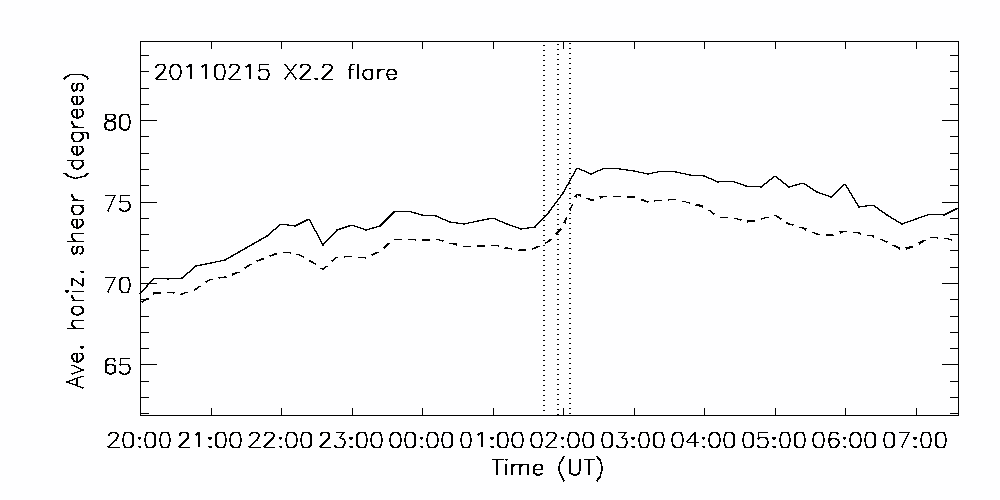}}
\resizebox{0.49\textwidth}{!}{\includegraphics*{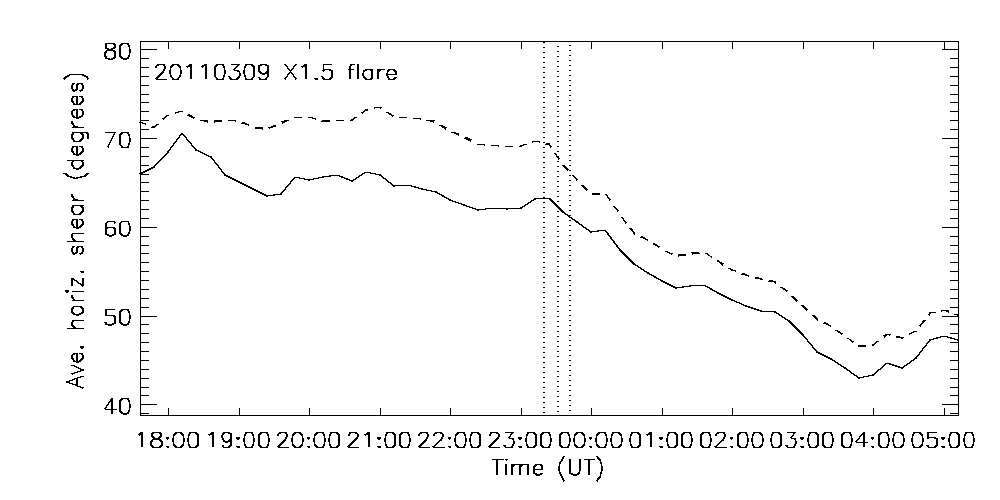}}
\resizebox{0.49\textwidth}{!}{\includegraphics*{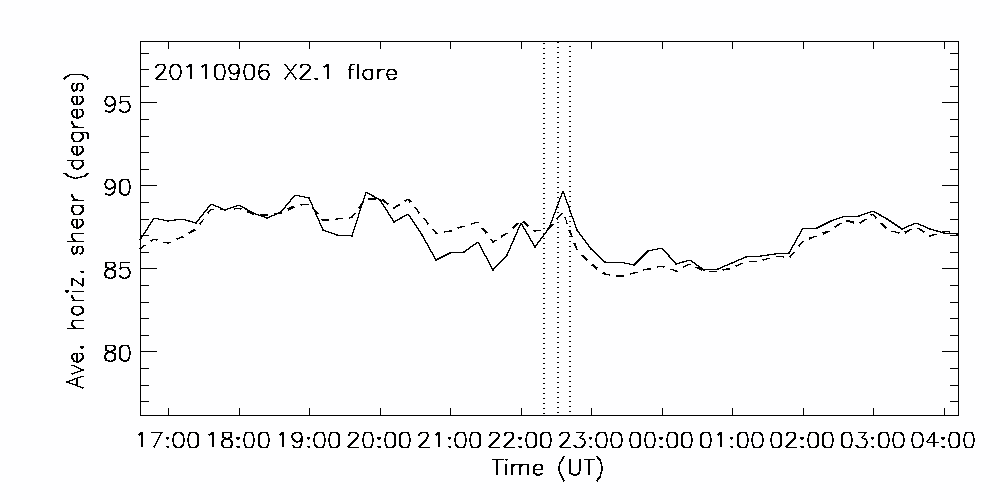}}
\resizebox{0.49\textwidth}{!}{\includegraphics*{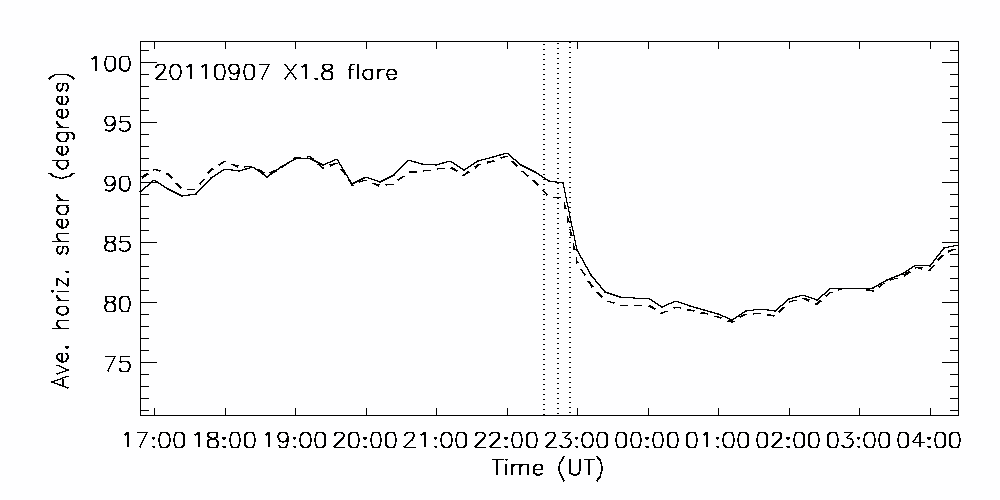}}
\resizebox{0.49\textwidth}{!}{\includegraphics*{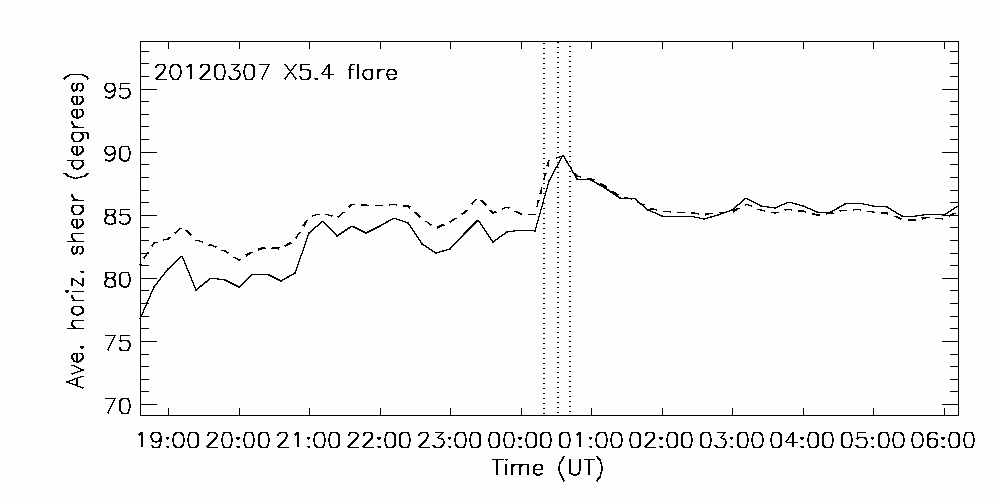}}
\end{center}
\caption{Shown here are the average horizontal shear angles (solid lines) and the field-weighted average horizontal shear angles (dashed lines) near each neutral line plotted against time. These quantities are averaged over the areas of integration indicated by the black rectangles in Figure~\ref{fig:br}. The vertical lines represent the GOES flare start, peak and end times. }
\label{fig:shearh}
\end{figure}

\begin{figure} 
\begin{center}
\resizebox{0.49\textwidth}{!}{\includegraphics*{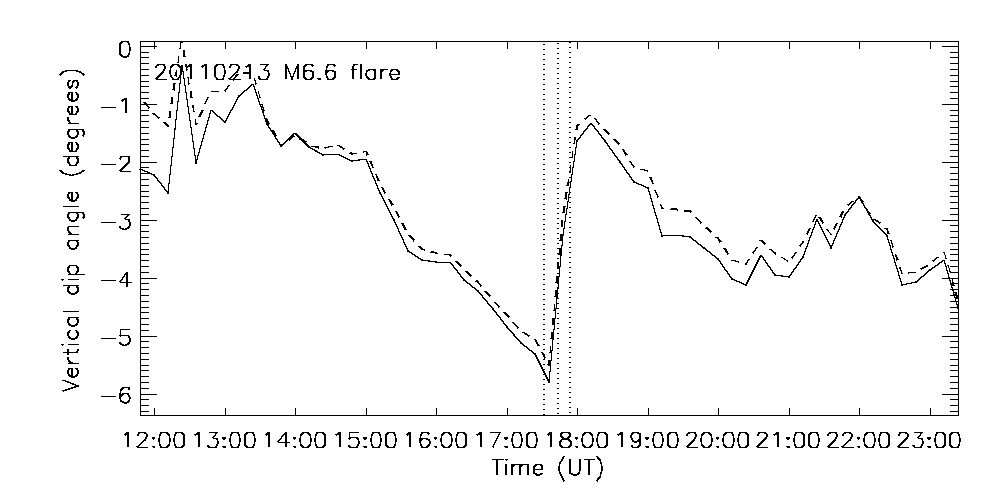}}
\resizebox{0.49\textwidth}{!}{\includegraphics*{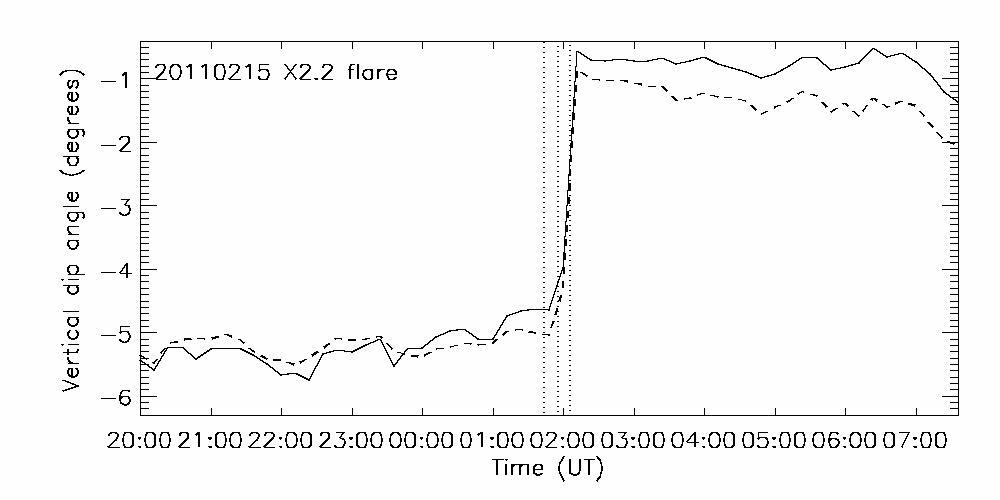}}
\resizebox{0.49\textwidth}{!}{\includegraphics*{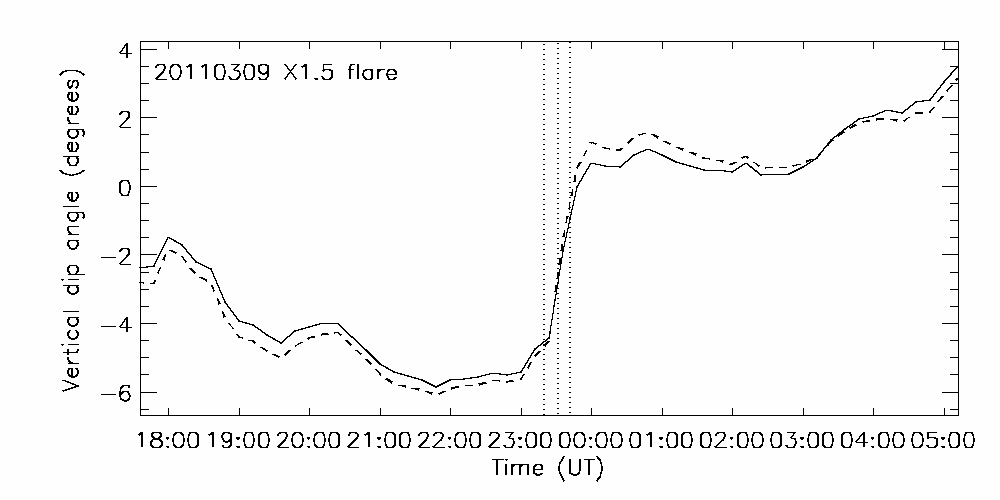}}
\resizebox{0.49\textwidth}{!}{\includegraphics*{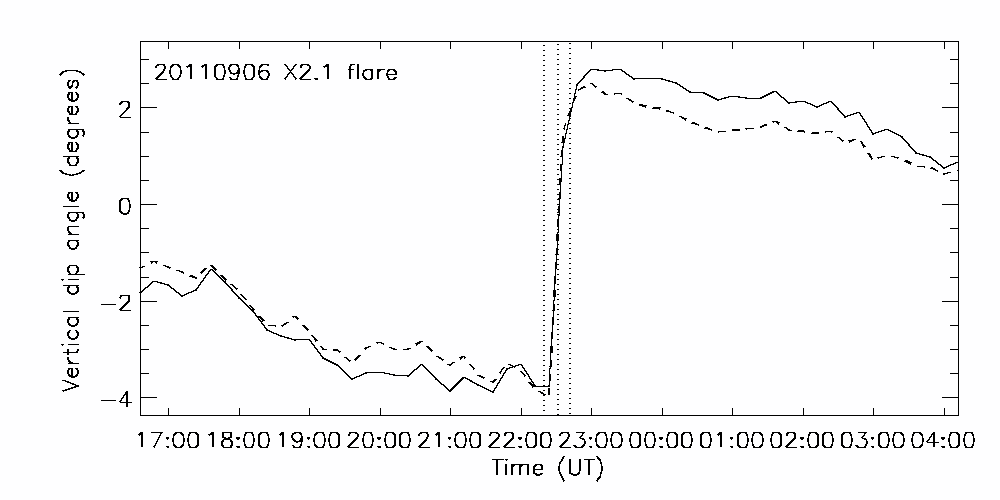}}
\resizebox{0.49\textwidth}{!}{\includegraphics*{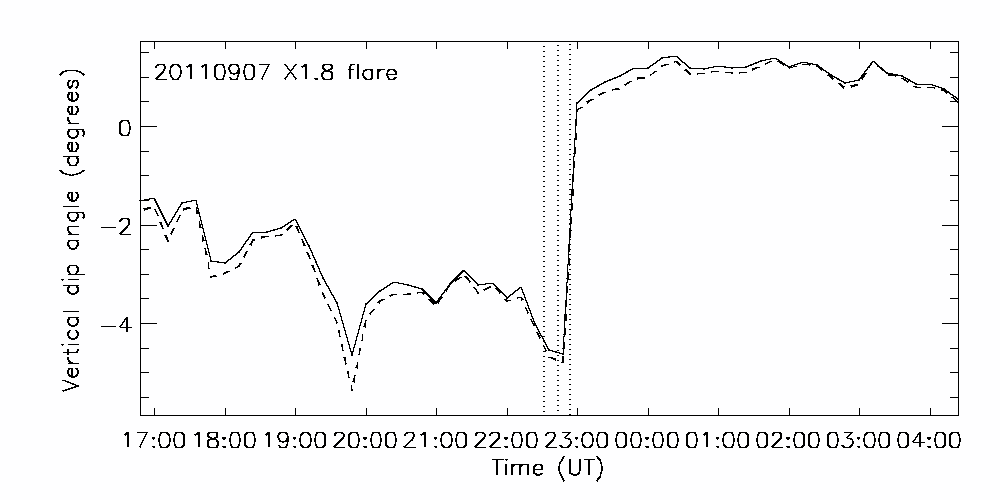}}
\resizebox{0.49\textwidth}{!}{\includegraphics*{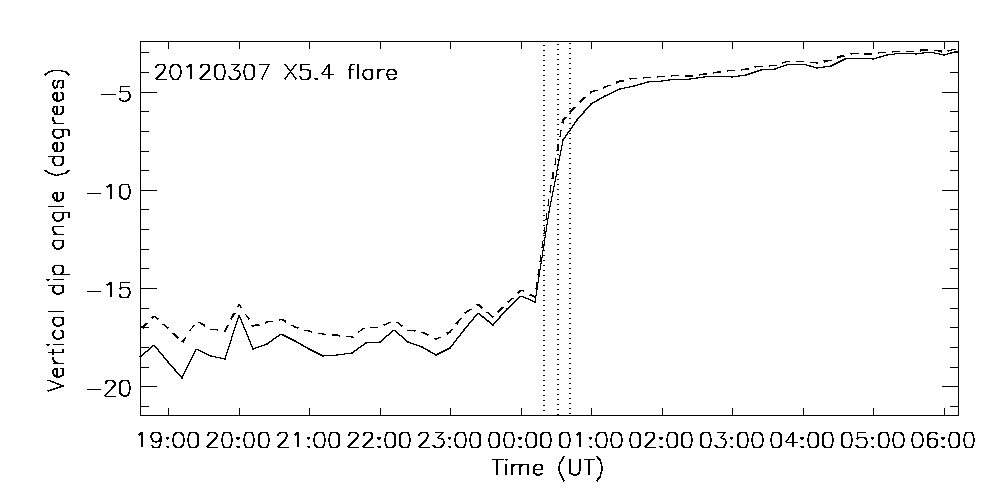}}
\end{center}
\caption{Shown here are the average vertical dip angles near each neutral line plotted against time. These angles are averaged over the areas of integration indicated by the black rectangles in Figure~\ref{fig:br}. The vertical lines represent the GOES flare start, peak and end times. }
\label{fig:shearv}
\end{figure}

Although potential fields do not accurately model the highly fluid-dominated magnetic field of the solar photosphere, they do represent Maxwell-stress-free states that are useful for diagnosing the stresses in observed non-potential fields. We compute the shear of the measured photospheric field with respect to a reference potential field as did Wang et al.~(1992), Gosain and Venkatakrishnan~(2012), Wang et al.~(2012) among other authors. The full and horizontal magnetic shear angles are defined by,

\begin{eqnarray}
S & = & \cos^{-1} [({\bf B}\cdot{\bf B}^\mathrm{p} ) / (B B^\mathrm{p})] ,\label{eq:shear} \\
S_\mathrm{h} & = & \cos^{-1} [({\bf B}_\mathrm{h}\cdot{\bf B}^\mathrm{p}_\mathrm{h} ) / (B_\mathrm{h} B^\mathrm{p}_\mathrm{h})] ,
\label{eq:shearh}
\end{eqnarray}

\noindent where ${\bf B}^\mathrm{p}$ is the unique potential field such that $B^\mathrm{p}_r = B_r$. In equations~(\ref{eq:shear}) and (\ref{eq:shearh}), $B^\mathrm{p} = [(B_r^\mathrm{p})^2+(B_{\theta}^\mathrm{p})^2+(B_{\phi}^\mathrm{p})^2]^{1/2}$, ${\bf B}^\mathrm{p}_\mathrm{h} = (B^\mathrm{p}_{\phi}, -B^\mathrm{p}_{\theta})$ and $B^\mathrm{p}_\mathrm{h} = [(B_{\theta}^\mathrm{p})^2+(B_{\phi}^\mathrm{p})^2]^{1/2}$. The temporal behavior of the average full and horizontal shear angles for each flare is plotted in Figures~\ref{fig:shear} and \ref{fig:shearh}, derived by averaging $S$ and $S_\mathrm{h}$ over the boxes around the neutral lines shown in Figure~\ref{fig:br}. Also shown are the field-strength-weighted full and horizontal shears, calculated in a manner analogous to Equation~(\ref{eq:tiltwt}). The average of the magnetic shear $S^B=BS$ is also plotted in Figure~\ref{fig:shear}, showing that this quantity is dominated by the behavior of $S$ during every flare. From Figures~\ref{fig:shear} and \ref{fig:shearh} it is clear that the full shear angles are smaller than the horizontal shear angles, but the changes in $S_\mathrm{h}$ are less significant than the changes in $S$. The plots in Figures~\ref{fig:shear} and \ref{fig:shearh} do not follow the dominant patterns of the azimuthal angular displacements in Figure~\ref{fig:azimdiffnlt}. There is some evidence that the full shear increased abruptly during the 2011 February 15 X2.2 and 2012 March 7 X5.4 flares. Three of the other flares did not show a significant, abrupt change in shear. The 2011 September 7 X1.8 flare field showed a marked permanent decrease in shear, particularly horizontal shear. However, this decrease began significantly before the flare start time and ended after the flare end time, making it difficult to attribute the decrease in shear to the flare alone. In fact at the height of the flare the steep decline of the shear may have been interrupted somewhat, but there seems to have been no significant increase of shear. The stepwise decrease in shear around the time of this flare appears to contradict the increasing azimuthal angular displacement show in Figure~\ref{fig:azimdiffnlt}. As seen in Figure~\ref{fig:dbr}, the field changes resulted in enhanced vertical fields on both sides of the neutral line that were distributed asymmetrically on both sides of the neutral line: on the north (positive) side the enhancements occurred at the east end of the neutral line while they occurred mostly at the west end on the south side. As a result, the potential field associated with the pre-flare measurements crosses the neutral line at a greater angle than the potential field based on measurements taken after the main field changes took place. Thus, even though the azimuthal displacement increased during this flare, the difference between the measured and potential horizontal fields actually decreased on average and so the shear angle decreased. In summary, the only significant and permanent shear changes directly attributable to the flares are the shear increases occurring during the 2011 February 15 X2.2 and 2012 March 7 X5.4 flares. As Table~\ref{fieldtable} shows, the changes in azimuthal displacement show a more consistent pattern than the shear changes.

We next compare the difference $\Delta\gamma =\gamma -\gamma^\mathrm{p}$ between the observed tilt angle $\gamma$ and the potential field tilt angle $\gamma^\mathrm{p}=\tan^{-1} ([(B^\mathrm{p}_{\theta})^2+(B^\mathrm{p}_{\phi})^2]^{1/2}/|B^\mathrm{p}_r|)$. We refer to these parameters as the average and field-strength-averaged dip angles. The average and field-strength-weighted average of $\Delta\gamma$ are plotted against time for each flare in Figure~\ref{fig:shearv}. A negative dip angle represents a field ${\bf B}$ more vertical on average than the reference field ${\bf B}^\mathrm{p}$. These parameters clearly show uniform behavior over the whole data set. In each case the dip angle abruptly increased by several degrees during the flare, from a negative value to a value close to zero. During the 2011 September 6 X2.1 flare the dip angle abruptly changed from about $-4^{\circ}$ to about $2^{\circ}$ before relaxing close to $0^{\circ}$ over the next five hours. Before the 2012 March 7 X5.4 flare the dip angle was between $-15^{\circ}$ and $-20^{\circ}$ and it abruptly rose to about $-5^{\circ}$ during the flare. Over the five-hour period before the 2011 February 13 M6.6 flare the dip angle steadily decreased from about $-1^{\circ}$ to about $-6^{\circ}$, before snapping back to about $-2^{\circ}$ during the flare. In all cases, therefore, the pre-flare neutral-line field was significantly less tilted (more vertical) than the corresponding potential field and, during each flare, this stress on the field was suddenly relaxed and the observed tilt angle more closely matched the tilt angle of the potential field. This pattern is consistent with the results of Gosain and Venkatakrishnan~(2010) based on Hinode/SOT vector data for the 2006 December 13 X3.4 flare.

As Table~\ref{fieldtable} shows, these results for the dip angles are more uniform than the tilt angle profiles in Figure~\ref{fig:tiltnlt} whose left panels show more modest signs of change. The September 7 X1.8 flare tilt angle profile is complicated by the increase of vertical flux shown in Figure~\ref{fig:frnlt}. The dip angle profile of this flare shows a much clearer stepwise change. This is because the sudden increase of vertical flux during the flare causes the potential field to become more vertical, with tilt angle closer to the tilt angle of the observed field. The abrupt relaxation of the field tilts towards the potential-field values is consistent with the interpretation of Hudson, Fisher and Welsch~(2008) the coronal field reconnecting, relaxing and contracting into a simpler and more tilted configuration.

\section{The electric current}
\label{s:current}

\begin{figure} 
\begin{center}
\resizebox{0.49\textwidth}{!}{\includegraphics*{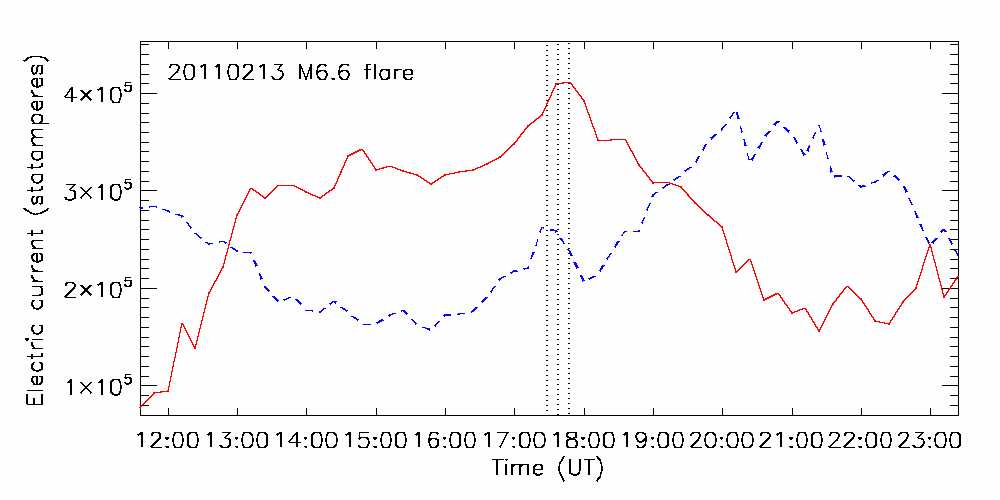}}
\resizebox{0.495\textwidth}{!}{\includegraphics*{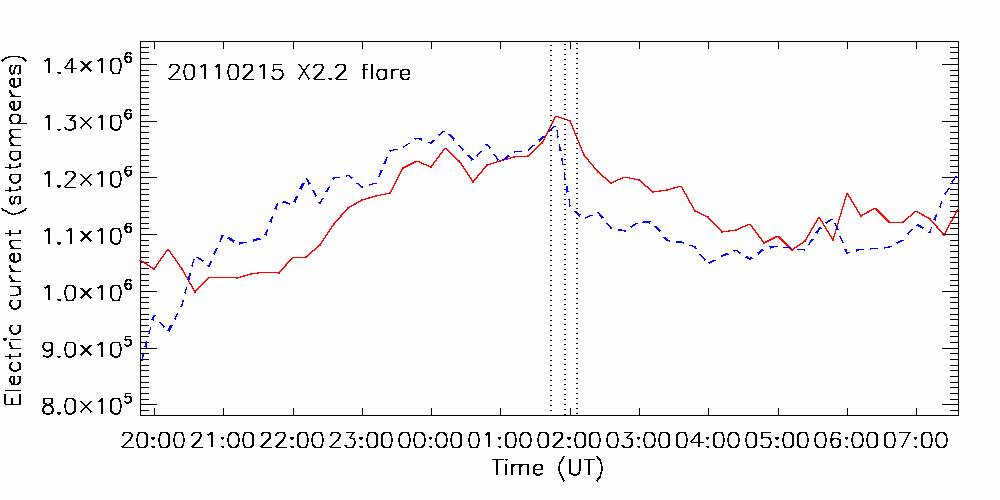}}
\resizebox{0.49\textwidth}{!}{\includegraphics*{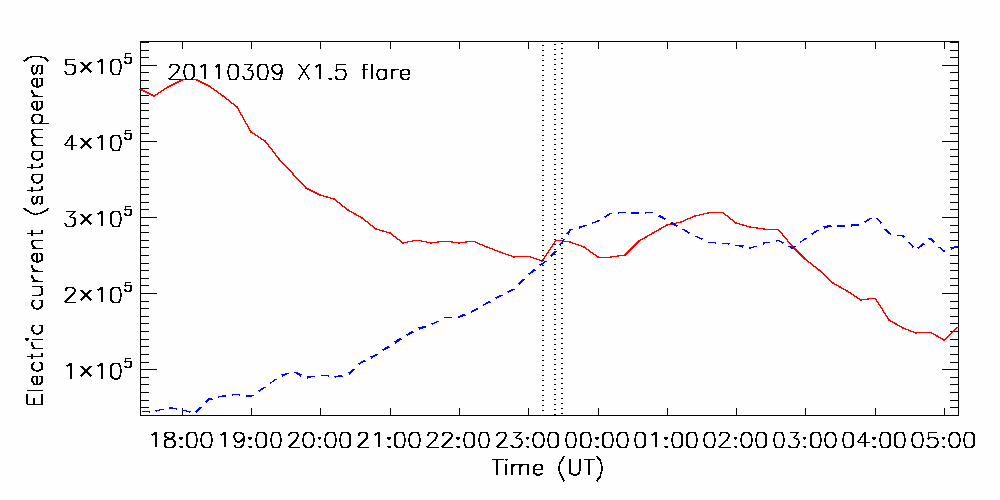}}
\resizebox{0.495\textwidth}{!}{\includegraphics*{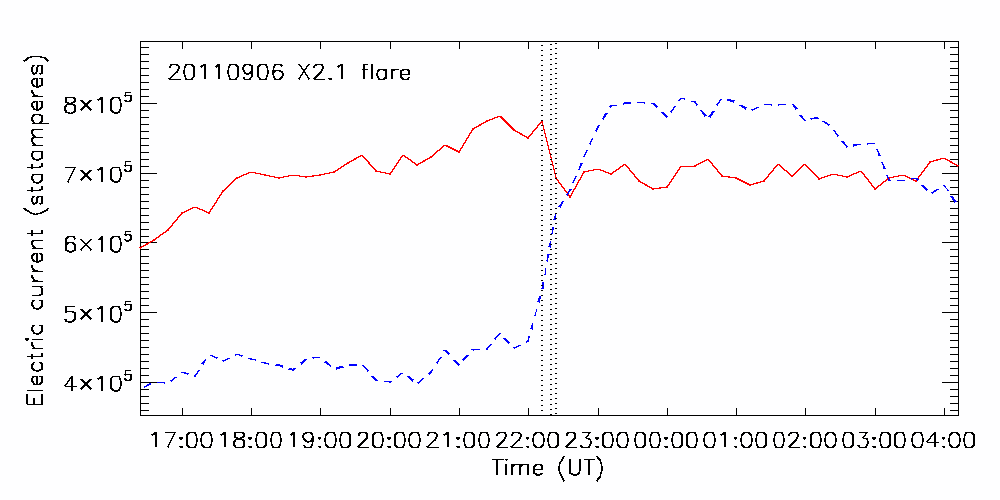}}
\resizebox{0.49\textwidth}{!}{\includegraphics*{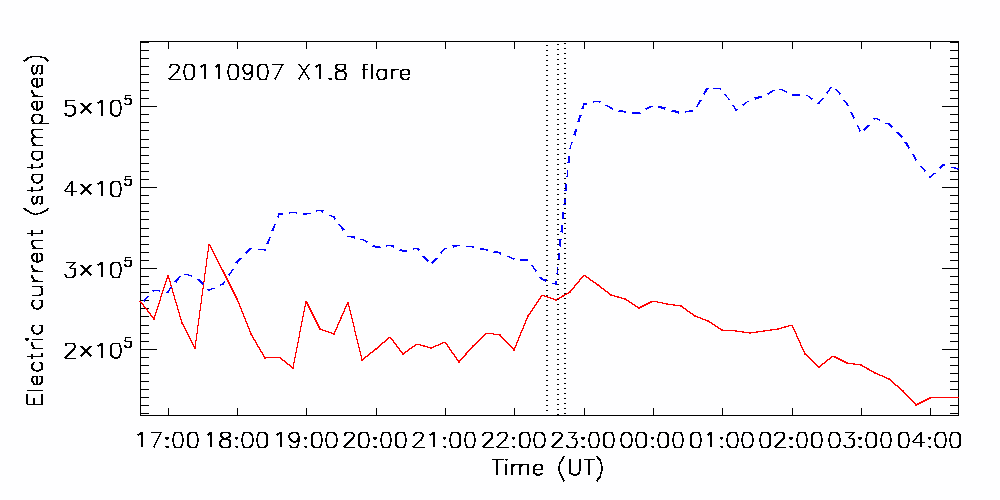}}
\resizebox{0.495\textwidth}{!}{\includegraphics*{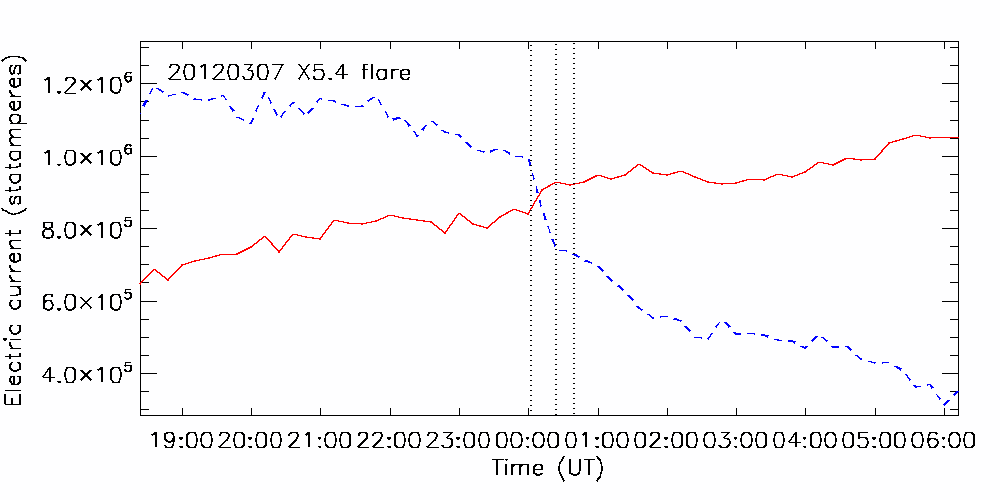}}
\end{center}
\caption{The integrated vertical electric current flux near each neutral line, $J_r^\mathrm{NL}$ is plotted here for each flare as a function of time. The red/blue solid/dashed lines represent positive/negative current. The vertical lines represent the GOES flare start, peak and end times. The areas of integration for the neutral-line calculations are indicated by the black rectangles in Figure~\ref{fig:br}. }
\label{fig:jrnlt}
\end{figure}

Figure~\ref{fig:jrnlt} shows the evolution of the integrated vertical current flux, $J_r^\mathrm{NL}$, near the neutral line for each flare. The current profiles in the figure are rather diverse but there are similarities between the electric current profiles of the two flares occurring in AR~11158, the 2011 February 13 and 15 flares, and also between those occurring in AR~11283, the 2011 September 6 and 7 flares. During both the 2011 September 6 and 7 flares the negative current increased abruptly and permanently, creating a more balanced current in one case and a more imbalanced current in the other. These changes appear to be related to the increases in positive vertical magnetic flux during both flares, shown in Figure~\ref{fig:frnlt}. Also the positive current decreased in a stepwise manner during the September 6 flare. The negative current decreased abruptly during the 2012 March 7 X5.4 flare. These last two changes were more significant than any changes in the vertical magnetic flux (Figure~\ref{fig:frnlt}). The vertical current reached local maxima during the 2011 February 13 and 15 flares but there was only an abrupt, permanent change in the negative current during the February 15 flare. These current maxima were not  accompanied by corresponding peaks in the magnetic flux profiles in Figure~\ref{fig:frnlt} and so may represent magnetic stresses that were built up before the flares and released during them. A similar interpretation may apply to the current reduction during the 2012 March 7 X5.4 flare. On the other hand, during the flares when the electric current was observed to increase abruptly near the neutral line of AR~11283, the 2011 September 6 and 7 flares, the full and horizontal magnetic shear angles (Figures~\ref{fig:shear} and \ref{fig:shearh}) either did not change permanently (September 6) or actually decreased (September 7). The interpretation of the magnetic changes during these flares is complicated by the abrupt and permanent vertical flux increases seen in Figure~\ref{fig:frnlt} which changed the reference potential field significantly as discussed in Section~\ref{s:potential}. This is likely related to the fact that the vertical magnetic flux increased near the neutral line during both flares. Recall from Section~\ref{s:potential} that the potential field associated with the September 7 pre-flare measurements crossed the neutral line at a greater angle than the potential field based on measurements taken after the main field changes took place, accounting for the reduction in magnetic shear there. The increase in current during the September 6 and 7 flares may therefore be mostly due to the increase in field strength near the neutral line of AR~11283 during each flare. Finally, the March 9 flare appears not to have a significant electric current signature. It is clear that the flare-related electric current changes, like the shear changes, do not follow a general pattern.

\section{The Lorentz force changes}
\label{s:lorentzfch}

\begin{table}
\scriptsize
\caption{Directions of Lorentz force changes.}\
\label{dftable}
\\
\begin{tabular}{lcccccccc}
\hline\hline
 & North & South & ${\bf B}_{\parallel}$ Axial & North $\delta{\bf F}_{\parallel}$ & South $\delta{\bf F}_{\parallel}$ & North $\delta{\bf F}_{\perp}$ & South $\delta{\bf F}_{\perp}$ \\
Date (UT) & Polarity & Polarity & Direction & Direction & Direction & Direction & Direction \\
\hline
2011 Feb 13 & $-$ & + & E & W & {\bf E} & S & N \\
2011 Feb 15 & $-$ & + & E & W & {\bf E} & S & $-$ \\
2011 Mar 9 & + & $-$ & W & {\bf W} & $-$ & S & N \\
2011 Sep 6 & + & $-$ & W & {\bf W} & E & $-$ & N \\
2011 Sep 7 & + & $-$ & W & {\bf W} & E & S & $-$ \\
2012 Mar 7 & + & $-$ & E & {\bf E} & W & $-$ & $-$ \\
\hline
\end{tabular}
\end{table}

We use the results of Fisher et al.~(2012) to estimate the changes in the Lorentz force vector acting on the volume below the photosphere as a result of each flare. Assuming that the photospheric vector field is observed over a photospheric area $A_\mathrm{ph}$ at two times, $t=0$ before the field changes begin, and $t=\delta t$ after the main field changes have occurred, the corresponding changes in the Lorentz force vector components between these two times are given by Equations~(17) and (18) of Fisher et al.~(2012):

\begin{equation}
\delta F_r = {1 \over 8 \pi} \int_{A_\mathrm{ph}} ( \delta B_r^2 - \delta B_\mathrm{h}^2 )\ \mathrm{d}A,
\label{eq:deltafr}
\end{equation}
and
\begin{equation}
\delta {\bf F}_\mathrm{h} = {1 \over 4 \pi} \int_{A_\mathrm{ph}} \delta ( B_r {\bf B}_\mathrm{h} )\ \mathrm{d}A,
\label{eq:deltafh}
\end{equation}
where at a fixed location in the photosphere
\begin{eqnarray}
\delta B_\mathrm{h}^2 & = & B_\mathrm{h}^2(\delta t) -B_\mathrm{h}^2(0)\ ,\\
\delta B_r^2 & = & B_r^2(\delta t) -B_r^2(0)\ ,\\
\delta ( B_r {\bf B}_\mathrm{h} ) & = & B_r(\delta t){\bf B}_\mathrm{h}(\delta t) - B_r(0) {\bf B}_\mathrm{h}(0)\ .
\end{eqnarray}

\noindent The Lorentz force acting on the atmosphere above the photosphere is equal and opposite to the force acting on the volume at and below the photosphere (Fisher et al.~2012).

\begin{figure} 
\begin{center}
\resizebox{0.49\textwidth}{!}{\includegraphics*{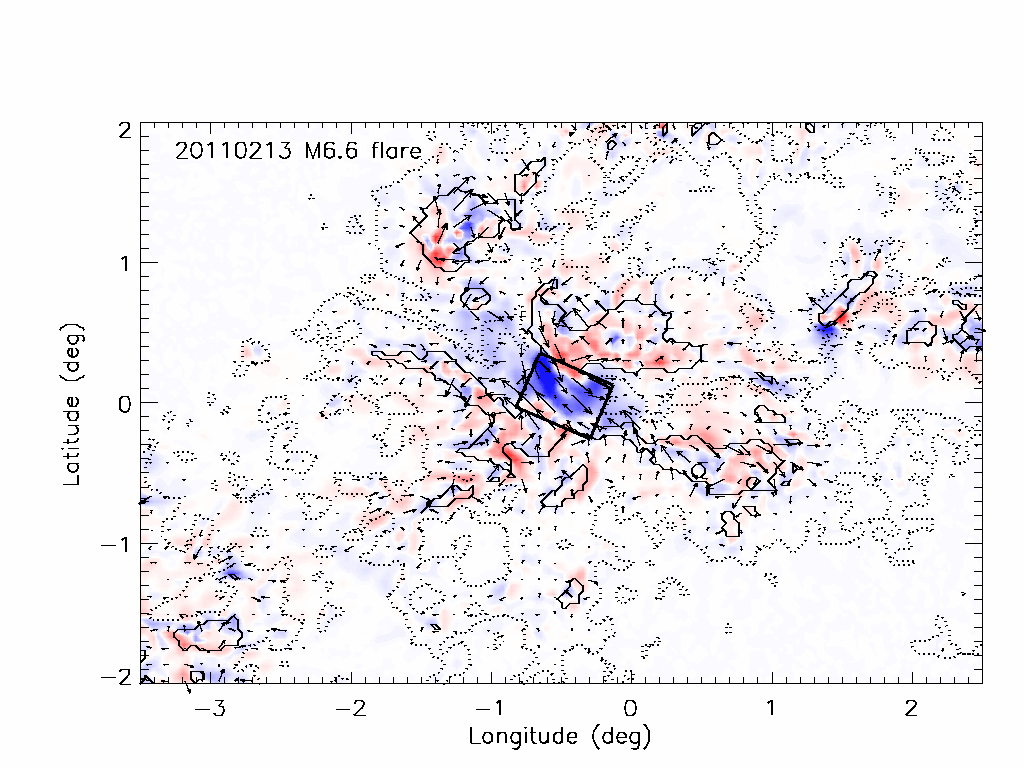}}
\resizebox{0.49\textwidth}{!}{\includegraphics*{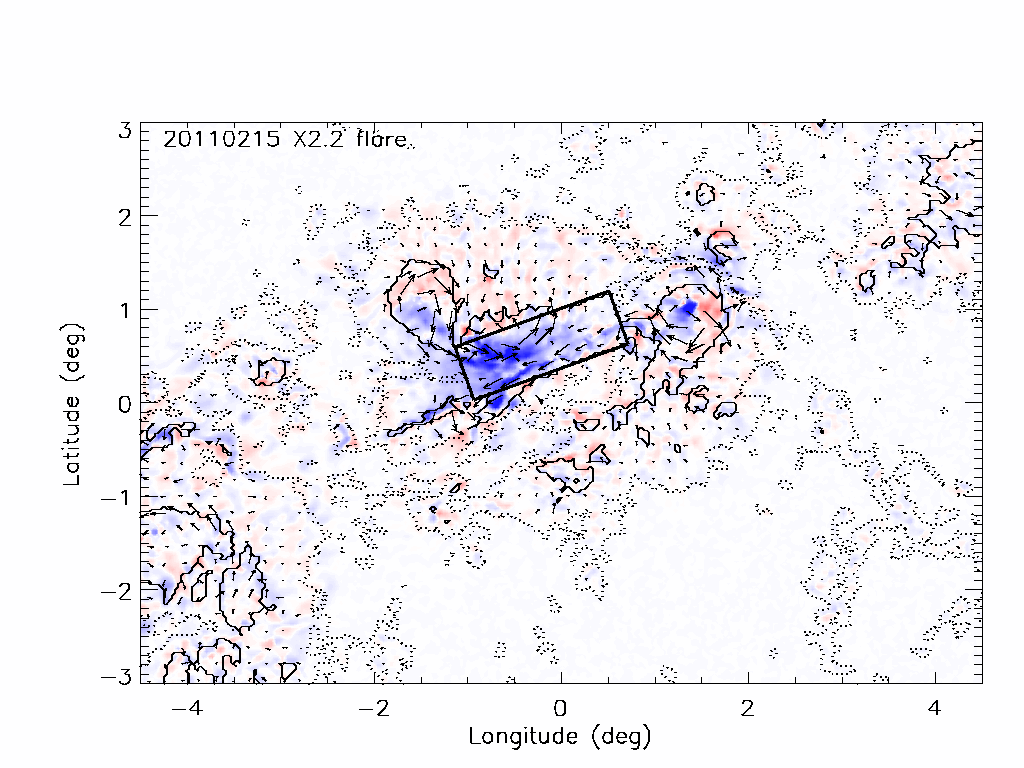}}
\resizebox{0.49\textwidth}{!}{\includegraphics*{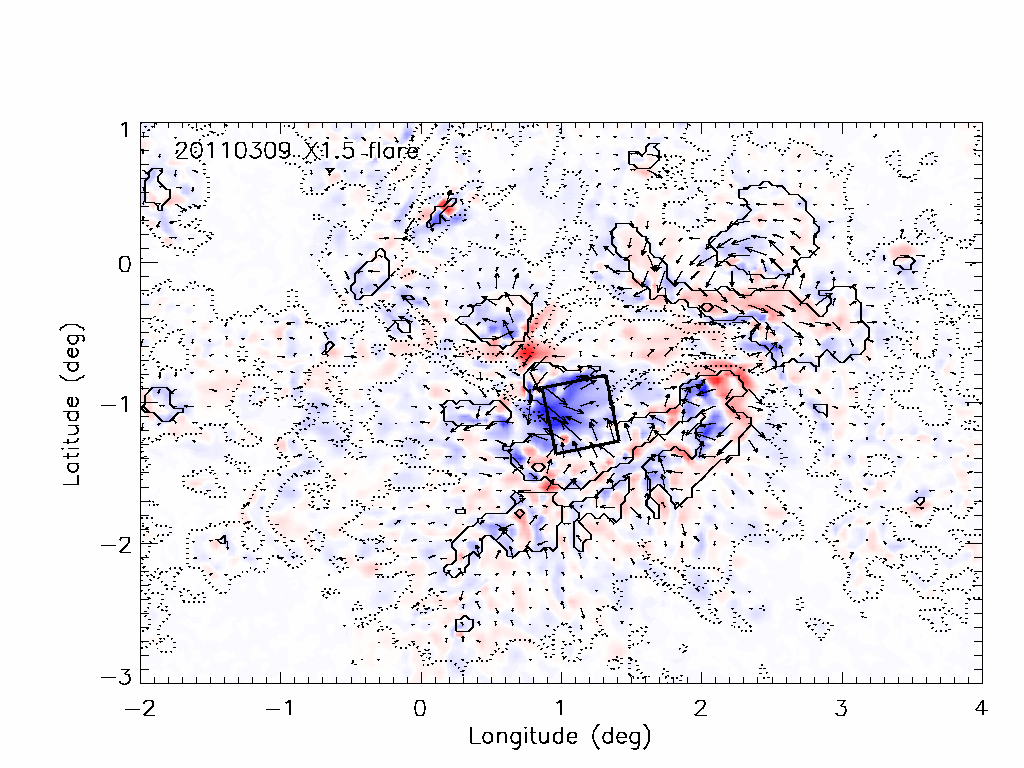}}
\resizebox{0.49\textwidth}{!}{\includegraphics*{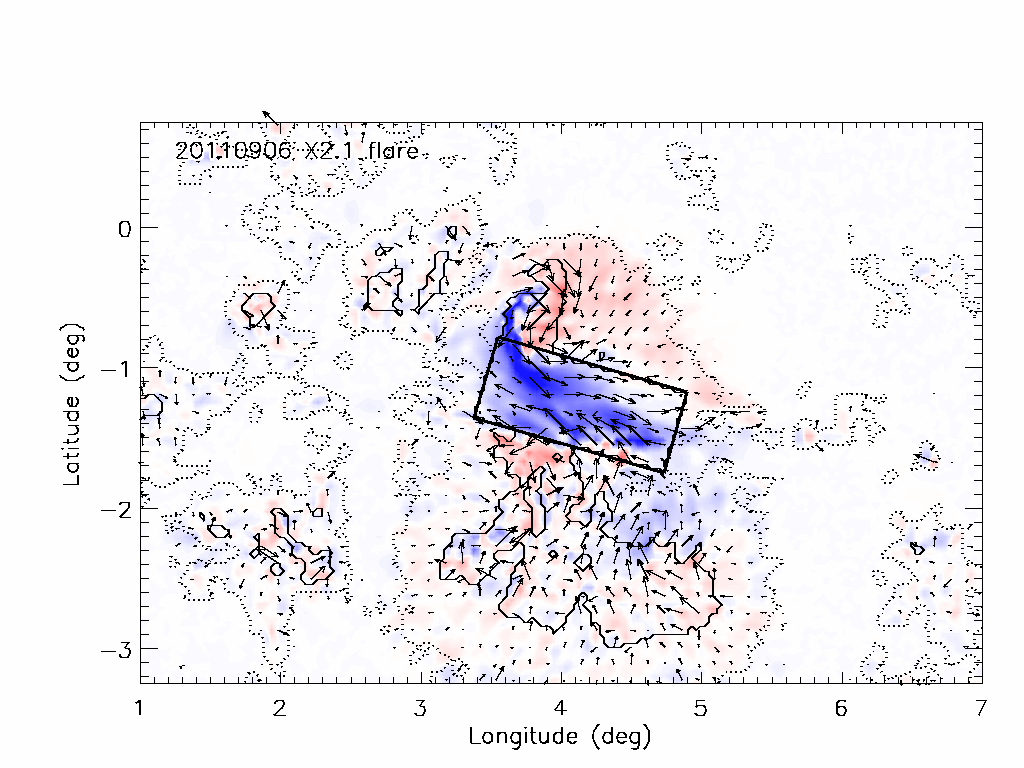}}
\resizebox{0.49\textwidth}{!}{\includegraphics*{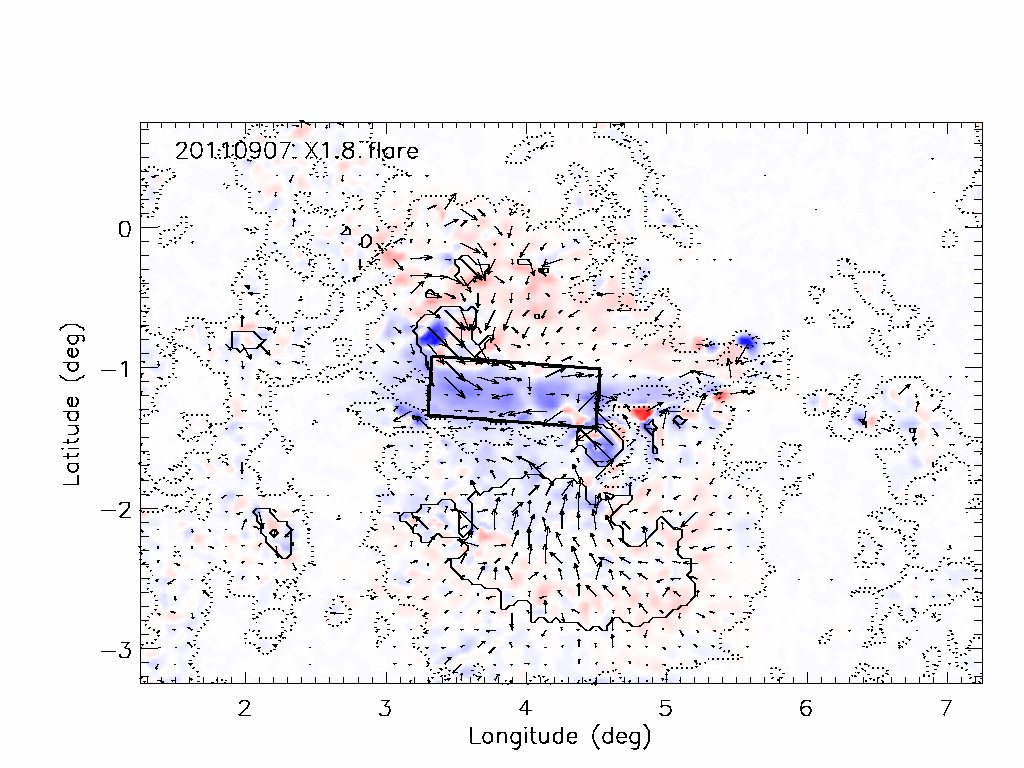}}
\resizebox{0.49\textwidth}{!}{\includegraphics*{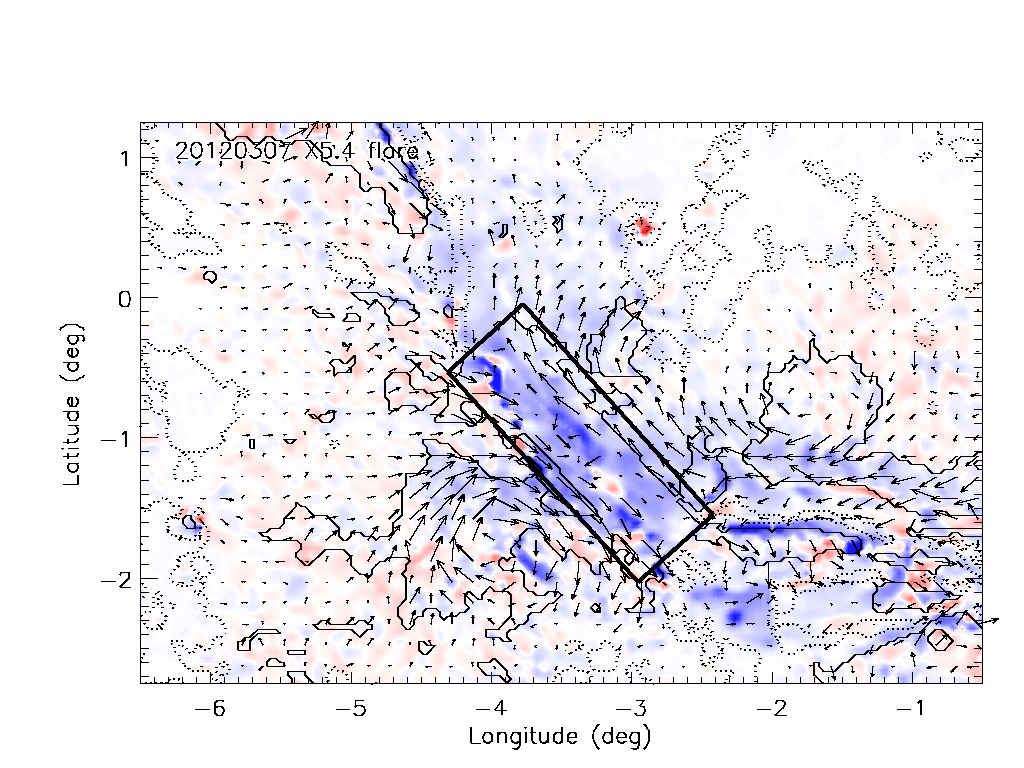}}
\end{center}
\caption{The Lorentz force vector changes during each flare. The vertical component, $\delta F_r$ is indicated by the color scale and the horizontal components by the arrows with saturation values $10^4$~dynes/cm$^2$ for the color scale and $2.5\times 10^3$~dynes/cm$^2$ for the arrows. Red/blue coloring represents positive/negative (upward/downward) Lorentz force change. The black rectangles mark the regions of major field change near the neutral lines. The solid and dotted contours indicate strong ($\mathrm{|}B_r\mathrm{|}>1000$~G) and quite strong  ($\mathrm{|}B_r\mathrm{|}>100$~G) fields, respectively.}
\label{fig:dfr}
\end{figure}

Figure~\ref{fig:dfr} shows for each flare a spatial map of the Lorentz force changes $F_r$ and ${\bf F}_\mathrm{h}$, derived by evaluating the integrals in Equations~(\ref{eq:deltafr},) and (\ref{eq:deltafh}) pixel by pixel. The sums of the distributions shown in Figure~\ref{fig:dfr} over the entire photospheric area gives the estimate for the total Lorentz force vector described by Fisher et al.~(2012). Petrie~(2012) argued that this calculation also gives a useful estimate of the spatial distributions of the Lorentz force vector components across the region for well-resolved changes in major magnetic structures. Alvarado-G\'omez et al.~(2012) gave an analogous argument for the corresponding volume integral within the solar interior.

Figure~\ref{fig:dfr} shows the spatial distributions of the changes in the Lorentz force components during the flares. As we saw in Figures~\ref{fig:frnlt} and \ref{fig:fparanlt}, the horizontal field changes $\delta{\bf B}_\mathrm{h}$ near the main neutral line increased the horizontal field strength, $\delta B_\mathrm{h}^2>0$, and were significantly greater than the vertical field changes $\delta B_r^2$. Equation~\ref{eq:deltafr} therefore leads us to expect the vertical Lorentz force changes to have been predominantly downward near the neutral lines. The plots in Figure~\ref{fig:dfr} confirm that near the main neutral lines the Lorentz forces clearly acted downwards into the photosphere during each flare. This behavior was anticipated to occur near neutral lines of flaring active regions by Hudson, Fisher and Welsch (2008) and Fisher et al.~(2012), and has been found in past estimates of Lorentz force changes by Wang and Liu~(2010), Petrie and Sudol~(2010) and Petrie~(2012). 

We now analyze the horizontal Lorentz force changes in a similar fashion. According to Equation~(\ref{eq:deltafh}), wherever the vertical field does not change significantly compared to the horizontal changes and is positive/negative, the horizontal Lorentz force changes $\delta{\bf F}_\mathrm{h}$ should be parallel/anti-parallel to the horizontal field changes $\delta {\bf B}_\mathrm{h}$. We already know from Figure~\ref{fig:dbr} that on both sides of the neutral line $\delta {\bf B}_\mathrm{h}$ pointed approximately parallel to the neutral line in all cases except the 2011 March 9 flare whose pattern is not so clear. Figure~\ref{fig:dfr} shows that, during the other five flares, the Lorentz force acted in opposite directions on each side of the neutral line, with the changes on the positive side pointing parallel to $\delta {\bf B}_\mathrm{h}$ and those on the negative side anti-parallel to $\delta {\bf B}_\mathrm{h}$ as expected.

\begin{figure} 
\begin{center}
\resizebox{0.5\textwidth}{!}{\includegraphics*{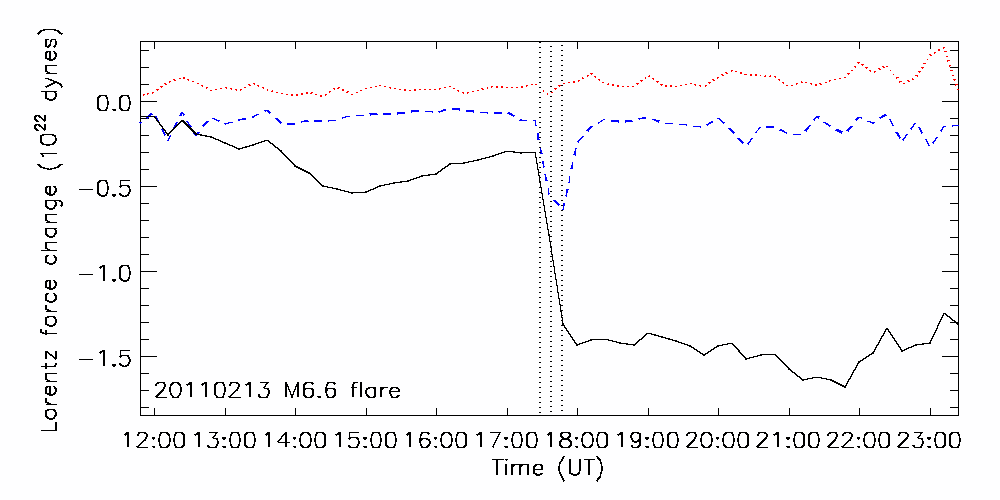}}
\resizebox{0.49\textwidth}{!}{\includegraphics*{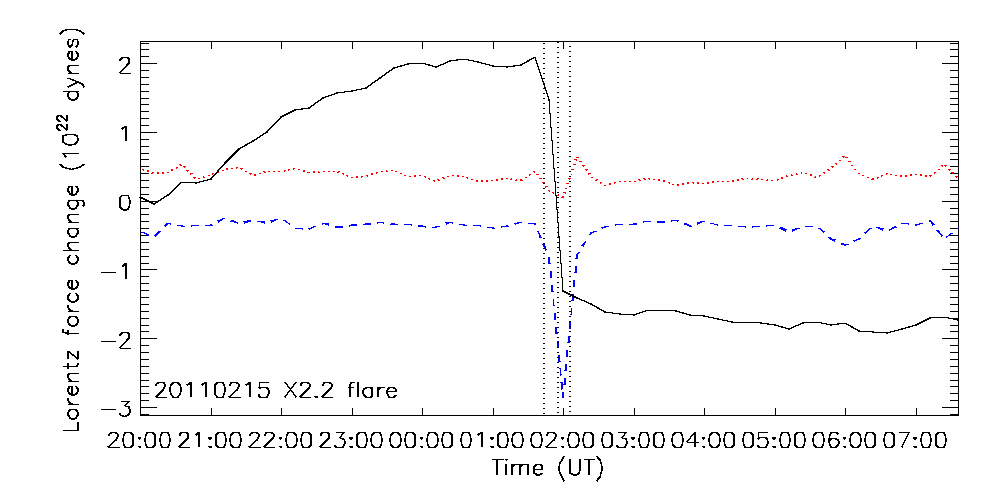}}
\resizebox{0.5\textwidth}{!}{\includegraphics*{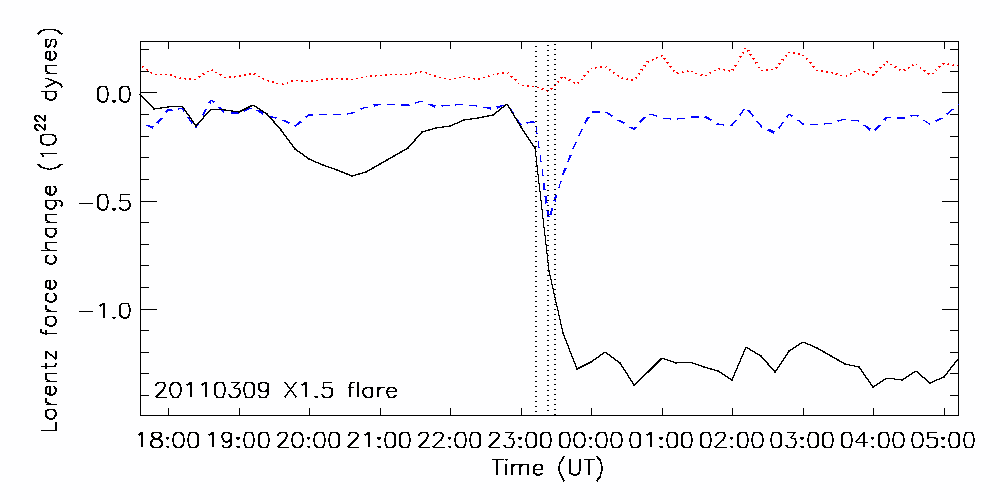}}
\resizebox{0.49\textwidth}{!}{\includegraphics*{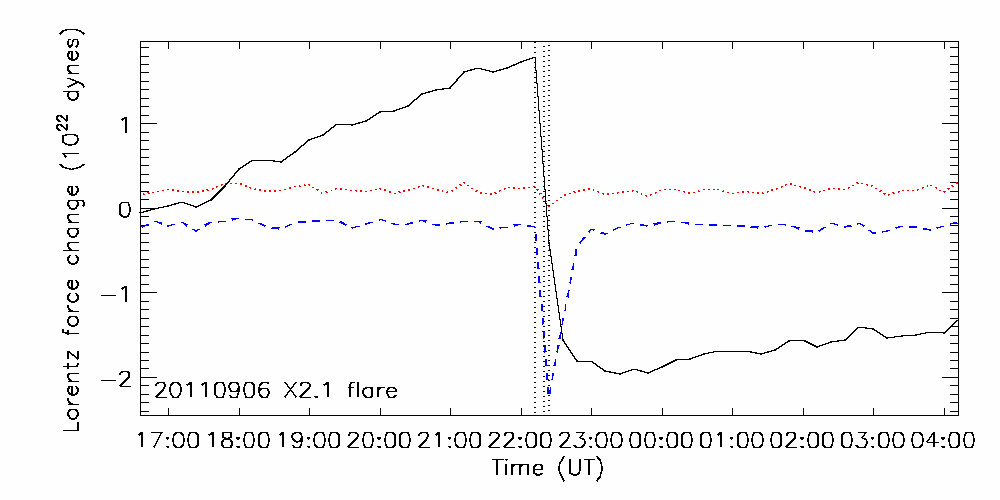}}
\resizebox{0.5\textwidth}{!}{\includegraphics*{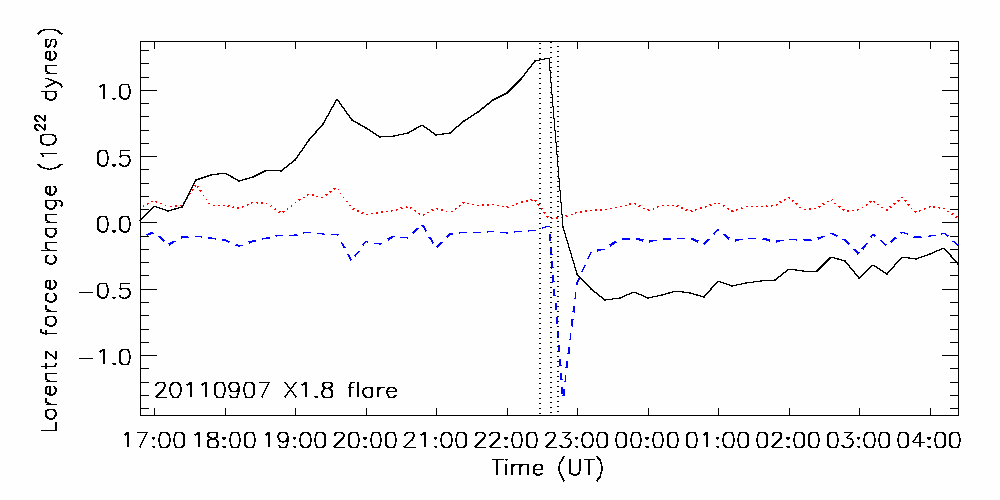}}
\resizebox{0.49\textwidth}{!}{\includegraphics*{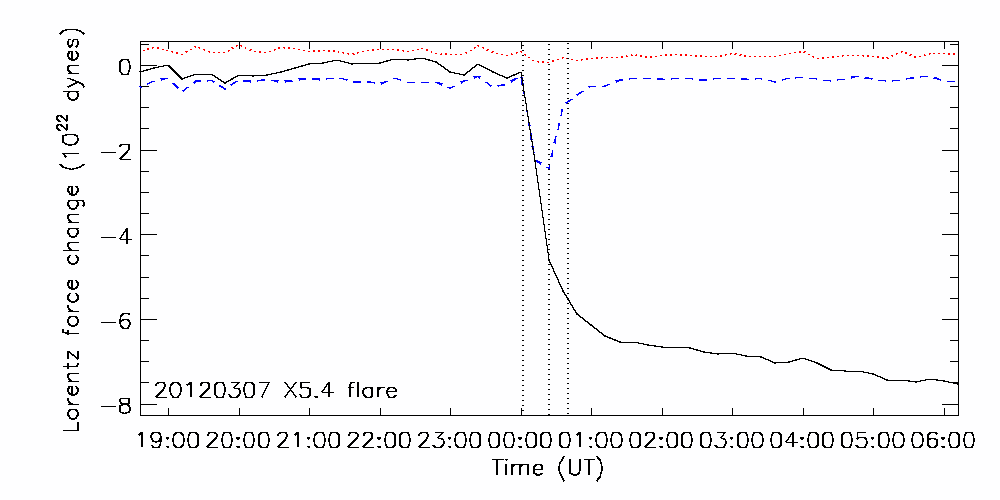}}
\end{center}
\caption{The Lorentz force vector components in the vertical direction, $\delta F_r^\mathrm{NL}$, are plotted as functions of time. The positive and negative running-difference changes are represented by the red dotted and blue dashed lines, respectively. The fixed-difference changes with respect to the first image are represented by the solid black lines. The areas of integration are indicated by the black rectangles in Figure~\ref{fig:br}. The red/blue dotted/dashed lines represent positive/negative force changes. The vertical lines represent the GOES flare start, peak and end times.}
\label{fig:dfrnlt}
\end{figure}

\begin{figure} 
\begin{center}
\resizebox{0.5\textwidth}{!}{\includegraphics*{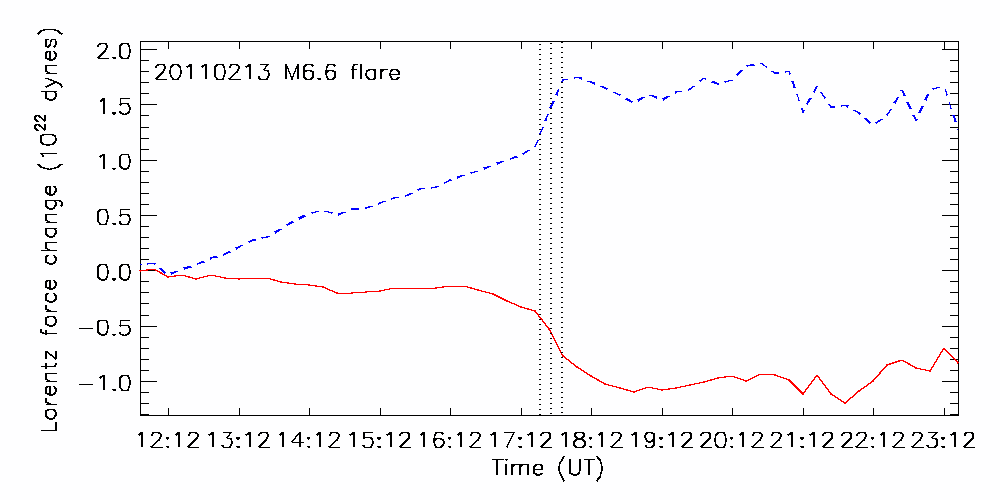}}
\resizebox{0.49\textwidth}{!}{\includegraphics*{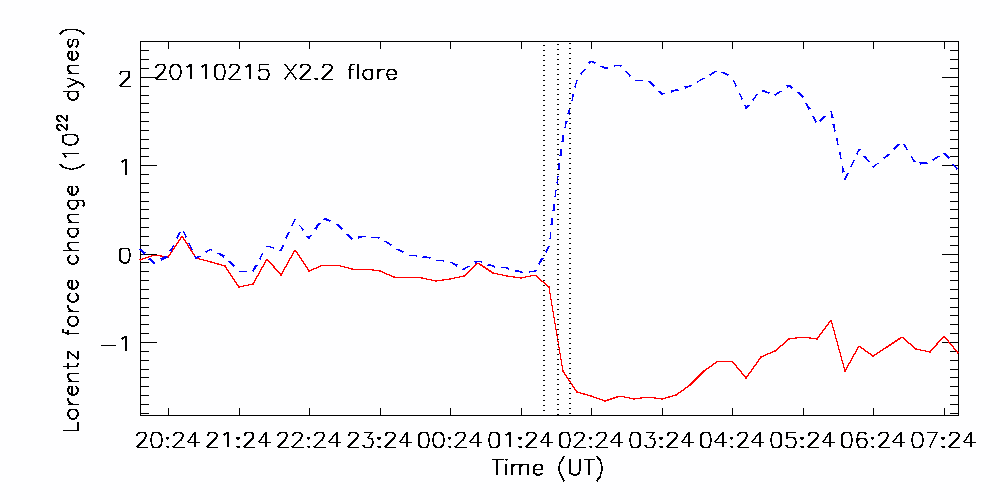}}
\resizebox{0.5\textwidth}{!}{\includegraphics*{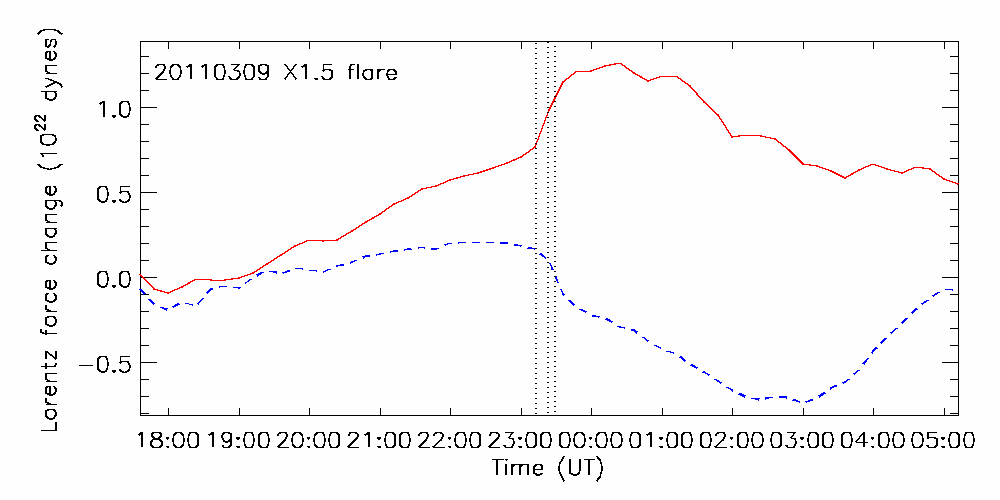}}
\resizebox{0.49\textwidth}{!}{\includegraphics*{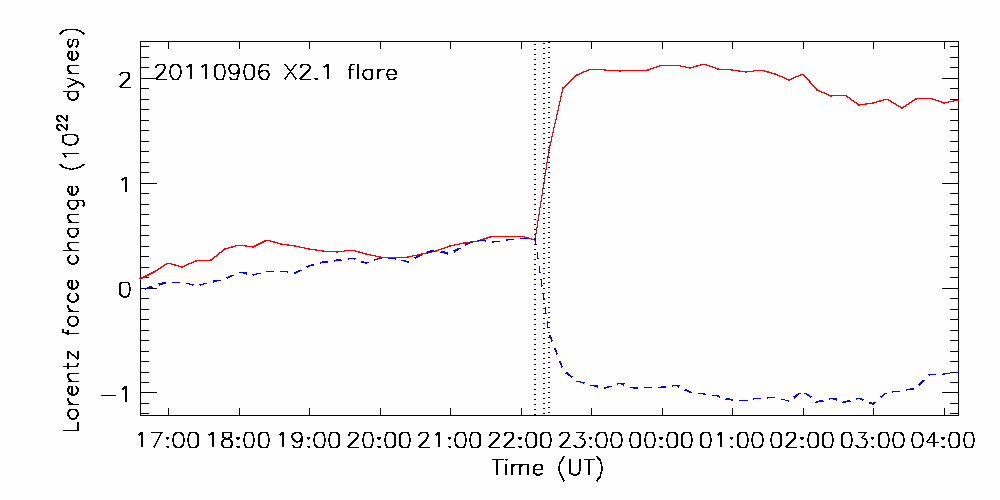}}
\resizebox{0.5\textwidth}{!}{\includegraphics*{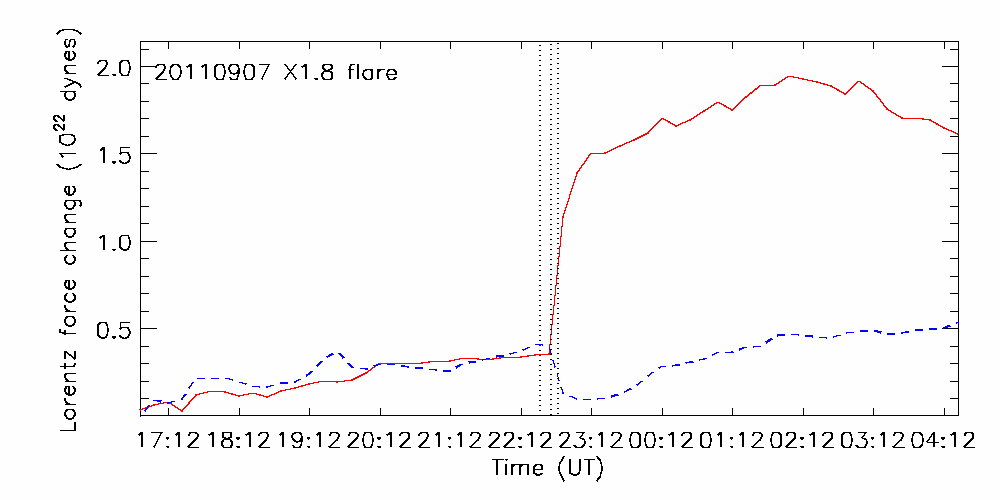}}
\resizebox{0.49\textwidth}{!}{\includegraphics*{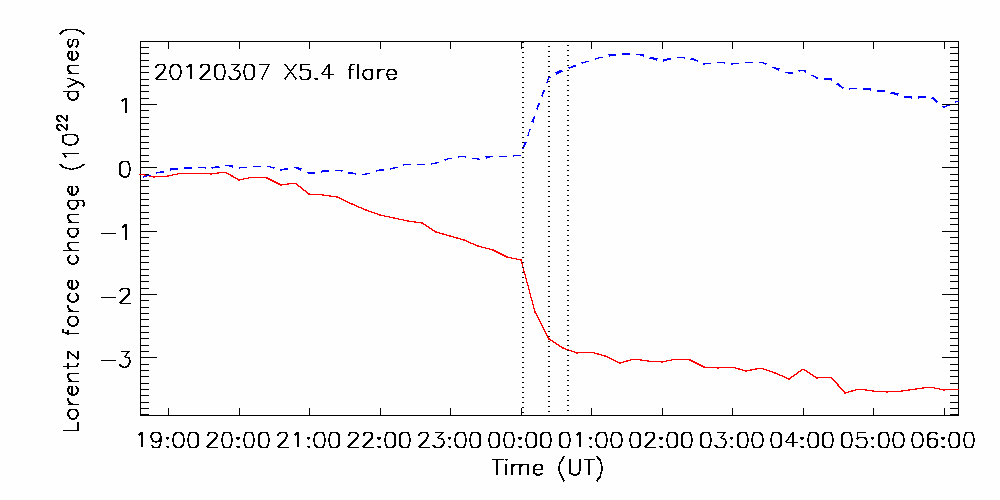}}
\end{center}
\caption{The Lorentz force vector components in the parallel direction, $\delta F_{\parallel}^\mathrm{NL}$, are plotted as functions of time. The red solid and blue dashed lines represent the force changes acting on the positive and negative sides of the neutral lines, respectively. The areas of integration are indicated by the black rectangles in Figure~\ref{fig:br}. Positive/negative force changes act in the approximately westward/eastward directions along the long edges of these rectangles, respectively. The vertical lines represent the GOES flare start, peak and end times.}
\label{fig:dfparasgnnlt}
\end{figure}

\begin{figure} 
\begin{center}
\resizebox{0.5\textwidth}{!}{\includegraphics*{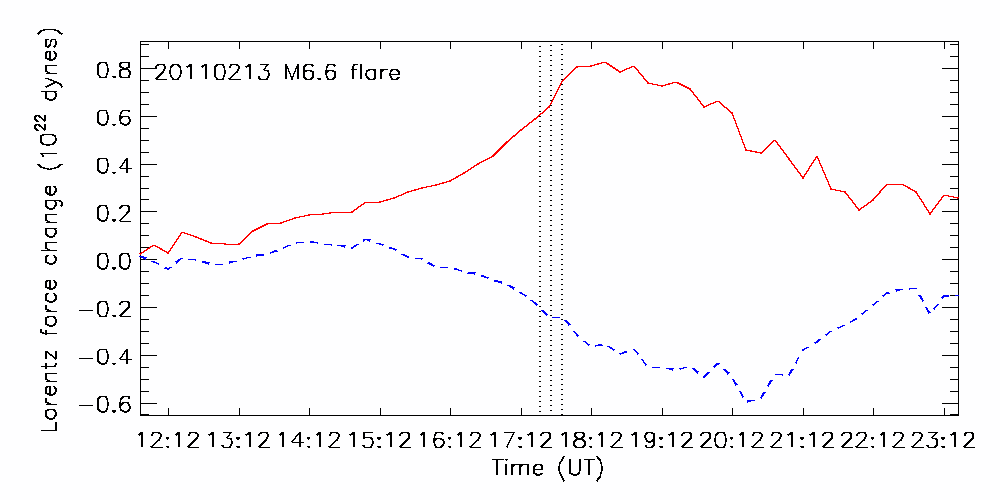}}
\resizebox{0.49\textwidth}{!}{\includegraphics*{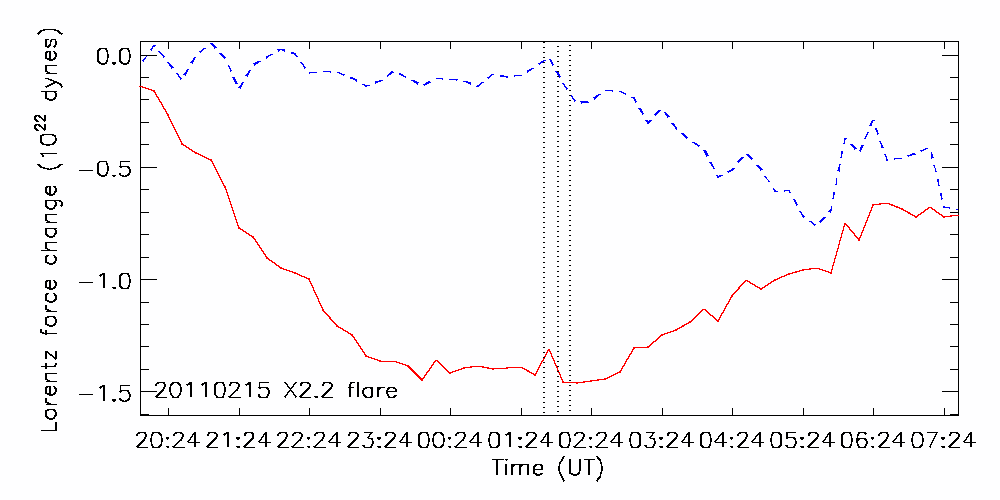}}
\resizebox{0.5\textwidth}{!}{\includegraphics*{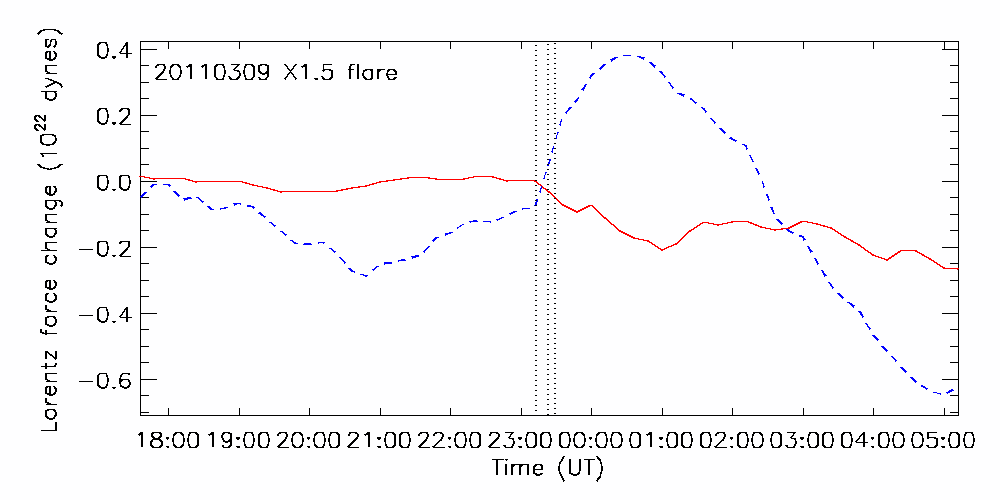}}
\resizebox{0.49\textwidth}{!}{\includegraphics*{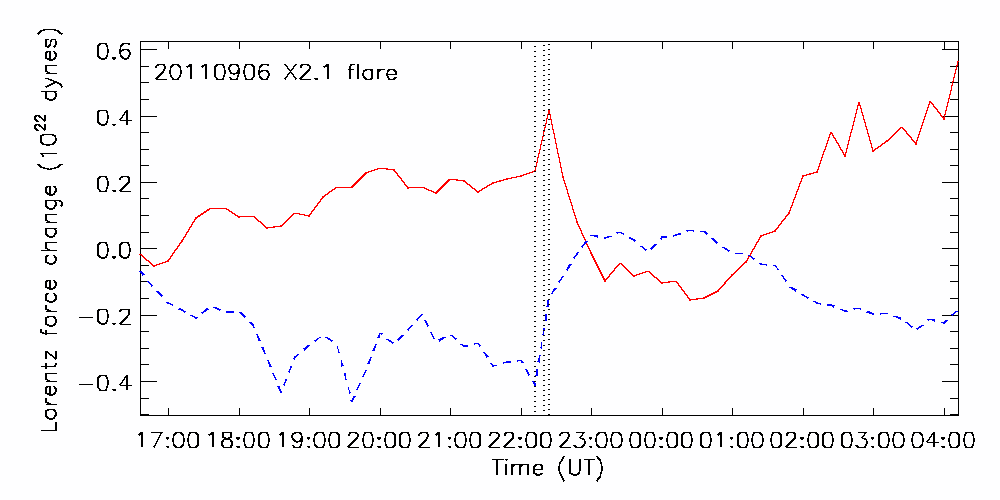}}
\resizebox{0.5\textwidth}{!}{\includegraphics*{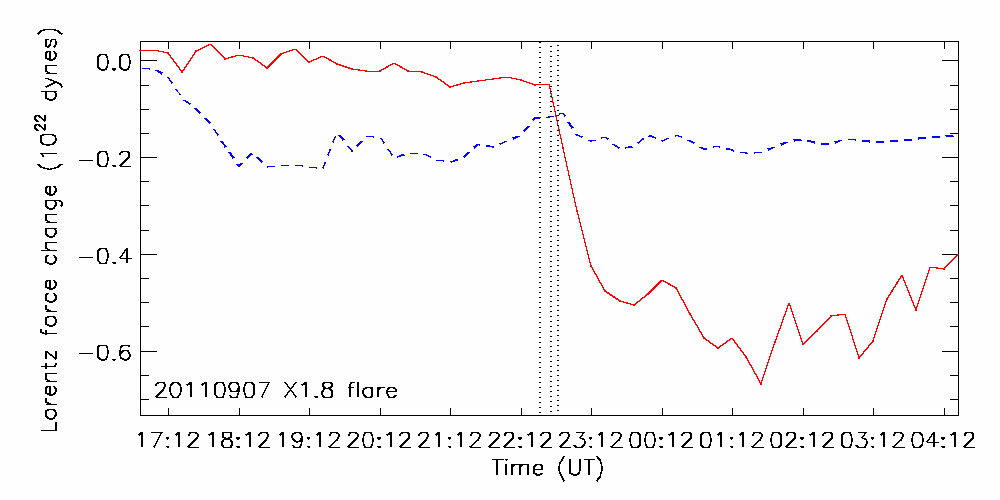}}
\resizebox{0.49\textwidth}{!}{\includegraphics*{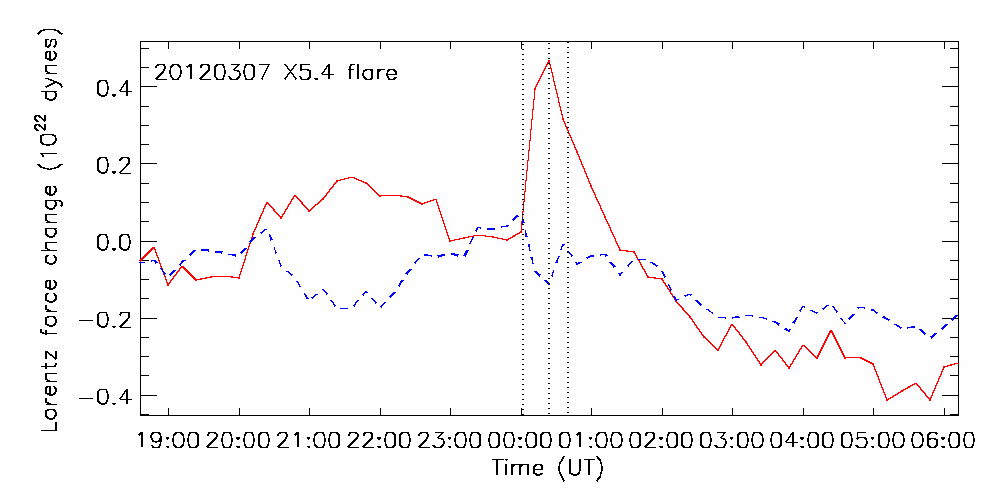}}
\end{center}
\caption{The Lorentz force vector components in the perpendicular direction, $\delta F_{\perp}^\mathrm{NL}$, are plotted as functions of time. The red solid and blue dashed lines represent the force changes acting on the positive and negative sides of the neutral lines, respectively. The areas of integration are indicated by the black rectangles in Figure~\ref{fig:br}. Positive/negative force changes act in the approximately northward/southward directions along the short edges of these rectangles, respectively. The vertical lines represent the GOES flare start, peak and end times.}
\label{fig:dfperpsgnnlt}
\end{figure}

Figure~\ref{fig:dfrnlt} shows the Lorentz force changes $\delta F_r^\mathrm{NL}$ in the vertical direction as functions of time. The red and blue curves in Figure~\ref{fig:dfrnlt} represent positive (upward) and negative (downward) force changes, respectively. They were derived by taking running differences between consecutive image pairs and were integrated over the areas $A_\mathrm{NL}$ represented by the black rectangles shown in Figures~\ref{fig:br}, \ref{fig:dbr} and \ref{fig:dfr}. The  running difference plots in Figure~\ref{fig:dfrnlt} show sharp, spiked signatures of the abrupt downward Lorentz force changes well above the noise levels during all six flares. It is clear from Equations~(\ref{eq:deltafr}) and (\ref{eq:deltafh}) that the Lorentz force changes should resemble the time-derivatives of the field changes. Step-like field changes of brief but finite duration, such as those seen in Figure~\ref{fig:dfr}, produce the brief spikes of Lorentz force change in Figure~\ref{fig:dfrnlt}. The fact that the dominant downward spikes in these plots are not matched by upward changes implies that the force changes were permanent. The black curves in Figure~\ref{fig:dfrnlt} show force changes derived by fixed differences with respect to the first image. These curves confirm that all of the flare-related force changes were permanent. They also show evidence of a steady net upward Lorentz force before three of the flares, the 2011 February 15 X2.2, September 6 X2.1 and September  7 X1.8 flares, perhaps associated with pre-flare inflation of the fields near the neutral lines. Gosain~(2012) described a pre-flare rise phase in a study of AIA observations of the 2011 February 15 X2.2 flare.

The sizes of these force changes, ranging from about $9\times 10^{21}$~dynes to about $6\times 10^{22}$~dynes, are comparable to those found in the previous estimates of flare-related Lorentz force changes by Wang and Liu~(2010) and Petrie and Sudol~(2010). For the 2002 July 26 M8.7 flare Wang and Liu~(2010) found a downward force change of $1.6\times 10^{22}$~dynes. Petrie and Sudol~(2010) found a range of longitudinal force change estimates up to about $2\times 10^{22}$~dynes. Petrie and Sudol's estimates are likely to have been underestimates because they included only information on the longitudinal field component.

Figures~\ref{fig:dfparasgnnlt} and \ref{fig:dfperpsgnnlt} show the Lorentz force vector changes, $\delta F_\parallel^\mathrm{NL}$ and $\delta F_\perp^\mathrm{NL}$, in the horizontal directions parallel and perpendicular to the neutral line as functions of time. The forces are integrated over the areas $A_\mathrm{NL}$ represented by the black rectangles shown in Figures~\ref{fig:br}, \ref{fig:dbr} and \ref{fig:dfr}. The parallel directions are the directions of the long edges of the rectangles, pointing approximately west. The perpendicular directions are the direction of the short edges of the rectangles, pointing approximately north. Figures~\ref{fig:dfparasgnnlt} shows significant and abrupt force changes at the time of every flare. The red and blue curves describe the parallel force changes occurring on the positive and negative sides of the neutral line, respectively. In each case the force changes are directed in opposite directions on the two sides of the neutral line. The corresponding perpendicular force changes, shown in Figure~\ref{fig:dfperpsgnnlt}, were smaller and less significant, and were generally not permanent.

Table~\ref{dftable} summarizes the directions of the parallel and perpendicular Lorentz force changes, $\delta F_\parallel^\mathrm{NL}$ and $\delta F_\perp^\mathrm{NL}$, associated with the six flares. In all cases where permanent changes of Lorentz force were detected on the northern/southern side of the neutral line, the perpendicular force change (columns 7 and 8 of Table~\ref{dftable}) was directed south/north, i.e., towards the neutral line. Both sides of the neutral line were observed to undergo significant, permanent perpendicular force change during only one flare, the 2011 March 9 X1.5 flare. There were no significant, permanent changes on either side of the neutral line during the 2012 March 7 X5.4 flare, and one side underwent permanent change during the four other flares. The parallel force changes (columns 5 and 6 of Table~\ref{dftable}) were more significant, and were detected in both polarities during all flares except the 2011 March 9 X1.5 flare, when the north side of the neutral line underwent significant, permanent change while the south side didn't. In each case when both sides of the neutral line experienced permanent parallel Lorentz force changes, these occurred in opposite directions on each side of the neutral line. In columns 5 and 6 of Table~\ref{dftable} the directions of the positive polarities are in bold typeface. These directions agree perfectly with the axial field directions (column 4). This implies that the positive polarity footpoints of the fields near the neutral line were pulled towards the corresponding negative footpoints. Meanwhile, the negative footpoints were being pulled towards the positive footpoints during five of the six flares, and underwent no permanent change during the remaining flare. Ignoring vertical movements, a convergence of footpoints along the direction parallel to the NL with no movement in the perpendicular direction would produce a decrease of shear, whereas a convergence of footpoints along the direction perpendicular to the NL with no movement in the parallel direction would produce an increase of shear. These interpretations ignore the effects of vertical movements such as loop collapse. The parallel force changes alone would therefore have relaxed the shear of the neutral line fields, whereas the perpendicular forces acted towards the neutral line and would have tended to increase the shear.

The perpendicular force changes were not generally significant compared to the background evolution. Furthermore, comparing the sizes of the force changes parallel and perpendicular to the neutral line, the ratio $\delta F_\perp^\mathrm{NL} / \delta F_\parallel^\mathrm{NL}$ is less than 0.1 for the large flares and less than 0.5 for the small flares. For comparison, the ratio $B_\perp^\mathrm{NL} / B_\parallel^\mathrm{NL}$ ranges from about 0.3 to about 0.5 (Figures~\ref{fig:fparanlt} and \ref{fig:fperpnlt}). The horizontal force changes therefore can't explain the changes of azimuthal difference and shear angles seen in Figures~\ref{fig:azimdiffnlt}, \ref{fig:shear} and \ref{fig:shearh}. They likely represent the contraction of the collapsing, line-tied loop fields, trying to pull the two sides of the neutral line towards each other.



\section{Conclusion}
\label{s:conclusion}

In 12 hour time series of 12-minute SDO/HMI vector field observations covering six major flares we have found consistent patterns in the field and Lorentz force vector changes near the main magnetic neutral lines of the flaring regions. We summarize the main results before drawing conclusions from them.

\begin{enumerate}

\item Near the main magnetic neutral lines, the field vectors became stronger and more horizontal during all six flares. This was almost entirely due to an increase in strength of the horizontal field components parallel to the neutral line in each case. The horizontal perpendicular and vertical components did not show comparably significant and permanent changes during the flares. Even during the 2011 September 7 flare, when the vertical field did change significantly and permanently, the change in the horizontal parallel field was more significant.

\item The magnetic tilt angle increased significantly during all flares with the arguable exception of the 2011 September 7 flare. The neutral line fields were less tilted than their corresponding potential fields before all six flares and relaxed abruptly and permanently closer to potential field tilt angles during every flare.

\item The horizontal fields became significantly and permanently more aligned with the neutral line during the four largest flares. The (full and horizontal) shear angle with respect to the reference potential field increased significantly during only two of the flares.

\item The electric current near the neutral line showed a marked change of behavior around the times of most of the flares but there was no consistent pattern in the character of this change.

\item The vertical Lorentz force had a large, abrupt, permanent downward change during each of the flares. The sizes of these force changes, ranging from about $9\times 10^{21}$~dynes to about $6\times 10^{22}$~dynes, are comparable to those found in the previous estimates of flare-related Lorentz force changes.

\item The horizontal Lorentz force component parallel to the neutral line showed significant, permanent changes acting in opposite directions on each side of neutral line during each of the six flares, comparable in size to the vertical force changes. In all cases the shearing forces were consistent with a decrease of shear near the neutral line, whereas the field itself became more aligned with the neutral line as a result of the increase in the horizontal field strength. The horizontal force changes perpendicular to the neutral line tended to act towards the neutral line but these were not as significant as the parallel changes.

\end{enumerate}

The abrupt and permanent increase in magnetic field strength near the neutral line during a flare, accompanied by corresponding increase in magnetic tilt and downward Lorentz force change, is most easily explained by the flux near the neutral line being compressed from above. Because the changes in the horizontal field were generally not accompanied by corresponding changes in the vertical field, the resulting changes of field tilt cannot have been the result of simple rotation of the magnetic vector towards the neutral line but must have come from the collapse of nearly horizontal magnetic field towards the photospheric magnetic neutral line from the surrounding volume. In each case the pre-flare field was less tilted than the reference potential field and abruptly collapsed during the flare to a tilt angle much closer to that of the potential field. This implies that, before each flare, the field was supported in a more vertical configuration and relaxed closer to a potential-field tilt angle during each flare. The large downward Lorentz force changes observed during all six flares are consistent with the interpretation of sheared loop collapse.

We find that the tilt angles of the observed fields with respect to the potential fields has a stronger increasing pattern than the tilt with respect to the vertical direction, whereas the behavior of the azimuthal angles of the observed fields is more consistent with respect to the neutral line than with respect to the potential fields. The observed field change during each flare was dominated by an abrupt, permanent and significant increase in the horizontal component parallel to the neutral line, and horizontal field became significantly more aligned with the neutral line during the four largest flares in the study. These results imply that the collapsing field was almost aligned with the neutral line in each case, and more aligned on average with the neutral line than the pre-flare neutral line field. On the other hand, the average shear angle increased during two flares and decreased during one, demonstrating that the flare process does not generally relieve magnetic stresses associated with magnetic shear. The lack of a general pattern in the vertical electric current changes also reflects this. The change in parallel horizontal fields and increased alignment between neutral line fields to neutral lines appear simply to be consequences of the relaxation of stresses associated with the non-potential tilt angles.

Wang~(2006) found an unshearing movement parallel to the neutral line in flare-related longitudinal magnetic field changes in all five events that he studied, consistent with overall release of shear. In our results, the un-shearing pattern of the generally dominant parallel horizontal Lorentz force changes would by themselves have reduced the shear of the field near the neutral lines during each flare, but the magnetic shear only decreased significantly during one of the six flares.  These un-shearing patterns can be interpreted as a signature of the neutral line fields contracting during the flare, pulling the two sides of the neutral line towards each other. It seems that the horizontal force changes did not prevail in un-shearing the neutral line fields because the line-tying effect of the dense photospheric plasma makes it much more difficult to move photospheric footpoints laterally than to change their tilt angle. This may explain the differing physical effects of the horizontal and vertical Lorentz forces: vertical, tilt-related magnetic stresses appear to be much more easily relieved by flares than horizontal, shear-related stresses.

Nonlinear force-free field extrapolations have presented a consistent picture of a sheared structure collapsing towards the neutral line, leaving a void above that is filled by more relaxed field (Jing et al.~2008, Sun et al.~2012, Liu et al.~2012). This modeling work suggests that if magnetic shear increases at low altitudes during flares it may decrease above a certain height, producing a net decrease of shear in the system as a whole. Our results suggest that the main flare-related field changes are caused by the release of magnetic stresses associated with non-potential tilt/dip angles, and that the consequent strengthening of the parallel horizontal field component may or may not increase photospheric magnetic shear. The scenario of sheared loop collapse developed above would lead to a reduction of shear higher in the atmosphere in agreement with the modeling results.

We have only studied the effects of major flares on fields near neutral lines in this paper. While clear patterns have emerged from this sample, it will be instructive to examine a large sample of high-cadence vector-field measurements from HMI and the National Solar Observatory's Synoptic Optical Long-term Investigations of the Sun (SOLIS) telescope to discover if these patterns are generally dominant, and whether an analogous pattern can be found in the flare-related magnetic changes observed in sunspots.


\acknowledgements
I thank the referee for a stimulating review that helped to develop the project. I thank Sanjay Gosain and Alexei Pevtsov for discussions. SDO is a mission for NASA's Living With a Star program. This work was supported by NSF Award No. 106205 to the National Solar Observatory.

\end{document}